\documentclass[acmtog,nonacm]{acmart}

\usepackage{booktabs} 

\usepackage[ruled,vlined,noend]{algorithm2e} 

\usepackage{enumitem}
\PassOptionsToPackage{table}{xcolor}
\usepackage{colortbl}
\usepackage{array}
\usepackage{hyperref}

\SetAlFnt{\small}
\SetAlCapFnt{\small}
\SetAlCapNameFnt{\small}
\SetAlCapHSkip{0pt}

\acmJournal{TOG}

\newcommand{\M}{\mathcal{M}}
\newcommand{\V}{\mathcal{V}}
\newcommand{\E}{\mathcal{E}}
\newcommand{\F}{\mathcal{F}}
\newcommand{\T}{\mathcal{T}}

\newcommand{\X}{\ensuremath{\mathbf{x}}}
\newcommand{\U}{\ensuremath{\mathbf{u}}}
\newcommand{\Sf}{\ensuremath{\mathbf{s}}}
\newcommand{\phii}{\ensuremath{\mathbf{\phi}}}

\newcommand{\hdg}[1]{\vspace{.7ex}\noindent\textbf{#1}\hspace{0.5ex}}

\begin{document}
\title{Learned Adaptive Mesh Generation}
\author{Zhiyuan Zhang}
\affiliation{
  \institution{University of Edinburgh}
  \country{UK}
}

\author{Amir Vaxman}
\affiliation{
  \institution{University of Edinburgh}
  \country{UK}
}

\author{Stefanos-Aldo Papanicolopulos}
\affiliation{
  \institution{University of Edinburgh}
  \country{UK}
}

\author{Kartic Subr}
\affiliation{
  \institution{University of Edinburgh}
  \country{UK}
}



\begin{teaserfigure}
	\begin{center}		
        \includegraphics[width=0.99\columnwidth]{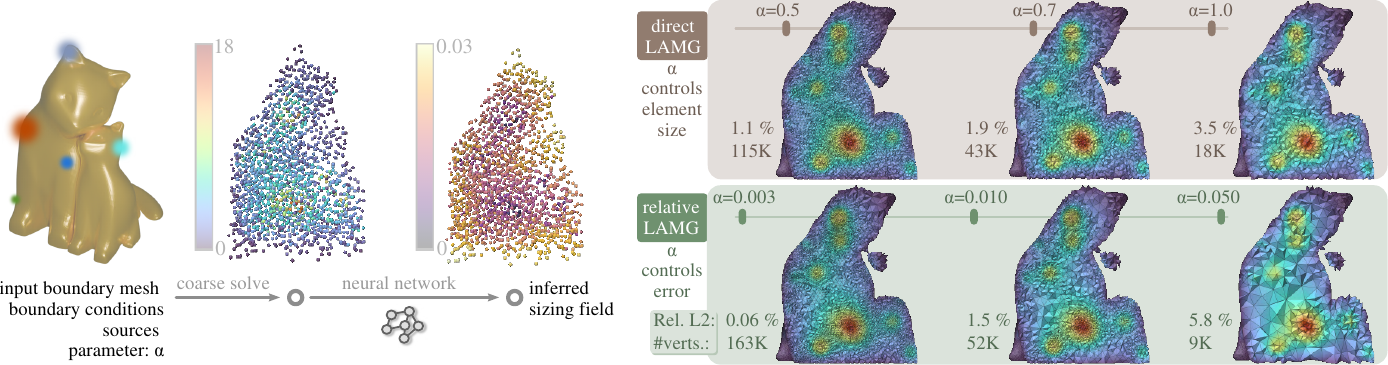} 
\caption{\label{fig:teaser}We propose a method called Learned Adaptive Mesh Generation (LAMG) for generating adaptive tetrahedral meshes to solve linear elliptic PDEs, specifically the Poisson equation, in 3D domains. Given a triangle mesh of the boundary of the domain, boundary conditions and internal source terms that define the PDE, we demonstrate that solving the PDE via FEM applied to the tetrahedral mesh produced by LAMG is more efficient than existing iterative refinement methods, for similar levels of error. An input parameter $\alpha$ controlls the approximation either by scaling the sizes of elements (direct LAMG) or by acting as an input error tolerance (relative LAMG). The figure illustrates our approach (left) and provides examples of solutions obtained using our method. 
%
}
	\end{center}
\end{teaserfigure}

\begin{abstract}
    Elliptic Partial Differential Equations (PDEs) play a central role in computing the equilibrium conditions of physical problems (heat, gravitation, electrostatics, etc.). Efficient solutions to elliptic PDEs are also relevant to computer graphics since they encode global smoothness with local control leading to stable, well-behaved solutions. The \emph{Poisson} equation is a linear elliptic PDE that serves as a prototypical candidate to assess newly-proposed solvers. Solving the Poisson equation on an arbitrary 3D domain, say a 3D scan of a turbine’s blade, is computationally expensive and scales quadratically with discretization. Traditional workflows in research and industry exploit variants of the finite element method (FEM), but some key benefits of using Monte Carlo (MC) methods have been identified. 
    
    Our key idea is to exploit a sparse and approximate solution (via FEM or MC) to the Poisson equation towards inferring an adaptive discretization in `one shot'. We achieve this by training a lightweight neural network that
    generalizes across shapes and boundary conditions. Our algorithm, Learned
    Adaptive Mesh Generation (LAMG), maps from a coarse solution to a sizing field that defines a local (adaptive) spatial resolution. This output space, rather than directly predicting a high-resolution solution, is a unique aspect of our approach. We use standard methods to generate tetrahedral meshes that respect the sizing field, and obtain the solution via one FEM computation on the adaptive mesh. That is, we our neural network serves as a surrogate model of a computationally expensive method that requires multiple (iterative) FEM solves.
    We demonstrate the versatility, controllability, robustness and efficiency of LAMG via systematic experimentation. 
\end{abstract}

%
%


\maketitle

\section{Introduction}\label{sec:intro}

The evolution of quantities over space and time in the real world is described mathematically using Partial Differential Equations (PDEs). 
Solutions to these equations yield spatio-temporal functions that characterize the quantity being modelled, e.g. light, heat, deformation, or stock prices.
There is a long history of computational schemes to solve various classes of PDEs across physics~\cite{griffiths2023introduction,baron1822theorie,mclean2012continuum,chyuan2003efficient}, engineering~\cite{durka1996structural,ZTZ:2013} and computer graphics~\cite{bridson2015fluid,liu2016towards,fedkiw2001visual}.
Still, it can be computationally expensive to solve even a seemingly simple PDE, such as the Poisson equation~\cite{poisson1826memoire}, when the spatial domain is defined by an arbitrary 3D mesh, say a 3D scan of a turbine blade. We propose an efficient approach to solve Poisson equations in 3D domains defined within an input boundary mesh. 
 
The Poisson equation is a classic second-order PDE that is used to model systems after they have attained equilibrium. It is classified as an \emph{elliptic PDE} since the coefficients of its second-order terms satisfy the same criteria as those in the equation of an ellipse. This structure leads to smooth, well-behaved solutions that are characteristic of steady-state solutions. There are two relevant categories of methods to solve Poisson equations in arbitrary 3D domains (meshes).  The \emph{finite element method} (FEM) discretizes the domain, and solves the problem using a global assembly of local basis elements. The Monte Carlo (MC) method~\cite{kakutani1944143, muller1956some, Sawhney:2020:MCG, sabelfeld1995integral} solves the equation at selected points in the domain by using random walks from these points to the boundary. 

FEM is seen as the \emph{de facto} standard method and is deeply embedded in most workflows requiring the solution of PDEs. Its naïve version faces several challenges such as sensitivity to the mesh, difficulties generating quality meshes in 3D, computational cost, and handling boundary conditions whose discontinuities do not align with the mesh. Several works~\cite{hughes2005isogeometric, shewchuk2002good,he2021mesh,cervera2022comparative} extend FEM to address (subsets of) these criteria. \emph{Adaptive Mesh Refinement} (AMR) addresses this by adaptively refining mesh elements (tetrahedra in 3D) where the approximation error is high. This is typically achieved using recovery-based error estimators, such as Zienkiewicz-Zhu (ZZ)~\cite{zienkiewicz1992superconvergent1,zienkiewicz1992superconvergent2}, which approximate the error by comparing the discontinuous gradients (fluxes) inherent to compatible FEM discretizations against a recovered smooth gradient field. Since the Poisson equation governs diffusion phenomena implies continuous fluxes, large gradient discrepancies across element boundaries serve as a robust indicator for refinement. While AMR can be tuned via an input error tolerance to yield accurate solutions, it often results in large meshes with irregular arrangements of elements. A major drawback is that this identification relies on \emph{a posteriori} error estimation, which necessitates repeated FEM solves in an iterative loop until the desired accuracy is attained.

Monte Carlo methods, on the other hand, are gaining popularity due to their robustness and independence of 3D discretization. They estimate the solution at a query point in the domain by performing many random walks from that point to the boundary. However, since they are inherently suited to \emph{point-sampled} estimates of the solution function, obtaining a dense solution over the continuous domain is computationally expensive. Monte Carlo methods are not part of the standard engineering toolbox. They are an active area of research and therefore their application has not been generalized across different types of PDEs.

Our primary goal is to design a method that computes the solution to Poisson equations over the continuous domain, using an adaptive mesh without the inefficiency of iterative refinement. The scalar \emph{sizing field}, or function that defines the local mesh resolution, is smooth for elliptic PDEs and is therefore amenable to learning. Despite its smoothness, learning to predict the sizing field directly from the definition of a Poisson equation---generalizing across 3D domains, boundary conditions and internal source terms---would be non-trivial and would rely on the training examples. Instead, our \emph{key insight} is to first rapidly compute a coarse solution to the PDE problem and then learn a mapping from the coarse solution to the required sizing field. We calculate the final solution to the PDE problem using FEM applied to an adaptive mesh that respects the predicted sizing field. 

Our method, Learned Adaptive Mesh Generation (LAMG) avoids exepnsive posteriori sequential FEM solving by predicting a sizing field in one shot using a lightweight neural network. Our results demonstrate the versatility, effectiveness, efficiency and robustness of our method. 
The input to our algorithm is a Poisson equation (with boundary conditions and source terms), an arbitrary 3D domain specified by a watertight boundary mesh and a variable that controls the computational budget (number of elements or input error tolerance). LAMG has the following features:
\begin{itemize}[topsep=.2em, leftmargin=1em]
    \item the sizing field is directly predicted from a coarse solution; 
    \item the coarse solution may be obtained via FEM or Monte Carlo; 
    \item an input tolerance, or target error, may be specified; 
    \item the network has a few thousand learnables (trained in $<2$h); and
    \item the final solution uses classical FEM and is therefore reliable.
\end{itemize}
We demonstrate the effectiveness of our method in calculating the steady state of heat diffusion in arbitrary meshes with non-trivial boundary conditions (Section~\ref{sec:exp}), linear elastic deformation (Section~\ref{sec:otherexp}) and analyze its strengths and weaknesses (Section~\ref{sec:discussion}).

\section{Background} \label{sec:relwork}

In this section, we review prior work on adaptive meshing and Monte Carlo methods to contextualize our contributions. We then establish the mathematical notation and governing formulations for the elliptic PDEs addressed in this paper.

\subsection{Related work}
\hdg{Finite Element Method}The Finite Element Method (FEM) is a mathematical method for obtaining approximate solutions to boundary value problems, working on a discrete mesh over the problem's domain. It is a well-established method, with extensive scientific literature (including major textbooks, e.g.~\cite{ZTZ:2013}), and is widely adopted across science and engineering, especially in fields such as solid mechanics. A large number of computer implementations exist, as libraries and/or as standalone code (see e.g. open-source repository~\cite{kostyfisik2023fea} for comparisons). 
%
FEM solves a weak, integral form of the governing equations, and uses mesh discretizations based on simple element shapes (e.g.\ tetrahedra or hexahedra in 3D), where the solution is interpolated using appropriate shape functions. While $p$-FEM approaches exist~\cite{Babuska1981}, where accuracy is increased by increasing the order of the interpolating shape functions within the element, in the prevalent $h$-FEM approach accuracy is obtained by appropriate reduction of the element sizes (and thus increase of the mesh density). The generation of appropriate meshes, especially in 3D problems, therefore plays an important role in both the accuracy and the computational efficiency of the method.

\hdg{Domain Discretization}FEM’s reliance on discretization makes mesh resolution a key bottleneck. Uniform meshes are simple but inefficient, motivating adaptive meshing that concentrates elements where needed. \emph{Adaptive Mesh Generation (AMG)} is a single-step procedure that generates a mesh guided by spatial resolution demands. These demands can be derived from geometric cues (e.g., surface curvature~\cite{Shewchuk:98, si2008adaptive,secco2021efficient}) and/or from estimated simulation error~\cite{yano2012optimization}. The common strategy involves building a background size field that defines target element sizes throughout the domain. Meshing tools like \textit{gmsh}~\cite{geuzaine2009gmsh}, TetWild~\cite{Hu:2018}, fTetWild~\cite{Hu:2020}, and TetGen~\cite{hang2015tetgen} use this field to produce a high-quality tetrahedral mesh. In this work, we use \textit{gmsh} for all of our meshing. Since AMG regenerates the entire mesh, it is often used for static configurations or as an initial mesh in time-dependent problems. In contrast to AMG, \emph{Adaptive Mesh Refinement (AMR)} is a dynamic and iterative process. AMR is a well-established topic (e.g., ~\cite{BERGER1984484, BERGER198964}), propelled by the need to increase resolution in places with high turbulence for hyperbolic equations. They usually targets adaptive grids, but has been applied to tetrahedral meshes as well~\cite{park2016unstructured}. Starting from a coarse mesh, AMR locally refines areas where the solution exhibits large errors or gradients. Traditional meshing is commonly done using local operations like edge and face splitting and edge flips. These often result in anisotropic elements and deteriorating mesh quality over multiple refinement cycles. Despite this, AMR remains attractive due to its precision and resource efficiency. Closer to our approach is AMR refinement with the ZZ estimator~\cite{zienkiewicz1992superconvergent1, zienkiewicz1992superconvergent2}, which estimate an a posteriori error measure using the predicted error in the elasticity energy. Alternatively, residual-based estimators~\cite{ainsworth1997posteriori} are often employed for their rigorous error bounds and numerical stability, with extensive adaptations developed for domain-specific applications~\cite{frey2005anisotropic}. Recent extensions to AMR includes AMG methods ~\cite{freymuth2024iterative}. Beyond purely spatial adaptation, other modern approaches seek to improve efficiency in different ways. For instance, some coupled approaches integrate remeshing directly into the timestep solve for transient problems~\cite{ferguson2023timestep, wen2025optimal}. Furthermore, other methods propose dynamically switching between different simulation models entirely to trade accuracy for speed~\cite{trusty2024trading}. 

\hdg{Learning-based methods} Given the cost of iterative AMR, there is growing interest in \emph{learning-based methods} to accelerate or replace traditional refinement pipelines. Pfaff \textit{et al.}~\shortcite{pfaff2020learning} learned a re-meshing for cloth simulation. This method directly learns the simulation which achieves visual-plausibility but cannot gracefully trade-off accuracy for speed. Some works use reinforcement learning for AMR~\cite{foucart2023deep, freymuth2023swarm,yang2023reinforcement} using sophisticated reward functions that are limited to simple 2D shapes. Some approaches overcome this using surrogate models in supervised-learning~\cite{chen2021output,zhang2020meshingnet,zhang2021meshingnet3d}. However, their simplified parameterization of the problem limits their work to single PDEs and a narrow range of shapes. Imitation learning has been explored to solve the problem~\cite{freymuth2024iterative}, but scaling the production of reference data is expensive since it relies on expert knowledge. Furthermore, Legeland \textit{et al.}~\shortcite{legeland2025non} proposed a non-iterative prediction of optimized mesh. The input to their network include the PDEs, so their method implicitly learns the solution to the PDE and therefore does not generalize across shapes and boundary conditions.

\hdg{Monte Carlo Methods}Monte-Carlo methods have been used to solve PDEs for decades~\cite{muller1956some,sabelfeld1995integral} but MC goemetry processing~\cite{Sawhney:2020:MCG} has recently gained popularity as an alternative to FEM~\cite{sugimoto2023practical, Sawhney:2020:MCG, rioux2022monte}. Rather than computing a solution on a discretization of the domain, this class of methods estimate the solution at each query point in the domain by stochastically sampling random walks to the boundary. They have minimal requirements on the input geometry, since they operate on a boundary mesh, but they are susceptible to slow convergence and noise depending on the PDE being solved.  The central idea has been extended to general boundary value problems~\cite{sabelfeld1995integral} and to different types of PDEs including fluid simulation, heat diffusion and elasticity~\cite{rioux2022monte, sabelfeld2016random,sabelfeld2002random} . We exploit the ability of these methods to yield sparse solutions in the domain, and use it as the input to our learning framework.

\subsection{Review and notation: Solving the elliptic PDEs}
Poisson's equation is the quintessential elliptic PDE because it defines the mathematical archetype for steady-state equilibrium. While other classes model change over time, it describes how a spatial field responds instantly to sources.
Mathematically, it serves as the simplest second-order elliptic operator, exhibiting characteristic properties like global dependence and maximum smoothness. Physically, it is the foundational model for gravity, electrostatics, and heat distribution. Consequently, it remains the primary benchmark for developing theoretical proofs and modern numerical solvers.

\paragraph{Poisson's Equation} We consider a continuous domain $\Omega \in \mathbb{R}^3$ of an arbitrary genus and number of boundary surfaces, denoted $\partial \Omega$. We focus on solving the Poisson problem with Dirichlet boundary conditions. Consider the Laplace-Beltrami operator $\Delta$. We solve:
\begin{align}
\nonumber \Delta u &= f \quad \text{on } \Omega, \\ 
u &= g \quad \text{on } \partial \Omega,
\end{align}
for provided $f:\Omega \setminus \partial \Omega \rightarrow \mathbb{R}$ and $g:\partial \Omega \rightarrow \mathbb{R}$. 

\paragraph{The Linear Elasticity Equation} For a continuous elastic body $\Omega$, we solve for a vector-valued displacement field $u: \Omega \rightarrow \mathbb{R}^3$. The governing equation under static equilibrium requires that the internal stresses, represented by the stress tensor $\sigma(u)$, balance any applied body forces $f: \Omega \rightarrow \mathbb{R}^3$. This is expressed as:
\begin{align}
\nonumber \nabla \cdot \sigma(u) + f &= 0 \quad \text{on } \Omega, \\
u &= g_D \quad \text{on } \partial\Omega_D, \\
\nonumber \sigma(u) \cdot n &= g_N \quad \text{on } \partial\Omega_N,
\end{align}
where the boundary $\partial\Omega$ is partitioned into a Dirichlet part $\partial\Omega_D$ with prescribed displacements $g_D$, and a Neumann part $\partial\Omega_N$ with prescribed traction forces $g_N$. The stress $\sigma$ is linearly related to the strain $\varepsilon(u) = \frac{1}{2}(\nabla u + (\nabla u)^T)$.

Henceforth, we collectively denote the full problem (the governing equation, domain, source, and boundary conditions) as $\mathfrak{B}$.

\paragraph{Piecewise-Linear (PL) Finite elements} For an FEM solver, we work with a tetrahedral mesh $\M = \left\{\V, \E, \F, \T\right\}$ that tessellates the domain $\Omega$. The mesh consists of a set of vertices $\V$, edges $\E$ connecting pairs of vertices, triangular faces $\F$ formed by triplets of vertices, and tetrahedral elements $\T$ defined by quadruplets of vertices. The core idea is to approximate the continuous solution $u$ using conforming piecewise-linear basis functions defined over the mesh elements $\T$. This procedure discretizes the continuous PDE, transforming it into a large system of linear equations to be solved for the unknown values at the vertices $\V$.

For the Poisson equation, this results in a system of the form $Lu = Mf$, where $L$ is the cotangent Laplacian matrix, $u$ contains the unknown scalar values at each vertex, and $Mf$ represents the discretized source term. 
For the linear elasticity equation, this yields a system of the form $Ku = F$, where $K$ is the global stiffness matrix representing the material's properties, $u$ is the vector of unknown displacements at each vertex, and $F$ is the vector of nodal forces.

\paragraph{Monte-Carlo (MC) Walk on Spheres (WoS) Poisson solver} We use the formulation of~\cite{Sawhney:2020:MCG}. In this approach, the domain $\Omega$ is not meshed. Instead, the Poisson equation is solved independently per point. To do this, from every point $x \in \Omega$, we approximate the solution by an estimator based on the walk-on-spheres (WoS) algorithm, which reaches a random point on the boundary. The estimator then uses the mean-value property and estimates the value at the point as an average of the samples.

\definecolor{mybrown}{RGB}{139,69,19} 

\begin{figure*}
  \begin{tabular}{m{.39\linewidth} p{.65\linewidth}|}
    \includegraphics[width=\linewidth]{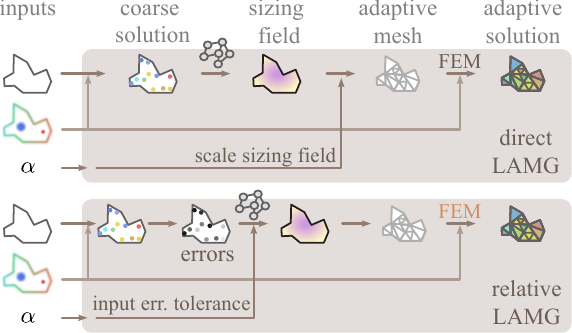}
	&
	\begin{tabular}{c}
		\arrayrulecolor{white}
		\begin{tabular}{|>{\columncolor{mybrown!10}}c|>{\columncolor{mybrown!4}}c|>{\columncolor{mybrown!10}}c|}
			 \hyperref[sec:method]{\textbf{Method}} &
			 \hyperref[sec:exp]{\textbf{Experiments (heat)}} &
			 \hyperref[sec:otherexp]{\textbf{Extended experiments}} 
			 \\ 
			 \hline
			 \hyperref[sec:approach]{overview of approach} &  
			 \hyperref[sec:baselines]{baselines, comparisons} &  
			 \hyperref[sec:wos]{using coarse WoS}   
			 \\
			 \hyperref[sec:directrelative]{direct/relative LAMG} &  
			 \hyperref[sec:training]{data \& training} &  
			 \hyperref[sec:elasticity-direct]{linear elastic deformation}   
			 \\
			 \hyperref[sec:nwarch]{network details} &  
			 \hyperref[sec:expdirect]{results w/ direct LAMG} &  
			 \hyperref[sec:heterogeneous]{heterogeneous materials}   
			 \\			 			 
			 &
			 \hyperref[sec:exprelative]{results w/ relative LAMG}   
			 &
		\end{tabular} 		
		\\	
		\\	
		\\	
		\begin{tabular}{|>{\columncolor{mybrown!10}}c|>{\columncolor{mybrown!4}}c|>{\columncolor{mybrown!10}}c|>{\columncolor{mybrown!10}}c|}
			 \textbf{Heat (direct)} &
			 \textbf{Heat (relative)} &
			 \textbf{Linear elasticity} &
			 \textbf{WoS} 
			 \\ 
			 \hline
			 Fig.~\ref{fig:direct:Poisson} (quant.)&  
			 Fig.~\ref{fig:relative:Poisson} (quant.) &  
			 Fig.~\ref{fig:direct_elasticity} (quant.) &  
			 Figs.~\ref{fig:wosvsfem},~\ref{fig:robustness} (direct)			
			 \\
			 Fig.~\ref{fig:qualitative} (qual.)&  
			 Fig.~\ref{fig:qualitative_[relative]} (qual.) &  
			 Fig.~\ref{fig:qual_elasticity} (qual.)&
			 Fig.~\ref{fig:relative:PoissonWOS} (relative)
		\end{tabular} 		
	\end{tabular}
  \end{tabular}
\caption{\label{fig:ovw} This figure serves as an overview to this paper. \emph{Left:} We propose two variants (direct and relative) that permit different control of the output error via parameter $\alpha$. The direct method controls the number of elements while the relative method allows controlling output error. \emph{Right:} Hyperlinked (clickable) shortcuts to the different sections (top) and figures (bottom) allows easy navigation of this article.}
\end{figure*} 

\section{Method}\label{sec:method}
The input to our method is a boundary mesh (watertight 2-manifold). 
The core method aims to learn a sizing field that mimics adaptive meshing using AMR in one shot using full supervision. At inference time, we use this output sizing field to obtain an adaptive mesh using \textit{gmsh}~\cite{geuzaine2009gmsh}. Finally, we solve the elliptic PDE over the 3D domain using an standard FEM solver MFEM~\cite{anderson2021mfem} on the adaptive mesh. 

Our method takes a PDE problem, a boundary mesh and an input parameter and outputs an adaptive mesh with the solution to PDE problem using piecewise linear elements over the mesh (see Section~\ref{sec:approach}). The input parameter is used to control computational efficiency. Its interpretation depends on which of the two variants, direct or relative method (described in Section~\ref{sec:directrelative}) is used. 


\subsection{Approach}\label{sec:approach}
Given a PDE problem $\mathfrak B$, a boundary mesh $\partial \M$ and a parameter $\alpha$ to control the computational budget, we first sample a sparse set of data points $X=\{\X_i\}_{i=1}^n$, where $\X_i \in \mathbb{R}^3$ is inside the watertight domain and use a coarse solution to the PDE to obtain a corresponding scalar feature $\phii_i\in \mathbb{R}$ at each $\X_i$. This feature is either directly the degrees of freedom of the coarse solution or from estimated error from the coarse mesh solution using ZZ error estimator. We term these two variants as \emph{Direct} and \emph{Relative} methods respectively. 

The sparse data $\{\X_i, \phii_i\}$ is input to our trained neural network, $h_\theta$, which predicts a sizing field $S=\{\Sf_i\}_{i=1}^n$ defined on each $\X_i$. Finally, this sizing field is used by \textit{gmsh} to generate a high-quality adaptive mesh $\M_a$, on which a single, final FEM solve is performed to get the accurate solution $\U_a$. This general procedure is summarized in Algorithm~\ref{alg:LAMG}. 

\begin{algorithm}[h]
	\caption{Adaptive FEM with LAMG}
	\label{alg:LAMG}

	\SetKwInOut{Input}{inputs}
	\SetKwInOut{Output}{output}
	
	\Input
	{		
		\par 
		\Indp
        $\partial \M$, $\;\mathfrak{B}$, $\;\alpha$ \tcp*[r]{Boundary mesh, PDE, parameter}
        }

	\Output
	{
		\par 
		\Indp
        $\U_a$ \tcp*[r]{Learned adaptive solution to $\mathfrak{B}$} 
	}
	\vspace{.25em}
	{\hrule}
	\vspace{.45em}

    $\{X, \Phi\} \gets$ GenerateCoarseSolution ($\mathfrak{B}, \;n$)  

    $S \gets$ InferSizingField ($X, \;\Phi, \;\alpha$) \tcp*[r]{s.f. per $\X_i$ }

    $\M_a \gets$ AdaptiveTetMesh ($\partial\M, \{\Sf_i\}$) \tcp*[r]{\textit{gmsh}}
    
    $\U_a \gets$ FEMSolve ($\mathfrak{B}, \M_a$) \tcp*[r]{Single  solve}
\end{algorithm}

The network used and the last two steps of Algorithm~\ref{alg:LAMG} remain common to both our variants. A neural network is used within the function $\texttt{InferSizingField}(\cdot)$. The network $h_\theta(X,\Phi)$  is a simple, lightweight learning framework using a classic encoder-decoder architecture with a Graph Neural Network (GNN) as processor~\cite{hamilton2018representationlearninggraphsmethods}, where $\Phi$ is first encoded into a higher-dimensional ($16D$) feature vector at each node. Then, a GNN uses message passing to diffuse information across the graph in the encoded space. Finally, a decoder maps the aggregated features into a scalar value at each node of the graph---the predicted \emph{sizing field}. 
It is trained using supervisory sizing field using classical AMR (details in Section~\ref{sec:training}).
Given $X = \{\X_i\}_1^n$  with dimensions $n\times3$ and $\Phi=\{\phii_i\}_1^n$ with dimensions $n\times p$ ($p=1$ or $p=2$ as explained in Section~\ref{sec:directrelative}), we construct an undirected graph $G = (X, E)$ where the nodes are at the samples (rows) of $X$ and an edge $e_{ij} \in E$ exists if $\X_j$ is in the $k-$nearest neighbors (kNN) of $\X_i$ and if $e_{ij}$ is fully contained in the domain. Thus, we have graph $G$ with nodes $X$ and a scalar value
at each node, represented collectively by $\Phi$. Finally, we remesh with \textit{gmsh} and solve on the adaptive mesh with FEM.

\subsection{Direct versus relative approach}\label{sec:directrelative}
We devised two variants of the learned mapping $h_\theta$, which we call the direct and relative methods, with mutually different advantages. The former estimates the sizing field from the coarse solution field (temperature at each $\X_i$) while the latter mimics the classical use of adaptive FEM approaches by using an \emph{input error tolerance} (usually called target error). We now explain how each of these two variants  implements the general approach of Algorithm~\ref{alg:LAMG} slightly differently.

\paragraph{Direct method}
The function $\texttt{GenerateCoarseSolution}(\cdot)$ in Algorithm~\ref{alg:LAMG} returns a scalar $\phi_i = \U_i$ at each $\X_i$ and assembled into an $n\times 1$ matrix $\Phi$. Then, the sizing field is computed via $S = \eta \cdot h_\theta(X,\Phi)$ where the scaling parameter is set to the input parameter ($\eta = \alpha$). Thus the computational budget can be controlled via $\alpha$; scaling the sizing field up ($\alpha>1$) leads to a larger sizing field and hence the adaptive elements are uniformly scaled up, leading to higher error and faster runtime. Conversely, setting $\alpha<1$ results in scaling down all elements, lowering error and increasing the runtime.  

The primary advantage of the direct method is that it is agnostic to how the coarse solution was computed for $\phi_i = \U_i$. For instance, if a Monte Carlo method (e.g., Walk on Spheres) was used to solve the PDE at a sparse set of locations, the solutions can directly be used to infer an adaptive sizing field. Although the sizing field can be scaled by varying $\alpha$ (or via some transfer function $\eta = \psi(\alpha)$), it does not permit direct control over the resulting error in $\U_a$ (see Algorithm~\ref{alg:LAMG}). 

\paragraph{Relative method}
Here, the function $\texttt{GenerateCoarseSolution}(\cdot)$ in Algorithm~\ref{alg:LAMG} returns a 2-element vector $\phi_i=[\epsilon_i, \epsilon_{\mathrm{tol}}]^T$, at each $\X_i$. $\epsilon_i$ is the estimated error computed by a ZZ estimator~\cite{zienkiewicz1992superconvergent1,zienkiewicz1992superconvergent2} at each tetrahedron's center and $\epsilon_{\mathrm{tol}}=\alpha$ specifies the input tolerance that drives the target or termination criteria for adaptive refinement. The target error is a standard input to classical AMR.

Since the relative method infers the sizing field from estimated errors, rather than directly from degrees of freedom, it is more robust. In addition, it allows straightforward control of the output error. However, since it needs estimated errors $\epsilon_i$ from the coarse solution, it is inherently reliant on using an FEM pipeline to compute the coarse solution. See Section~\ref{sec:discussion} for a workaround, and how WoS may also be used. 

In summary, the direct method is useful when the application demands control over computational budget (number of elements, by scaling $S$ via $\alpha$) while the relative method is suited to applications where controlling error (setting $\alpha$ as the tolerance) is more important. A major benefit of LAMG is its versatility to be trained in either setting, with minimal changes to the overall procedure or architecture. In either variant, the mapping learned by the network does not operate on the entire PDE problem, but on properties of its solution field. This simplifies the network architecture and the training procedure.

\subsection{Network architecture and loss function}\label{sec:nwarch}
We use similar architectures for the direct and relative methods. The encoder and decoder are simple 2-layer fully connected feedforward networks, or multilayer perceptrons (MLP), with a ReLu and a linear layer. They map into and out of a 16-dimensional embedded space in which a GNN operates.
Let the input feature matrix be $\Phi \in \mathbb{R}^{n \times d}$, where $d=1$ for the direct method and $d=2$ for the indirect method. 
The encoder layers map the dimensions of $\Phi$ from $n\times d  \rightarrow n\times 8 \rightarrow n\times 16$. A GNN with $2$ flat linear layers and an aggregation function  operates in embedded space. 

The GNN processor  refines node embeddings over $L$ message-passing layers. In each layer,  message $m_{j \to i}$ passed from a neighboring node $j$ to a central node $i$ is an inverse-distance-weighted approximation of the feature gradient:
$
m_{j \to i} = w_{ij} \cdot |z_j - z_i|
$
where $z_i$ is the feature vector of node $i$ and $w_{ij} \propto 1/|| \X_i - \X_j||_2$ is the weight of the edge between them. 
For the direct method, we choose a simple mean as the aggregation function. For the relative method, we concatenate the mean and max of the total weighted messages, to encourage the network to detect and retain sharp local error spikes that drive adaptive refinement. 
The decoder finally maps the refined node embeddings to the final scalar sizing field $S \in \mathbb{R}^{n \times 1}$. The set of all learnable weights constitutes the model parameters $\theta$.

\paragraph{Loss function} 
We use the same loss function for both methods. Mesh element sizes typically follow a long-tailed distribution: the vast majority of the domain is covered by "average-sized" elements, while the critical adaptive behavior occurs at the extremes (extremely fine meshes for high-error regions, or coarse meshes for low-error regions). A standard Mean Squared Error (MSE) loss would be dominated by the average cases, causing the network to regress to the mean which would fail to capture the sharp refinements required for accurate simulation.

To address this, we operate in log-space and employ a dynamic weighted loss that penalizes errors in rare but important regions. Given training samples as tuples $(\X_i, \phi_i, \hat{s}_i)$ where $\hat{s}$ is the supervisory value for the sizing field at $\X_i$, we first compute $\hat{y}_i = \log{(\hat{s}_i)}$ and train the network to predict $y_i = \log{(s_i)}$. For each sample in a batch, we calculate its rarity score via 
$\kappa_i = 1 + \lambda (y_i - \mu_{b})^4$ where $\mu_b$ is the median of the batch $y_i$ and $\lambda$ is a sensitivity hyperparameter (fixed to $10.0$). The fourth power creates a "U-shaped" weighting profile where common, near-median values are treated with standard importance, while significantly larger penalties are imposed on outliers (fine/coarse elements), strongly encouraging the network to prioritize them. 
The loss function we use is
$$\mathcal{L} = \frac{1}{N} \sum_{i=1}^{N} \kappa_i \cdot \tilde{\mathcal{L}}_1(\hat{y}_i, y_i)$$
where $\tilde{\mathcal{L}}_1$  a weighted smooth L1 (Huber) loss rather than MSE for its stability. This behaves like a quadratic for small errors (smooth gradients) and linearly for large errors, preventing gradient explosions during the early phases of training.

\section{Experiments: LAMG for steady-state heat}\label{sec:exp}
In this section, we first explain the experimental setup before presenting results of direct LAMG (Section~\ref{sec:expdirect}) followed by relative LAMG (Section~\ref{sec:exprelative}) for solving for steady-state heat given boundary conditions and source terms. Within each of the latter two subsections, we present quantitative results and qualitative results separately. First, we explain the methods we compare against, the error metric used and details of how our network was trained. 

\subsection{Methods used in comparisons}\label{sec:baselines}

\quad \textit{AMR} The standard Adaptive Mesh Refinement (AMR) strategy operates in a loop—solve, estimate error, mark, and refine—until the target error tolerance is met. It serves as the source of ground truth for training and the primary benchmark for solution accuracy.

\textit{AMG} This baseline represents the theoretical upper bound of our method's performance. It uses converged AMR to estimate the \textit{reference sizing field}, followed by the generation of an adaptive mesh that respects the sizing field. Since it dissociates mesh generation from the learning error, it depicts the best possible mesh LAMG could produce given perfect information.

\textit{Walk on Spheres (WoS)} A grid-free, high-precision baseline that allows sparse estimates. Since WoS estimates the solution pointwise via Monte Carlo random walks without requiring a mesh, it provides an accuracy check for the Poisson problem, independent of discretization. It is expensive when dense estimation is required.

\textit{One-Step Heuristics} To assess the value of the learned adaptation, we compare against 1-step AMR and 1-step AMG. These baselines perform a single iteration of refinement based on the initial coarse error estimate. They serve to establish a lower bound for performance, demonstrating the limited error reduction achievable without either iterative refinement or our learned non-local correction.

\subsection{Dataset, training and measuring error} \label{sec:training}
We generate a training dataset of Poisson's PDE problems, each defined by a boundary mesh and corresponding boundary conditions.
The meshes are sampled from Thingi10k~\cite{zhou2016thingi10k} and the boundary conditions and source terms are generated as 3D Gaussian mixtures that can be evaluated at each $\X_i$. We sampled from 30 to 40 Gaussians for boundary conditions and 20 to 30 Gaussians for source terms, resulting in a diverse patterns with combinations of high- and low-frequencies. The Gaussians are randomized strictly within the domain bounds; for Dirichlet boundaries and source terms, we independently sample the standard deviations ($\sigma_x, \sigma_y, \sigma_z$) along each axis to produce non-spherical, axis-aligned shapes. Boundary condition amplitudes are sampled in range $[1, 3]$ with sigmas in the range $[0.05, 1.0]$. In contrast, source term amplitudes are sampled from a much higher range of $[3000, 5000]$ with significantly narrower sigmas of $[0.03, 0.04]$. This disparity is necessary because the source term acts as a reactive input that is smoothed by the diffusion operator ($\Delta u$) in the Poisson equation; essentially, extremely sharp and high-magnitude sources are required to drive solution gradients comparable to those imposed by the boundary conditions.

For each problem, we compute a reference Adaptive Mesh Refinement (AMR) solution to serve as the ground truth, which we validated against a converged uniform dense solution. To construct the training dataset, we perform a coarse solve on the initial mesh to derive the input feature field $\phi$. We generate training tuples $(\mathbf{X}_i, \phi_i, \hat{s}_i)$ by sampling this coarse field: depending on the chosen feature type (e.g., solution DOFs vs. element-wise error estimates), $\mathbf{X}_i$ corresponds to either a mesh vertex or an element barycenter/quadrature point. For each sample point $\mathbf{X}_i$, we locate the corresponding tetrahedron within the high-fidelity reference AMR mesh and extract its volume $v$. The ground-truth size target is then computed as $\hat{s}_i = \sqrt[3]{6v\sqrt{2}}$, converting the reference volume $v$ into the equivalent edge length of a regular tetrahedron.
Given a solution field $\U_a$ and a reference solution $\hat{\U}$, we measure and report the relative error of the solution as 
$
e = {\| \U_a - \hat{\U} \|_2}/{\| \U_a \|_2}.
$

\subsubsection{Training data for the direct method}
We trained the direct model ($h_{\theta4}$) with the result temperature field obtained from coarse FEM solution and the size field is obtained by running AMR from a coarse mesh(\textasciitilde 4k vertices) to a final high-resolution state (typically 15k-20k vertices). We used 5000 datasets generated from 500 different geometries selected from Thingi10k~\cite{zhou2016thingi10k} for 5345 trainable parameters. The training takes around 2 hours on single Nvidia RTX 2080.

\begin{figure}[htbp!]
	\centering
	\setlength{\tabcolsep}{2pt} 
	\renewcommand{\arraystretch}{0.8} 
	\begin{tabular}{@{}c@{}c@{}}
		\multicolumn{2}{c}{\includegraphics[width=0.90\columnwidth]{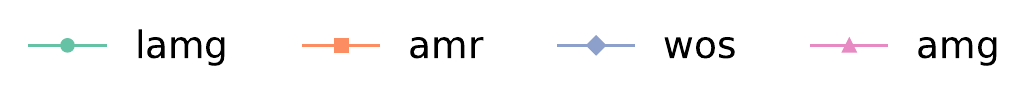}}\\
		\includegraphics[width=0.45\columnwidth]{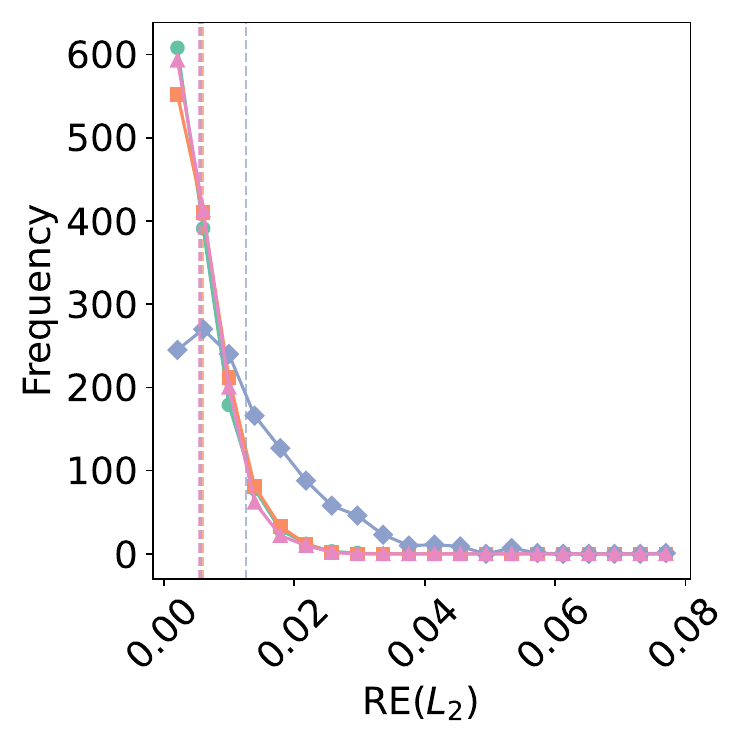} & 
		\includegraphics[width=0.45\columnwidth]{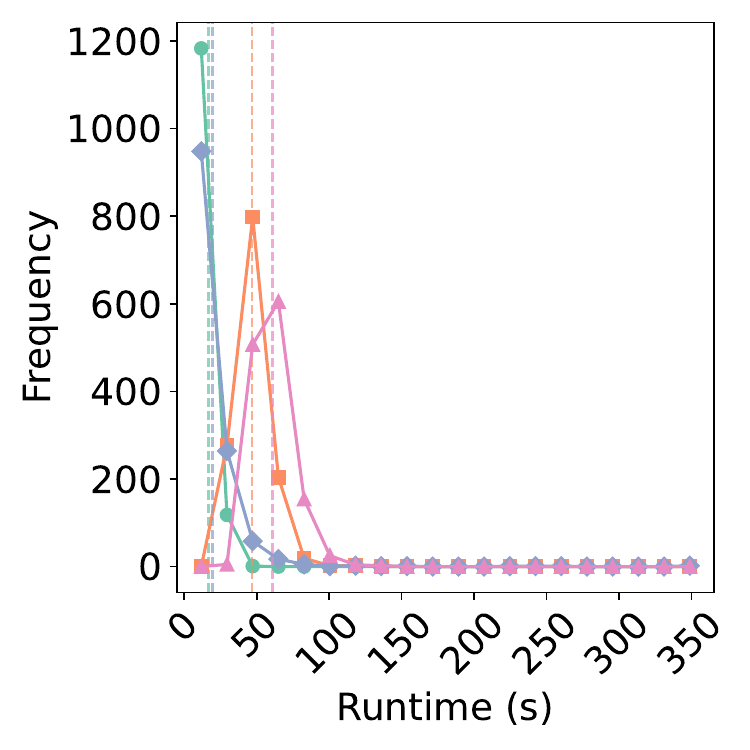}     \\
	\end{tabular}
    \vspace{-1mm}
    \caption{Performance comparison for the Poisson problem, plotting relative error (top row) and execution time (bottom row) for our method (LAMG) against AMR, AMG, and WoS baselines. Performance distributions on a large, held-out test set of 1303 unseen shapes (model $h_{\theta4}$). Across all scenarios, the plots show that for a similar error, LAMG is consistently faster.  
}
\label{fig:direct:Poisson}
\end{figure}

\subsubsection{Training data for the relative method}
The input initial error is obtained via a coarse FEM solve along with the ZZ-error~\cite{zienkiewicz1992superconvergent1,zienkiewicz1992superconvergent2} estimates at barycenters of the tetrahedra (quadrature points for linear elements). The reference sizing field is obtained by running AMR with a target error $\epsilon_{\mathrm{tol}}$ sampled from a log-uniform distribution ranging from $5 \times 10^{-2}$ to $2 \times 10^{-3}$. This sampling strategy ensures a balanced distribution of training samples across varying mesh densities; since the finite element error typically follows a power-law convergence with respect to the number of degrees of freedom, sampling $\epsilon_{\mathrm{tol}}$ uniformly in the logarithmic space results in a uniform coverage of the solution complexity spectrum (from pre-asymptotic to highly refined regimes). We used 1400 training samples generated from 200 different shapes from Thingi10k for our relative model ($h_{\theta R}$) with 1407 trainable parameters. The training takes around 20 minutes on single RTX 2080.

\begin{figure*}[t!]
    \centering
    
    \begingroup
        \setlength{\tabcolsep}{2pt} 
        \renewcommand{\arraystretch}{0.8}
        \begin{tabular}{ccccccc}
            mesh & reference & sizing field & $\eta=1$ & $\eta=0.8$ & AMR \\ 
            
            \includegraphics[height=1.2cm]{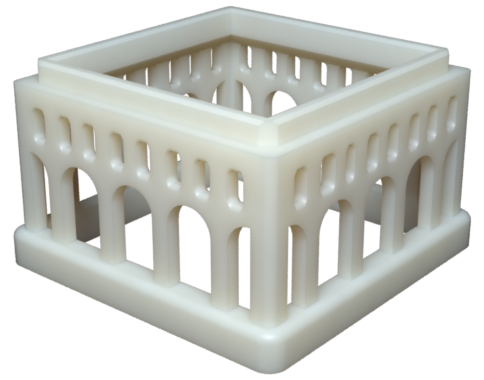} & 
            \includegraphics[width=0.15\linewidth]{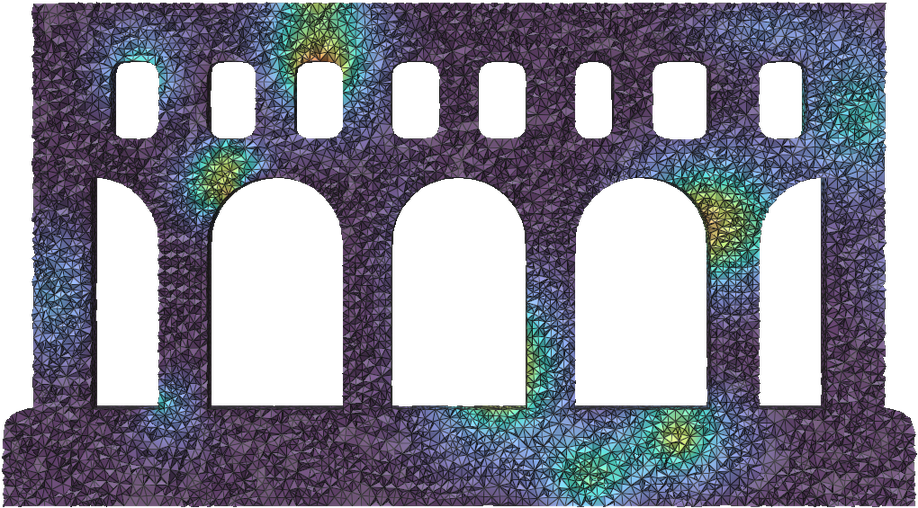} & 
            \includegraphics[width=0.15\linewidth]{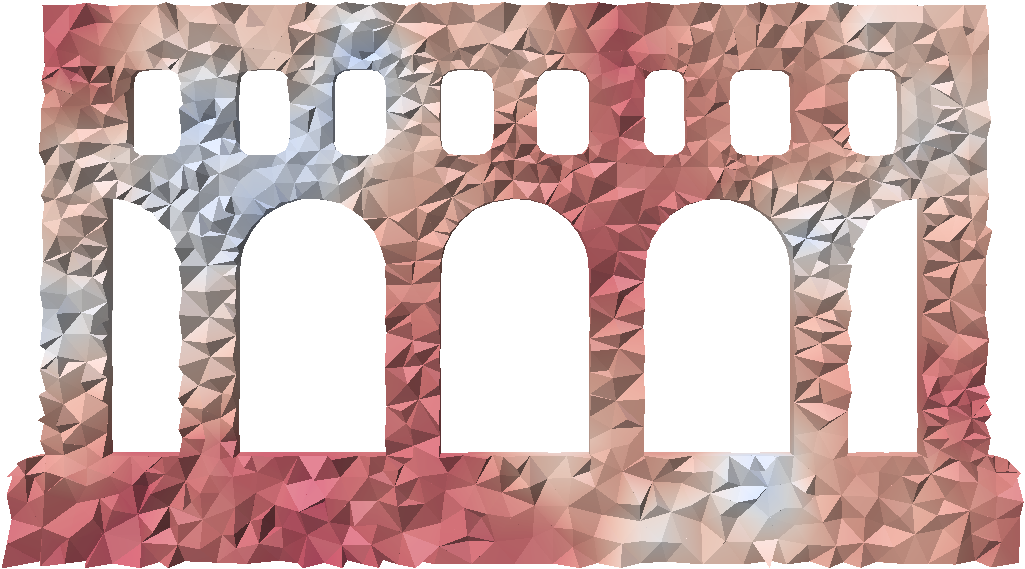} & 
            \includegraphics[width=0.15\linewidth]{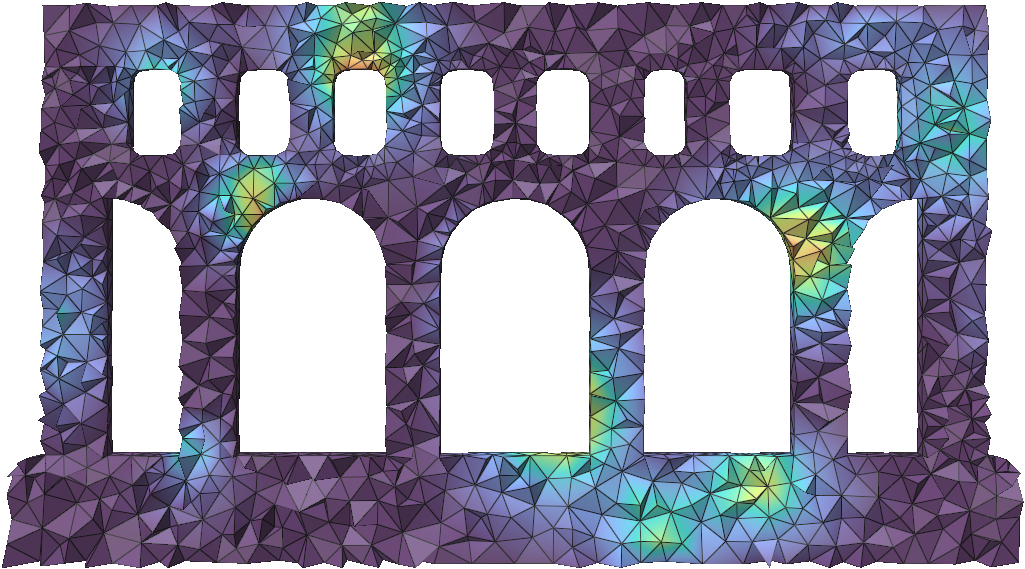} & 
            \includegraphics[width=0.15\linewidth]{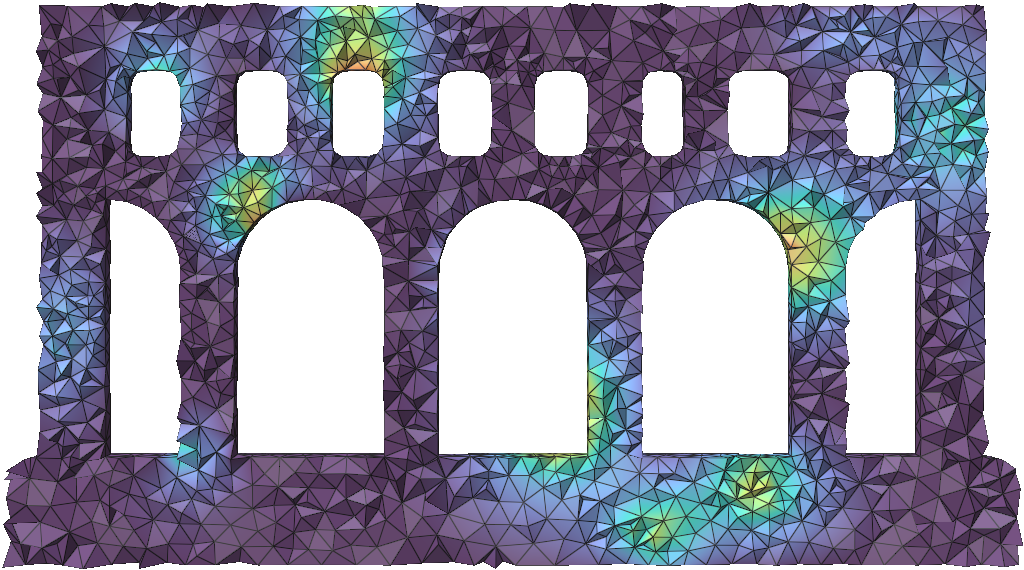} & 
            \includegraphics[width=0.15\linewidth]{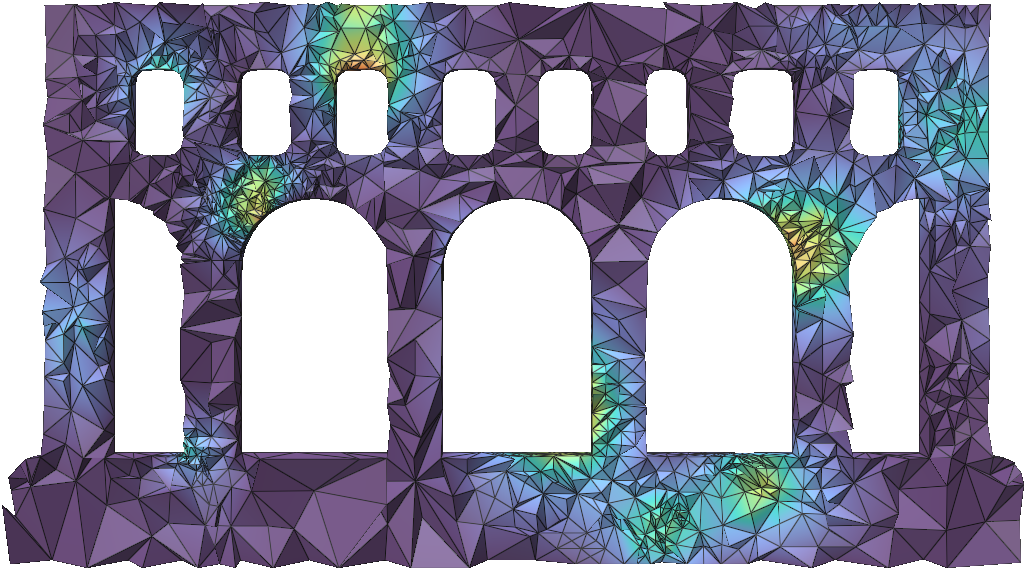} \\ 

            \includegraphics[height=1.5cm]{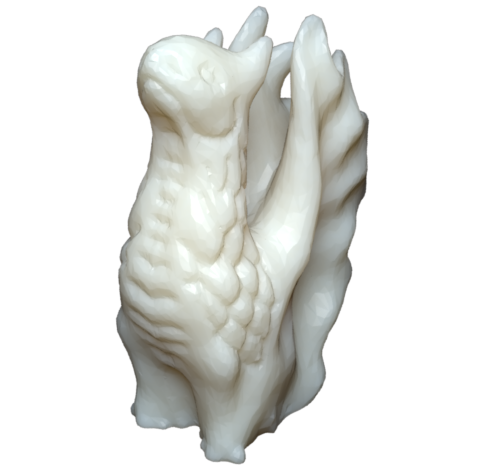} & 
            \includegraphics[width=0.15\linewidth]{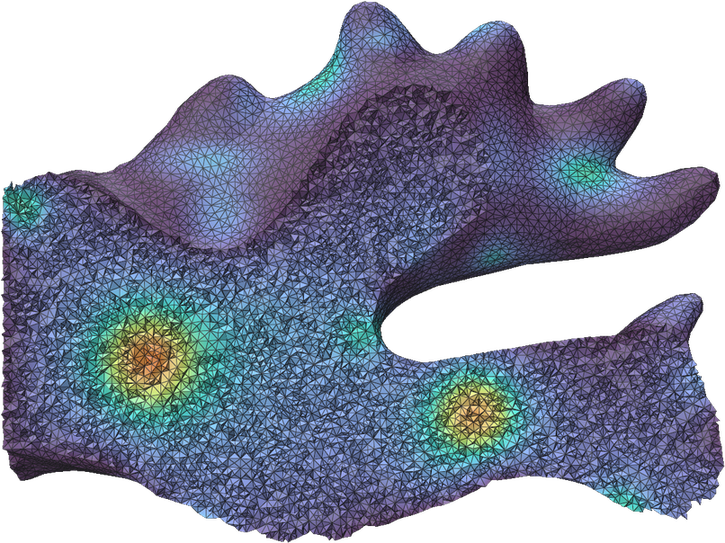} & 
            \includegraphics[width=0.15\linewidth]{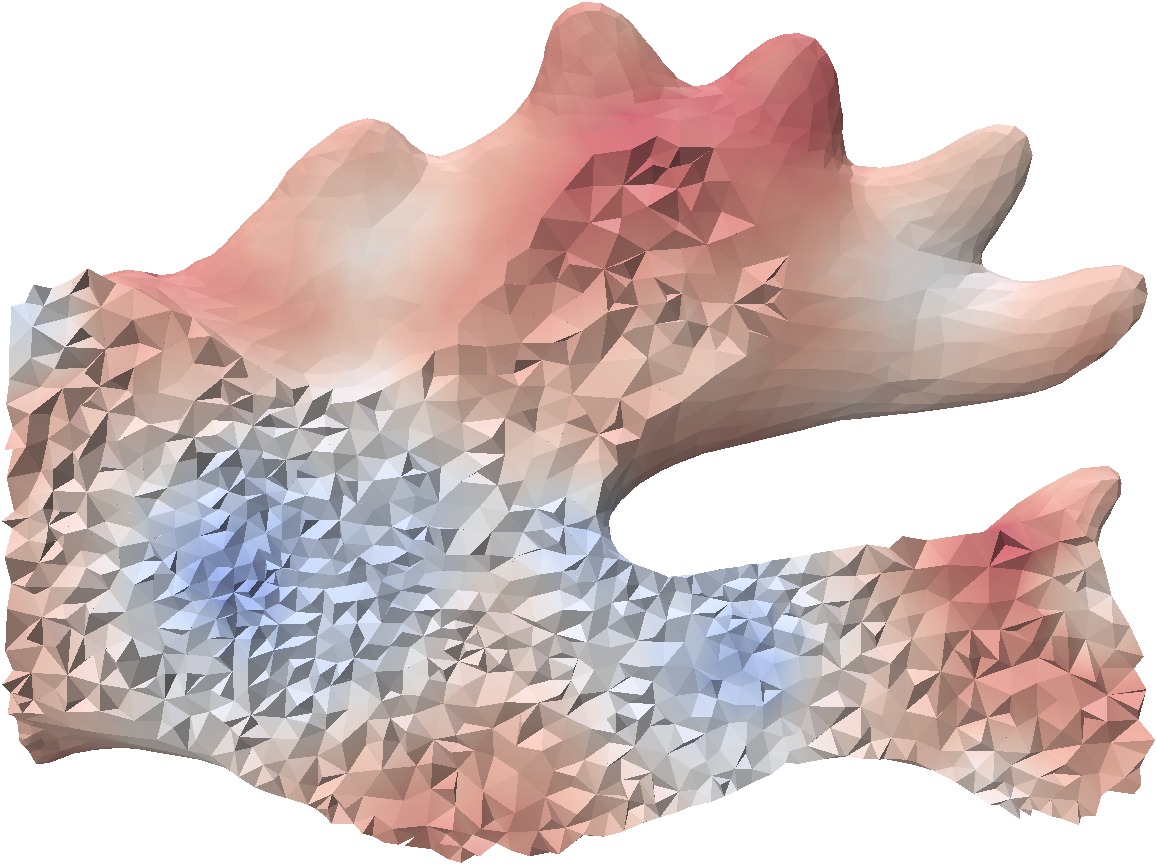} & 
            \includegraphics[width=0.15\linewidth]{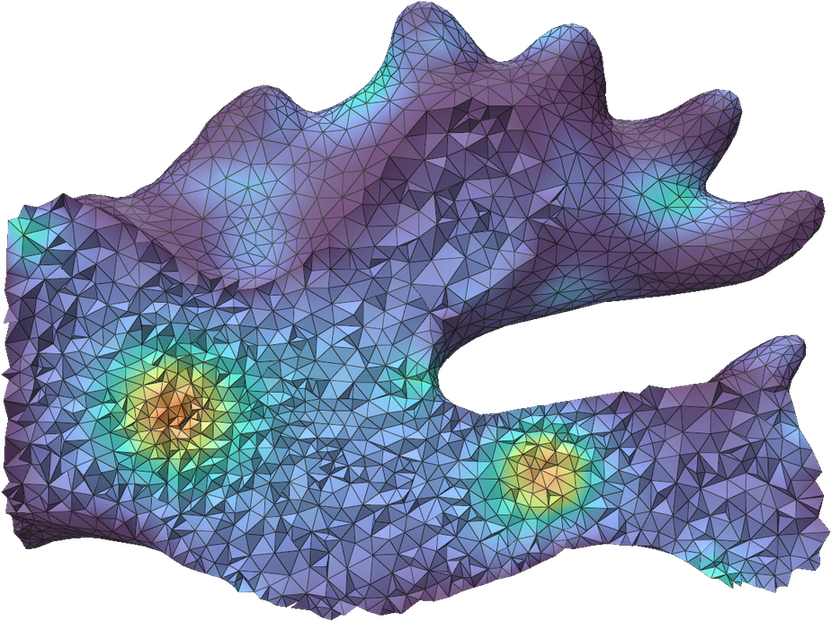} & 
            \includegraphics[width=0.15\linewidth]{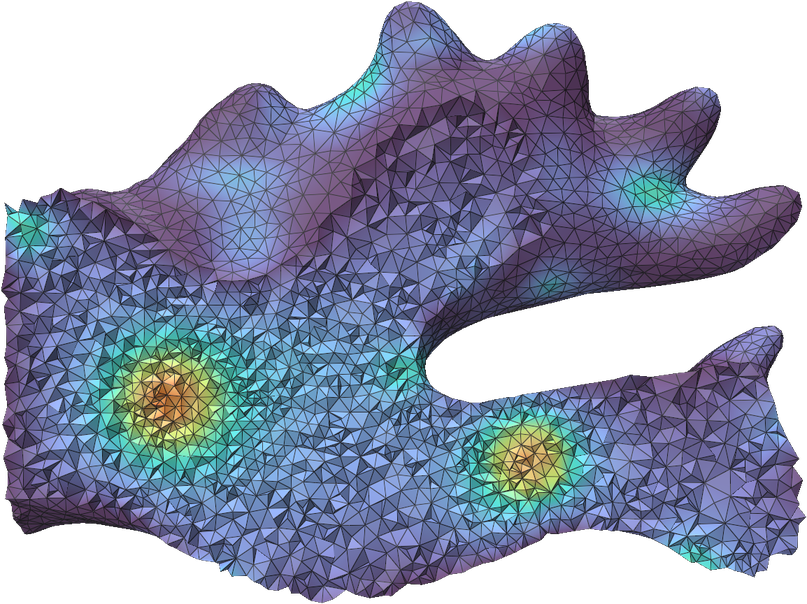} & 
            \includegraphics[width=0.15\linewidth]{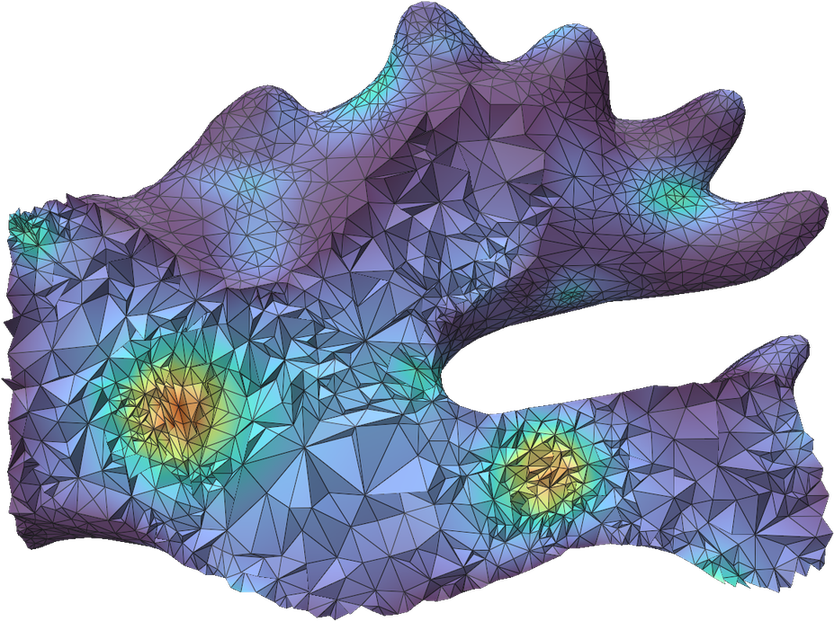} \\ 

            \includegraphics[height=1.0cm]{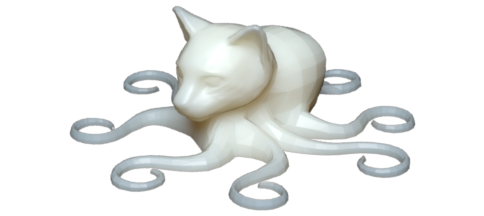} & 
            \includegraphics[width=0.15\linewidth]{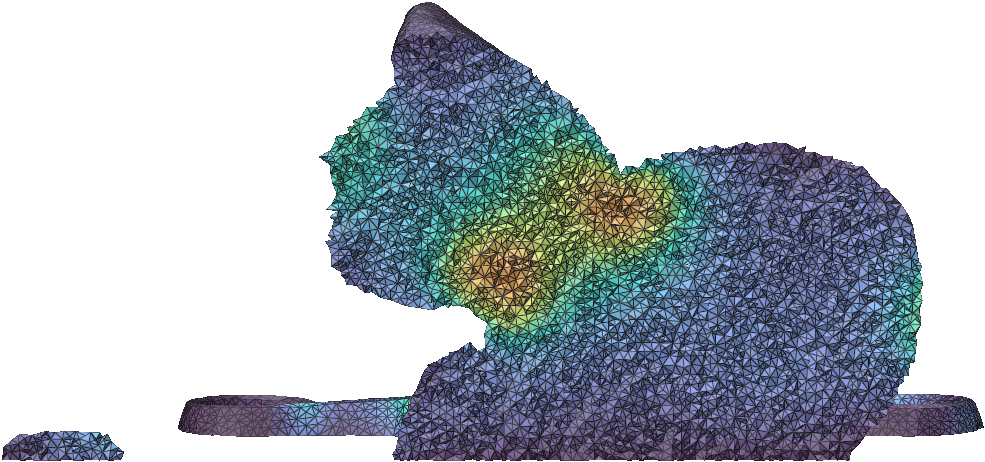} & 
            \includegraphics[width=0.15\linewidth]{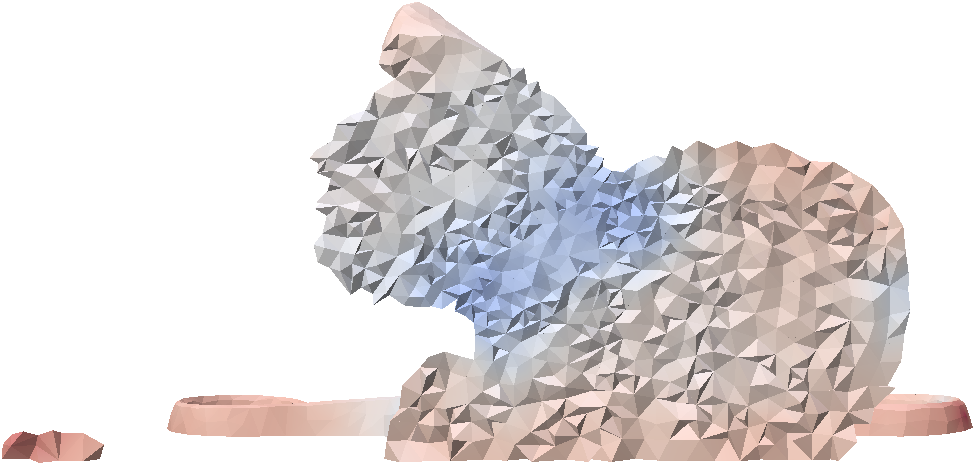} & 
            \includegraphics[width=0.15\linewidth]{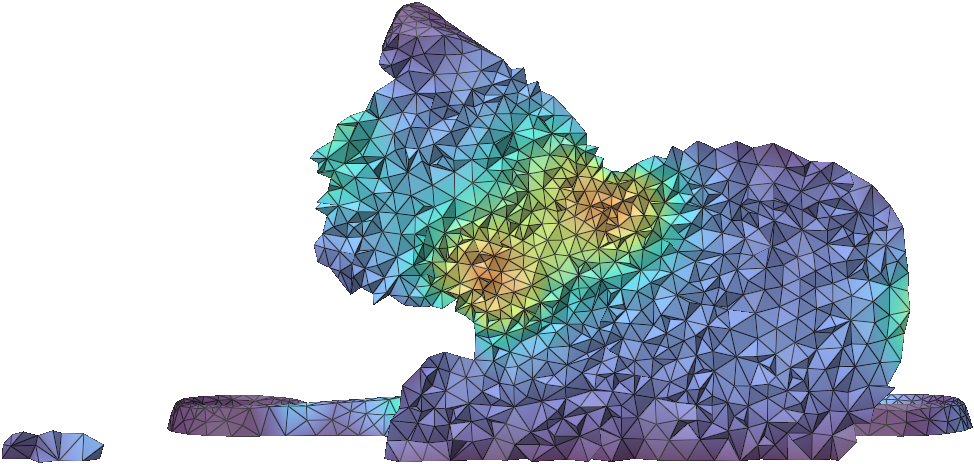} & 
            \includegraphics[width=0.15\linewidth]{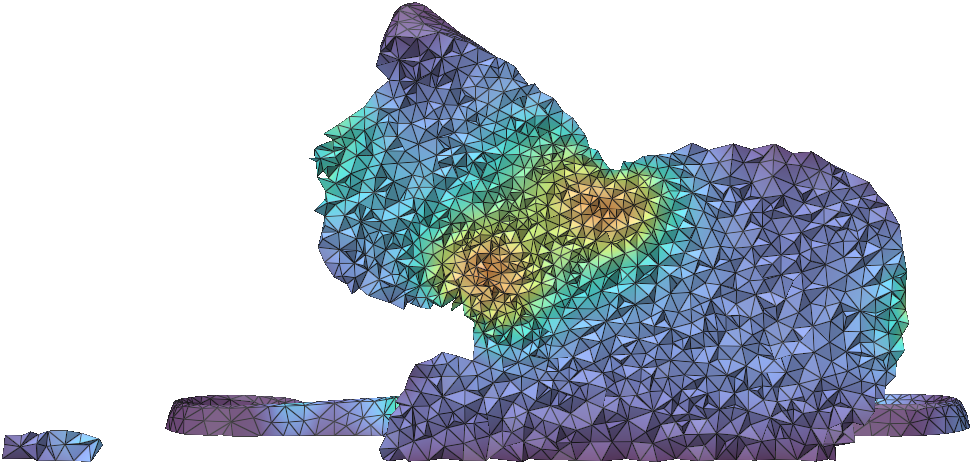} & 
            \includegraphics[width=0.15\linewidth]{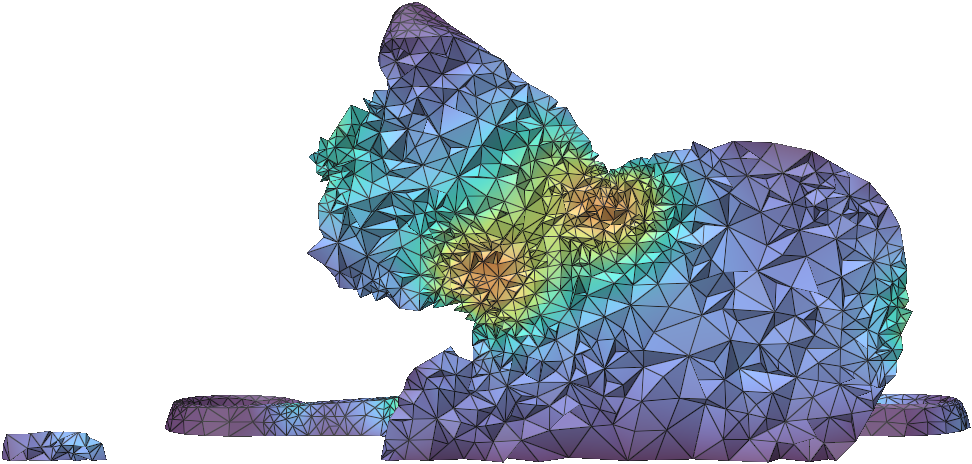} \\

            \includegraphics[height=1.5cm]{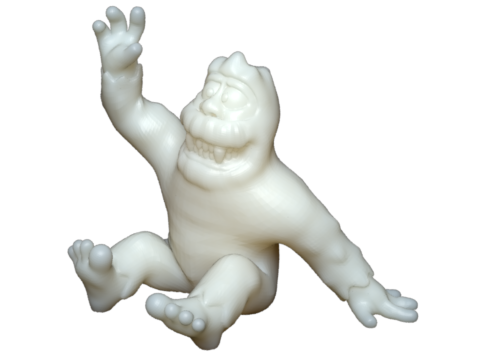} & 
            \includegraphics[width=0.15\linewidth]{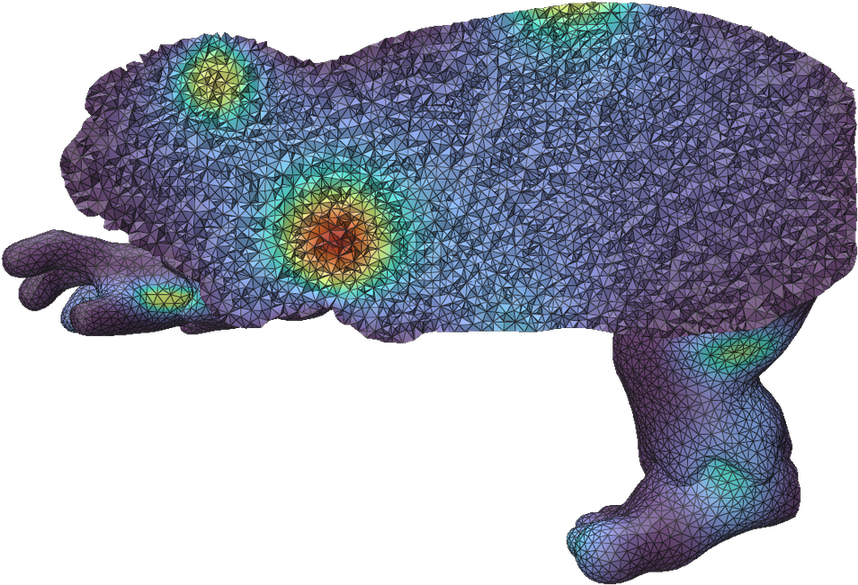} & 
            \includegraphics[width=0.15\linewidth]{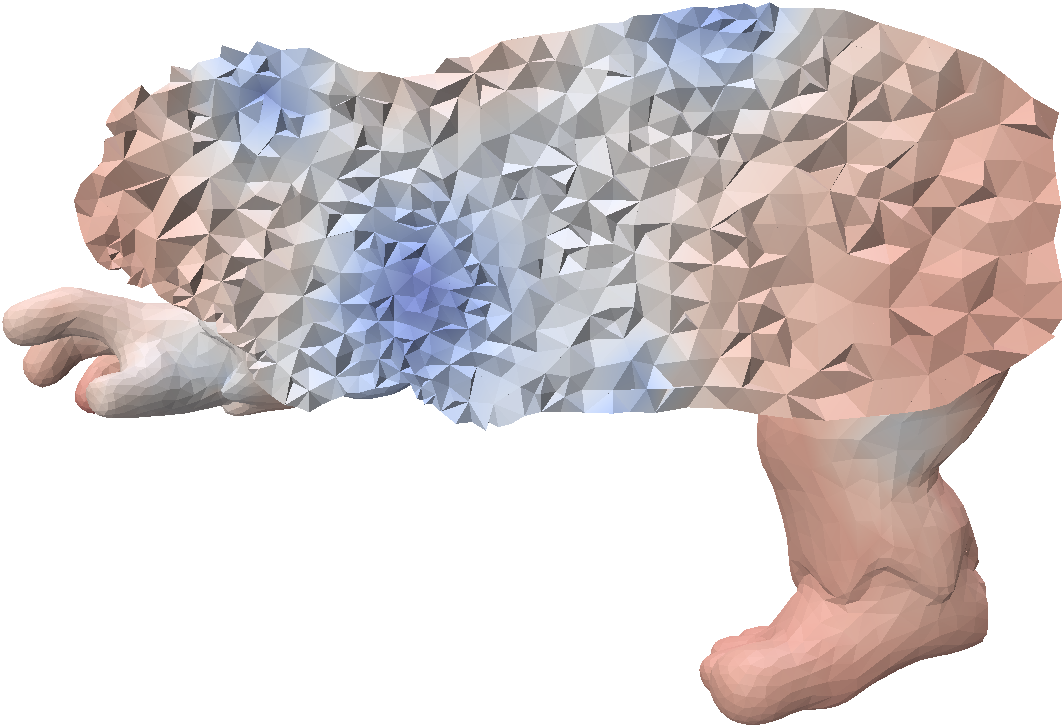} & 
            \includegraphics[width=0.15\linewidth]{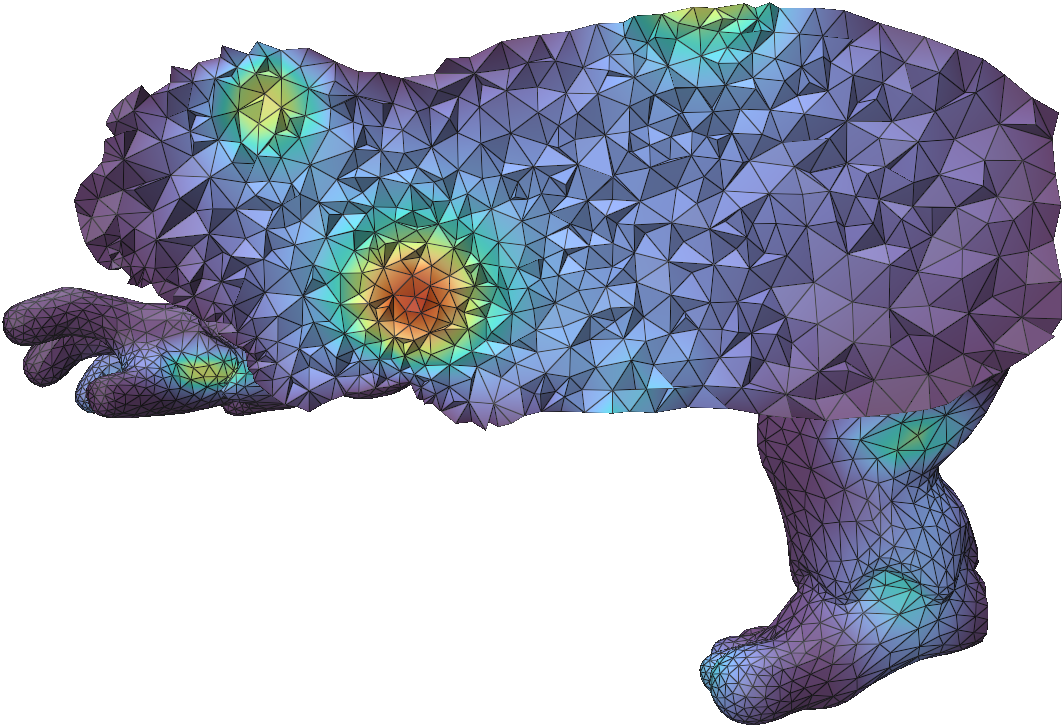} & 
            \includegraphics[width=0.15\linewidth]{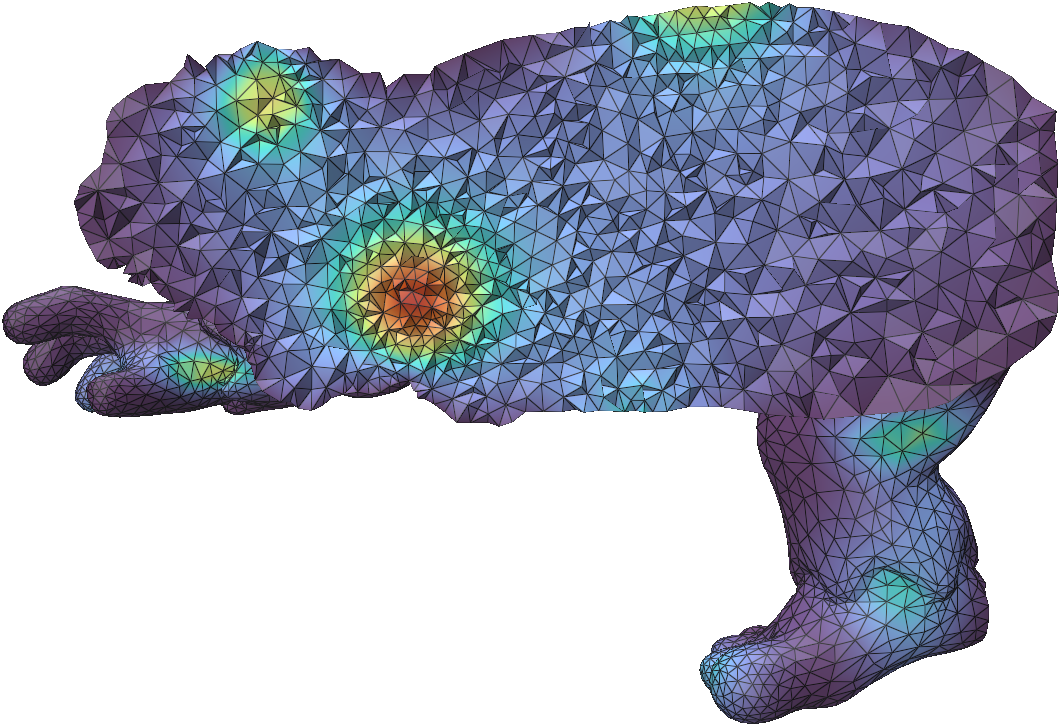} & 
            \includegraphics[width=0.15\linewidth]{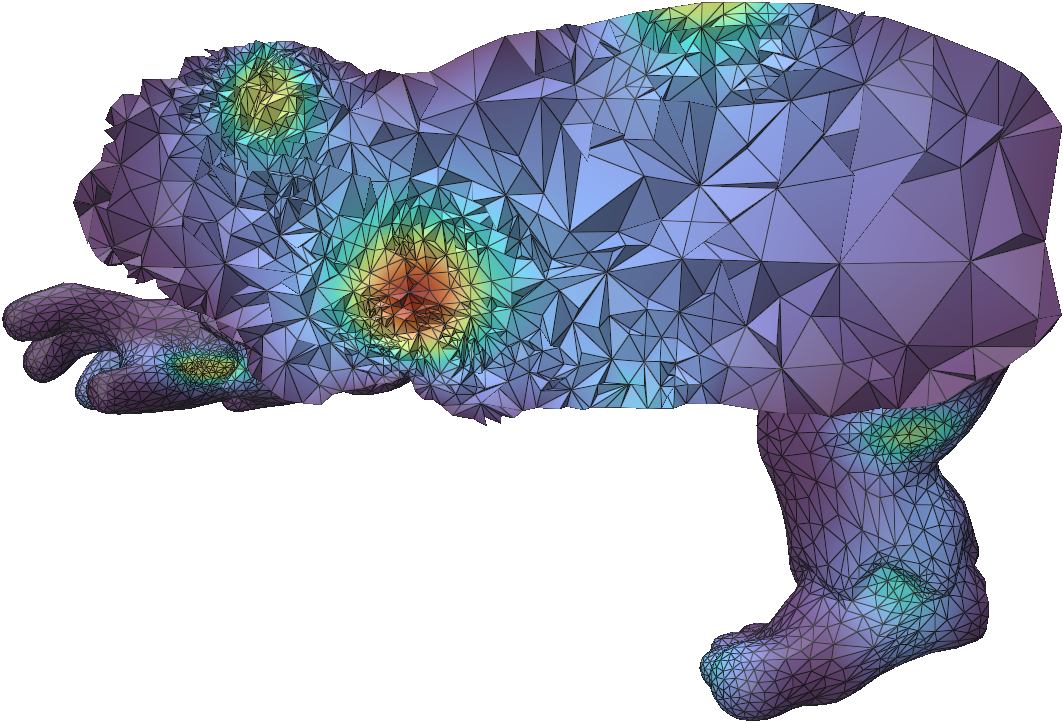} \\
        \end{tabular}
        
        \vspace{-2mm}
        \caption{
        Qualitative results on the Poisson problem for several unseen Thingi10k geometries. Each row presents a different challenging case. From left to right: input mesh, high-resolution reference solution, predicted sizing field, our final mesh ($\eta=1, 0.8$), and the iterative AMR baseline. These visuals demonstrate our one-shot method produces quality comparable to iterative AMR.
        }
        \label{fig:qualitative}
    \endgroup

    \begingroup
        \renewcommand{\arraystretch}{0.9} 
        \setlength{\tabcolsep}{4pt}      
        

        \begin{tabular}{l |c |c |c |c |c| c|c|c|c|c}
            Mesh & $\eta_{1}$ $L_2$& $\eta_{0.8}$ $L_2$ & AMR $L_2$ & $\eta_1$ \#Vert & $\eta_{0.8}$ \#Vert & AMR \#Vert & $\eta_1$ Time (s) &$\eta_{0.8}$ Time (s) & AMR Time (s)&Speedup \\
            \hline
            Architecture & 0.07533 & 0.05522 & 0.05156 & 15230 & 21137 & 23914 & 133.5 & 151.0 & 559.9& 3.71 \\
            Dragon       & 0.03575 & 0.02408 & 0.04462 & 19226 & 29100 & 20351 & 123.4 & 136.7 & 295.3& 2.16\\
            Octocat      & 0.02504 & 0.01578 & 0.01828 & 13344 & 22433 & 20381 & 94.0  & 120.5 & 247.0& 2.05\\
            SnowMonster  & 0.03947 & 0.03029 & 0.03184 & 15892 & 22743 & 26628 & 143.1 & 154.2 & 425.9& 2.76
        \end{tabular}
        
        \captionof{table}{
        Quantitative performance for the qualitative Poisson examples shown in Figure~\ref{fig:qualitative}. The results confirm that the high-quality visual results from our method are achieved with a significant speedup over the AMR baseline for a similar final error. The last column shows the speed up of LAMG over AMR.
        }
        \label{tab:qual_poisson_perf}
    \endgroup
    \vspace{-4mm}

\end{figure*}

\subsection{Direct method for steady-state heat}\label{sec:expdirect}
\subsubsection{Quantitative Results}
We validate our method on a large set of 1303 geometries from the Thingi10k~\cite{zhou2016thingi10k} dataset, which is a held-out test set of 1303 unseen shapes from the training of $h_{\theta4}$. We visualize these results as frequency polygons in Figure~\ref{fig:direct:Poisson}. The error distributions for all methods are nearly identical, with similar means except for the Monte Carlo solver (WoS) which is slightly higher. The runtime distributions, however, show that LAMG's performance is as good as WoS, mostly below 20 seconds, whereas AMR and WoS average approximately 45 seconds, and AMG is the slowest at around 55 seconds.

\subsubsection{Qualitative Results}

Figure~\ref{fig:qualitative} visualizes results on four complex geometries from our held-out Thingi10k test set. These examples highlight LAMG's performance across geometric challenges such as high-genus, thin features and sharp concavities. The sizing field captures high-gradient regions in the reference solution, leading to one-shot adaptive meshes that concentrate elements in those areas. LAMG is comparable in quality to meshes produced by the slower, iterative AMR baseline. The corresponding quantitative results are presented in Table~\ref{tab:qual_poisson_perf}, confirming that  high-quality meshes are generated with a speedup over AMR for similar error level.

\subsection{Relative method for steady-state heat}\label{sec:exprelative}
\subsubsection{Quantitative results}
\begin{figure}[htbp!]
	\centering
	\renewcommand{\arraystretch}{0.8} 
	\begin{tabular}{@{}c@{}c@{}}
		\multicolumn{2}{c}{\includegraphics[width=1.0\columnwidth]{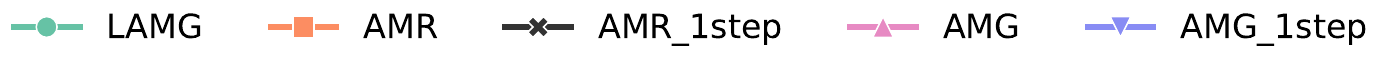}}\\
		\includegraphics[width=0.50\columnwidth]{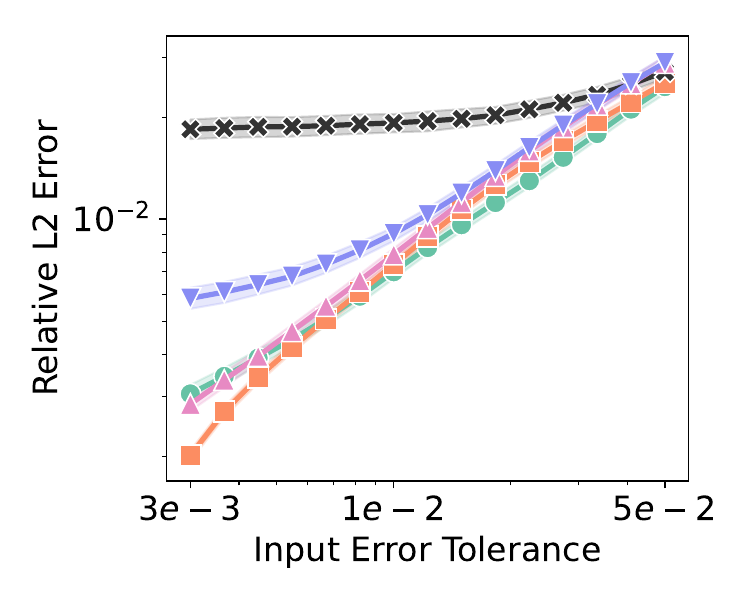} & 
		\includegraphics[width=0.50\columnwidth]{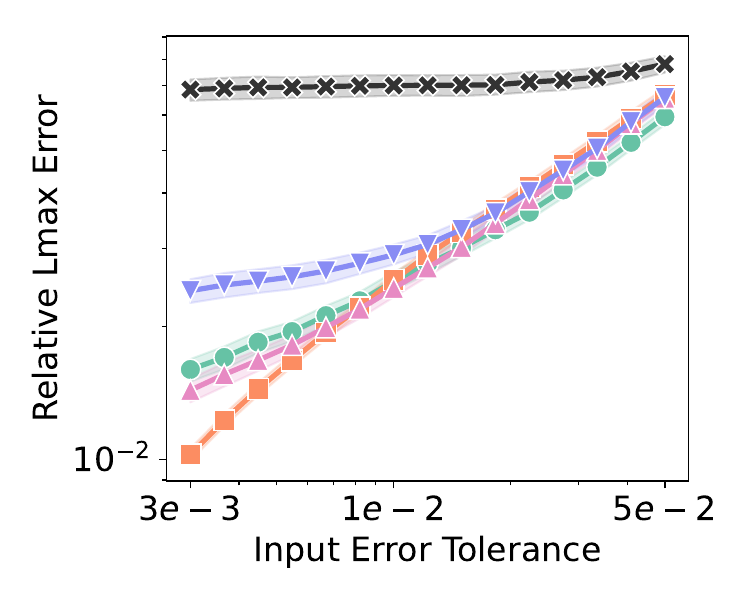}     \\
        (a) & (b)\\
        \includegraphics[width=0.50\columnwidth]{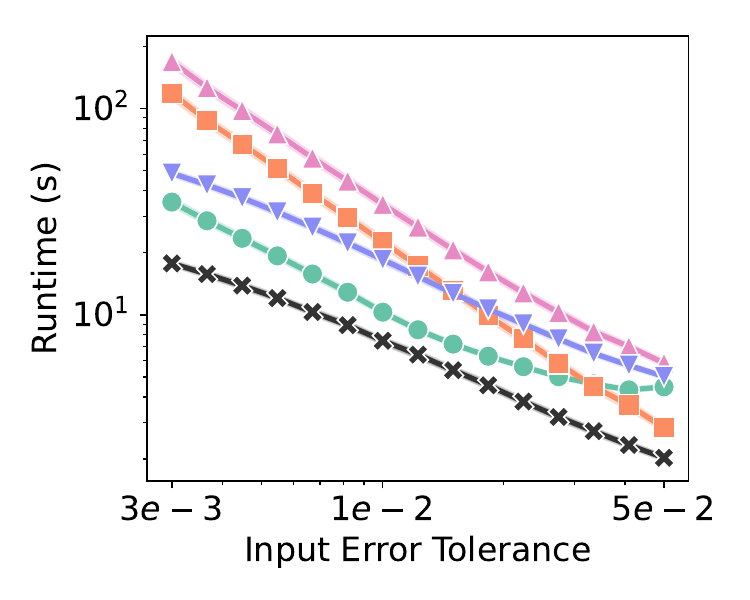} & 
		\includegraphics[width=0.50\columnwidth]{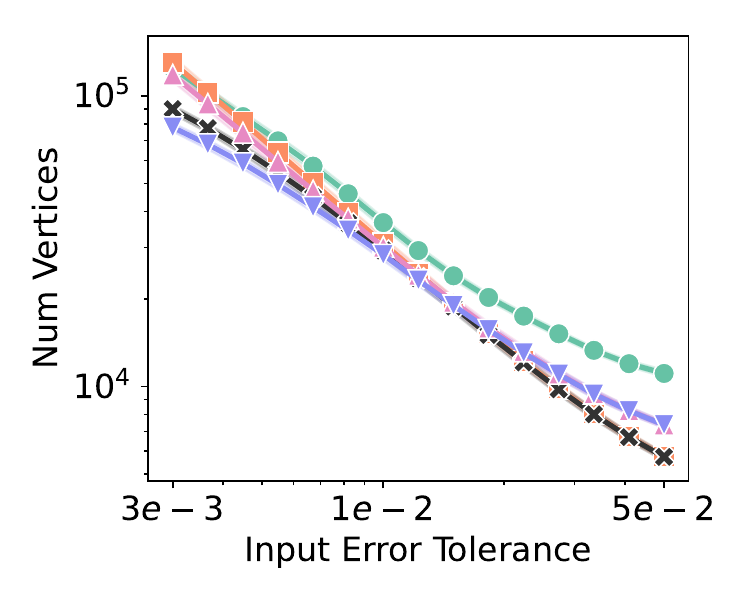}     \\
        (c) & (d)\\
        \includegraphics[width=0.50\columnwidth]{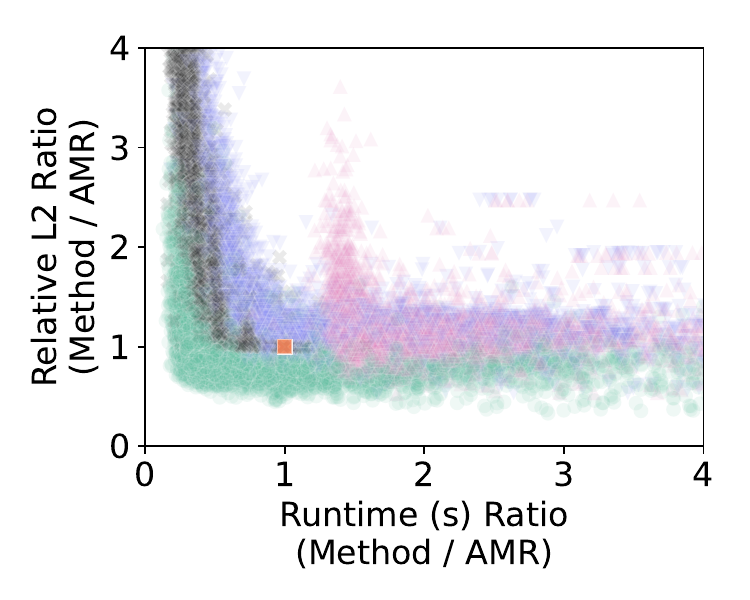} & 
		\includegraphics[width=0.50\columnwidth]{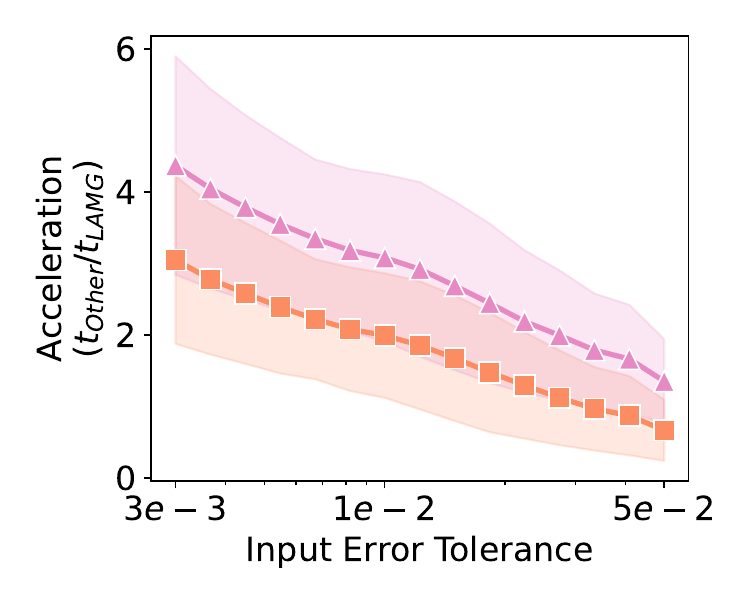}     \\
        (e) & (f)\\
	\end{tabular}
    \vspace{-1mm}
    \caption{Relative performance of LAMG and AMG compared to standard AMR across 600 test shapes with 9100 different PDEs setups.
The top row shows relative L2 and $L_{\infty}$ errors; both LAMG (green) and AMG (pink) match the accuracy of the adaptive baseline (orange), whereas 1-step heuristics degrade rapidly.
The middle row illustrates efficiency: LAMG runtime decreases significantly as tolerance tightens, achieving $<0.5\times$ the runtime of AMR at a input error tolerance of $3\times 10^{-3}$.
The bottom row details the trade-offs: the scatter plot (left) places most LAMG results in the high-efficiency/high-accuracy quadrant (left of the AMR baseline at $(1,1)$), while the runtime ratio (right) confirms that AMR is $2\times$--$4\times$ slower than LAMG for strict error targets. 
}
\label{fig:relative:Poisson}
\vspace{-4mm}
\end{figure}

Figure~\ref{fig:relative:Poisson} presents results of relative LAMG with different input error tolerances. We validated it on 600 different shapes using input error tolerances from $5 \times 10^{-2}$ to $3 \times 10^{-3}$. 

\paragraph{Accuracy (top row)} The first row displays the relative L2 (a) and relative Lmax (b) error. Both LAMG (green) and AMG (pink) result in similar accuracy, which is slight worse than AMR. However the error is near $1\times 10^{-3}$, indicating that they achieve accuracy comparable to the fully adaptive AMR baseline across the entire range of input error tolerances. For comparison, we also plot "1-step" heuristics (AMR\_1step and AMG\_1step); these degrade rapidly as the input error tolerance tightens, highlighting the necessity of the full iterative methods.

\paragraph{Efficiency (middle Row):} The second row analyzes runtime and complexity. (c) We observe a significant trend in the Runtime: while LAMG is slower than AMR for approximate calculation (high tolerance $5 \times 10^{-2}$), it is faster as the input error tolerance is reduced. Notably, at the strictest input error tolerance of $3 \times 10^{-3}$, LAMG achieves the desired accuracy in less than half the computation time of AMR. (d) at large input error tolerance, LAMG uses more vertices than AMR.

\paragraph{Trade-offs (Bottom Row)} The bottom row visualizes the performance trade-offs. The scatter plot (e) charts Relative L2 Error ratio vs Runtime Ratio where the ratios are with respect to AMR. Points to the left of $1$ and below $1$ are desirable.  Most LAMG data points (green) cluster to the left of the orange square at $(1,1)$, indicating that LAMG typically delivers similar accuracy at a lower computational cost. The distinct "long tail" extending to the right reflects cases where LAMG, relying on coarse solutions, overestimates the problem complexity, resulting in conservative but  slower meshing. Finally, the runtime comparison (f) quantifies the speedup: at a input error tolerance of $3 \times 10^{-3}$, the baseline AMR method takes approximately $2\times$ to $4\times$ longer to compute than LAMG.

\subsubsection{Qualitative Results}
\begin{figure*}[t!]
    \centering
    
    \begingroup
        \setlength{\tabcolsep}{2pt} 
        \renewcommand{\arraystretch}{0.8}
        \begin{tabular}{ccccccc}
            LAMG $\epsilon=5\times 10^{-2}$ & AMR $\epsilon=5\times 10^{-2}$ & LAMG $\epsilon=1\times 10^{-2}$ & AMR $\epsilon=1\times 10^{-2}$ &LAMG $\epsilon=3\times 10^{-3}$ & AMR $\epsilon=3\times 10^{-3}$ \\ 
            
            \includegraphics[width=0.15\linewidth]{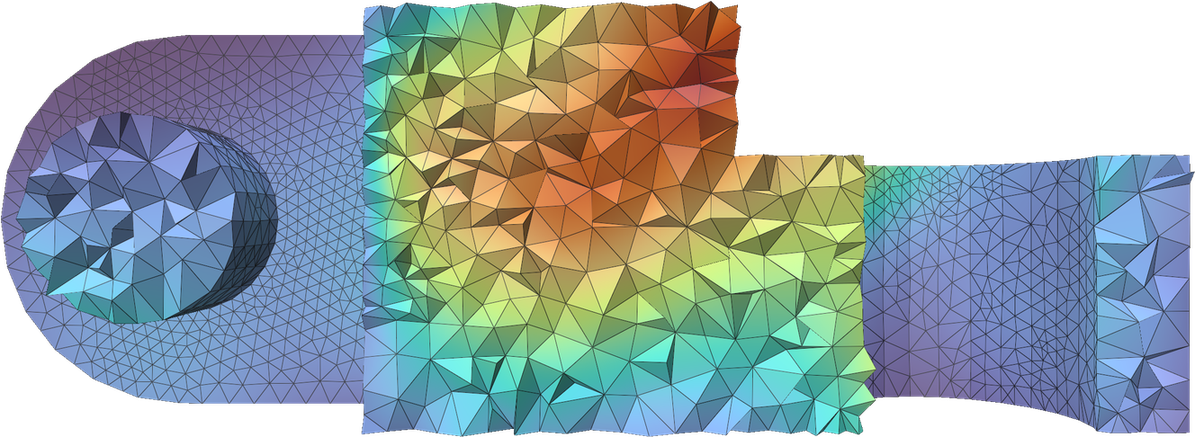} & 
            \includegraphics[width=0.15\linewidth]{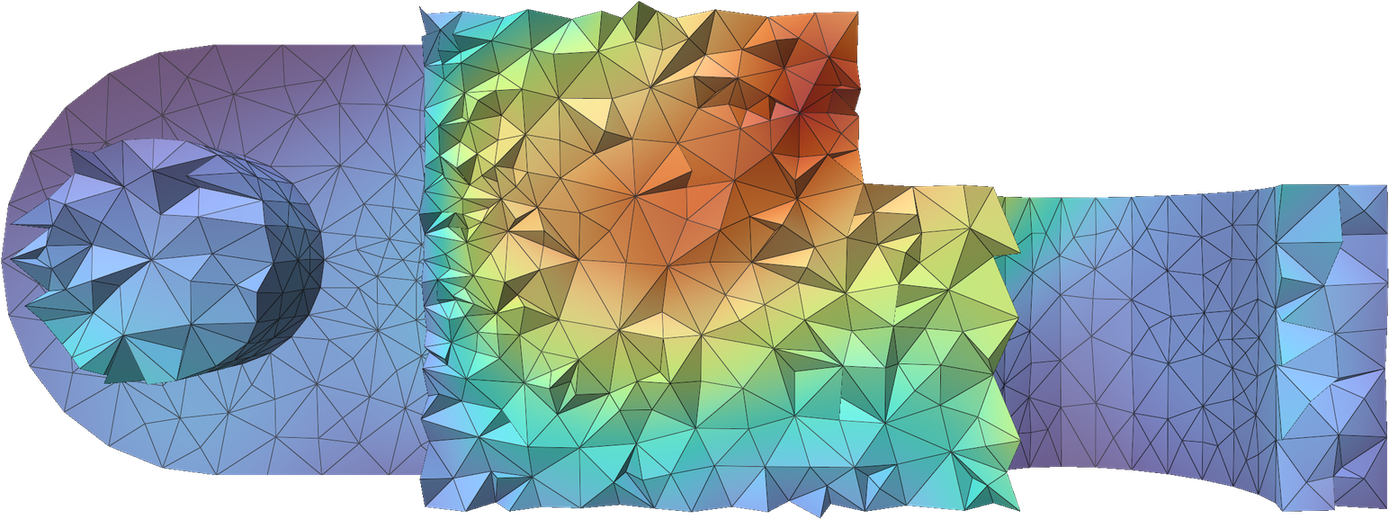} & 
            \includegraphics[width=0.15\linewidth]{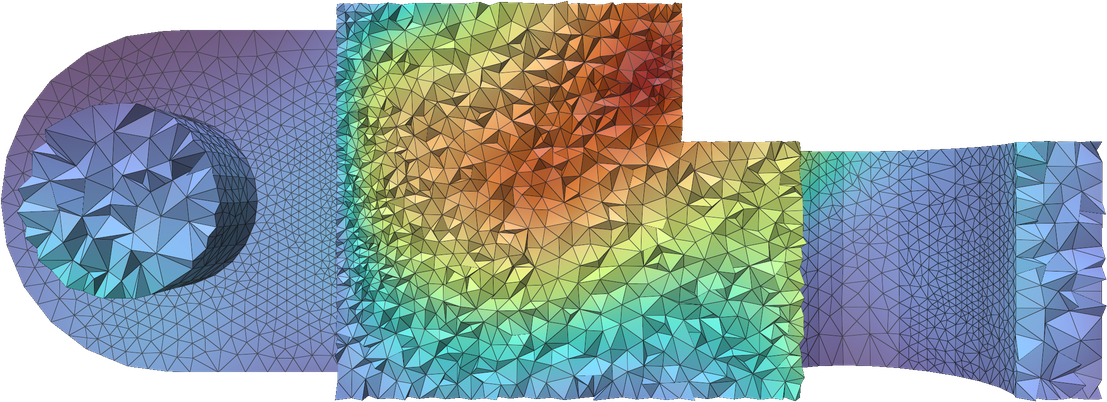} & 
            \includegraphics[width=0.15\linewidth]{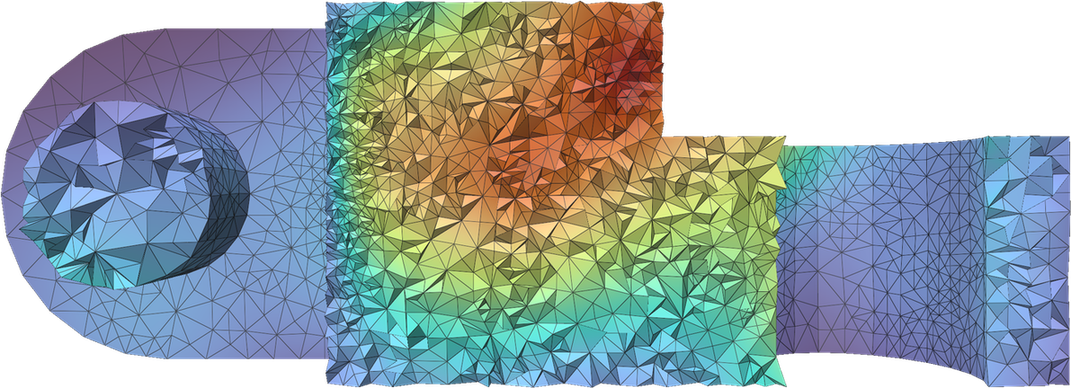} & 
            \includegraphics[width=0.15\linewidth]{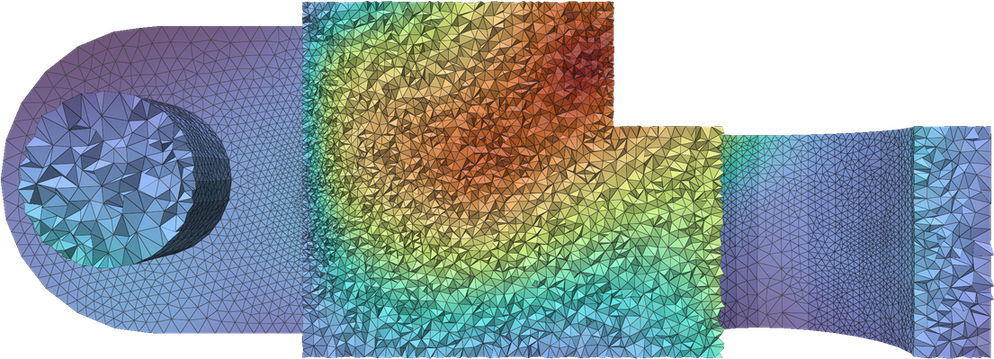} & 
            \includegraphics[width=0.15\linewidth]{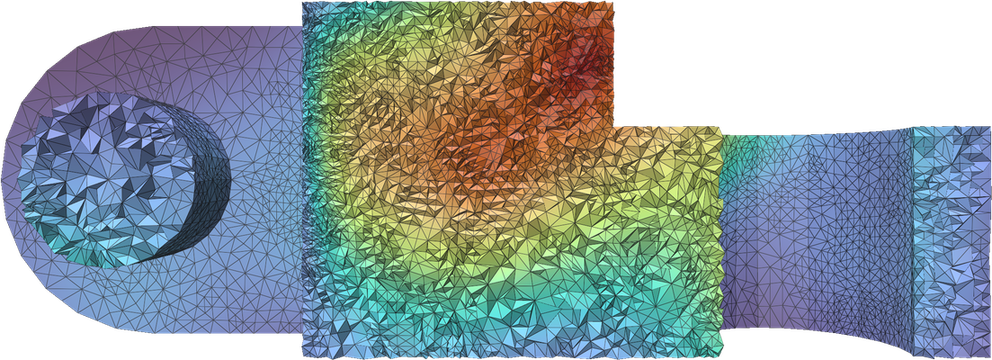} \\ 

            \includegraphics[width=0.15\linewidth]{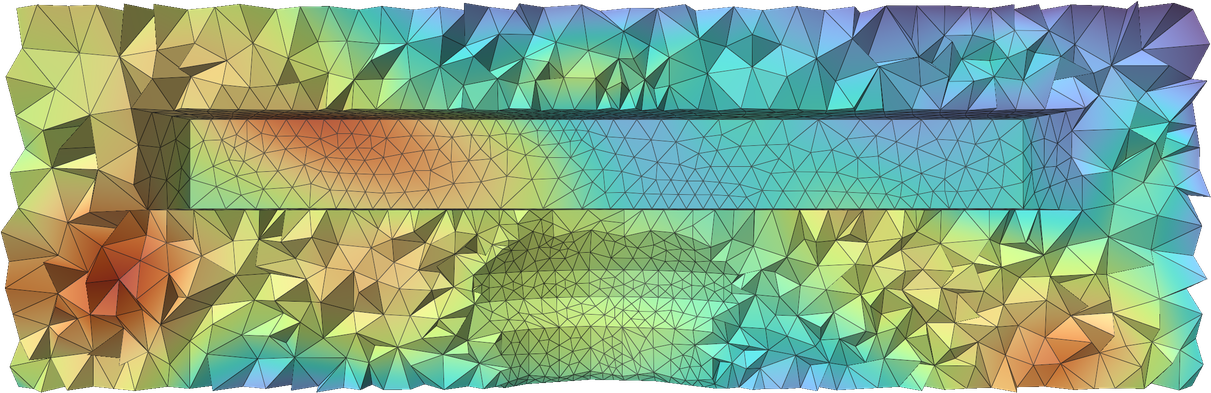} & 
            \includegraphics[width=0.15\linewidth]{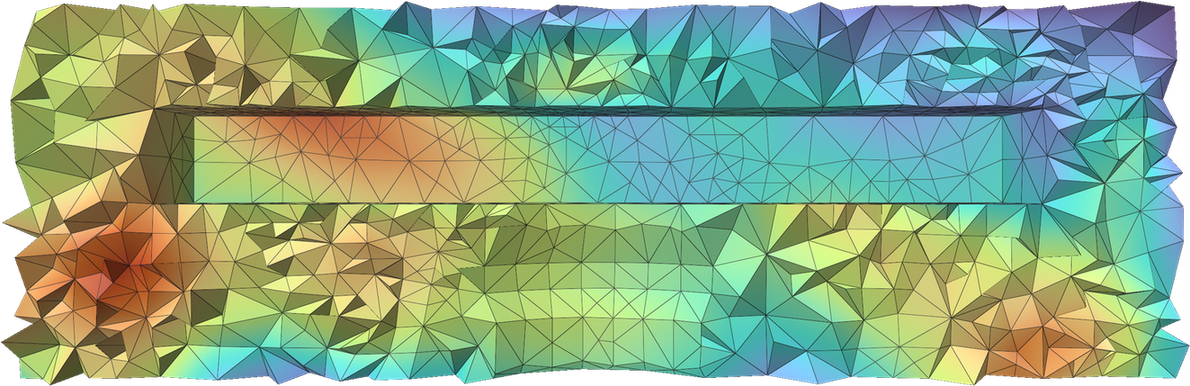} & 
            \includegraphics[width=0.15\linewidth]{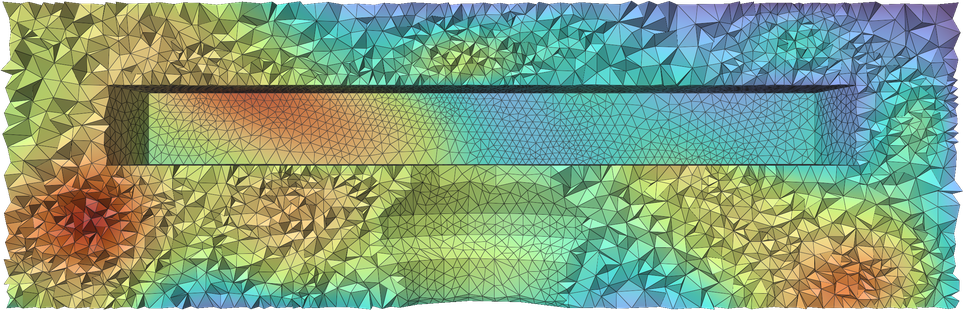} & 
            \includegraphics[width=0.15\linewidth]{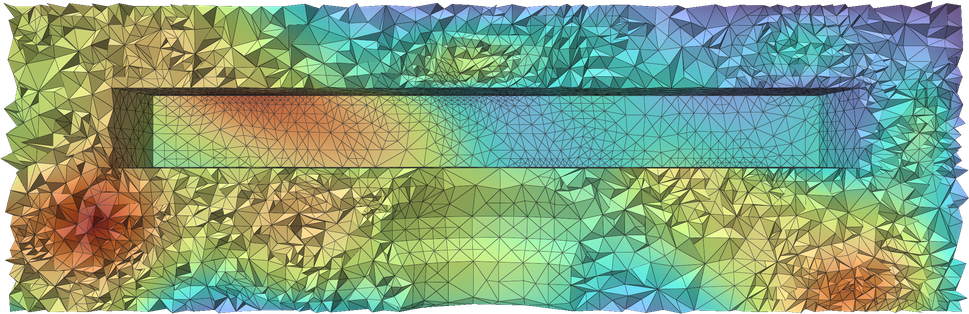} & 
            \includegraphics[width=0.15\linewidth]{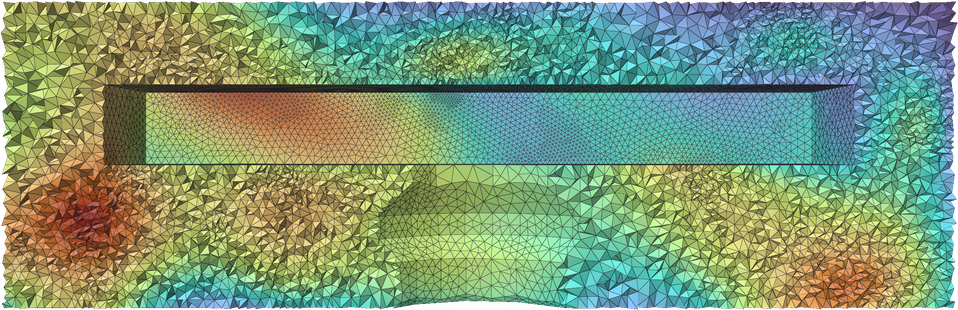} & 
            \includegraphics[width=0.15\linewidth]{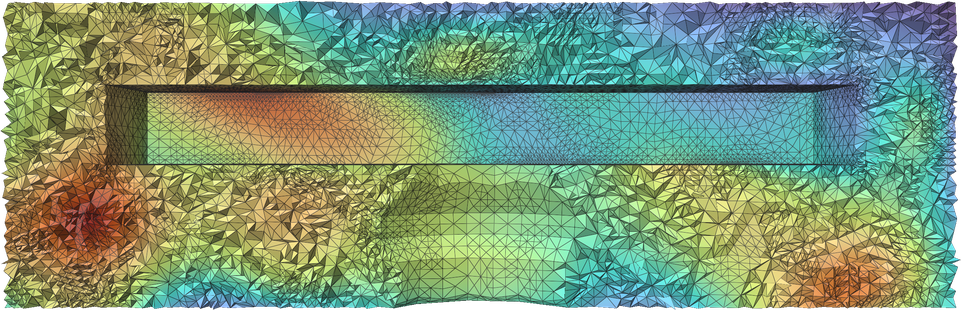} \\ 
            
            \includegraphics[width=0.15\linewidth]{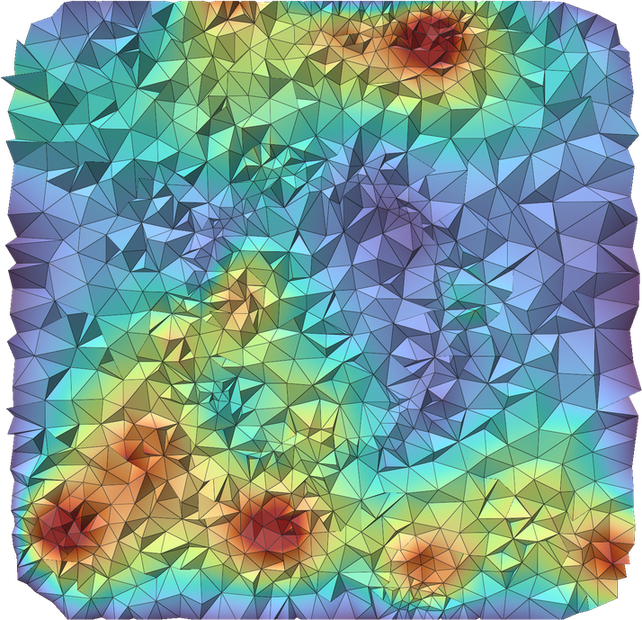} & 
            \includegraphics[width=0.15\linewidth]{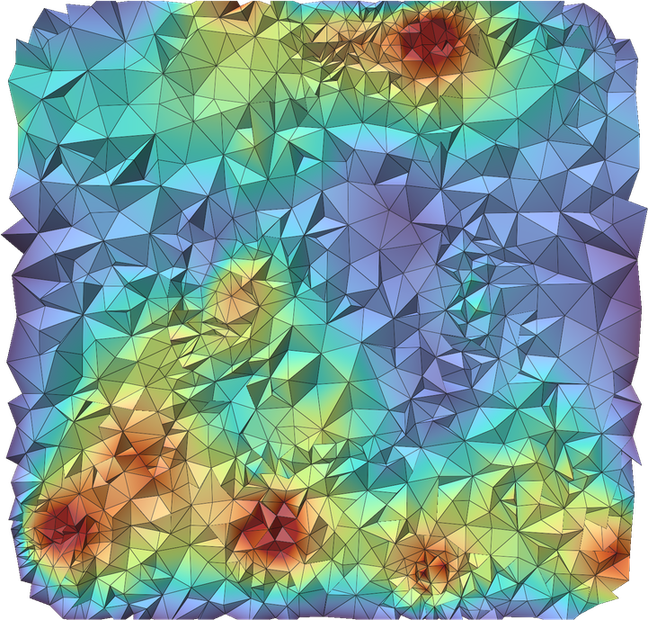} & 
            \includegraphics[width=0.15\linewidth]{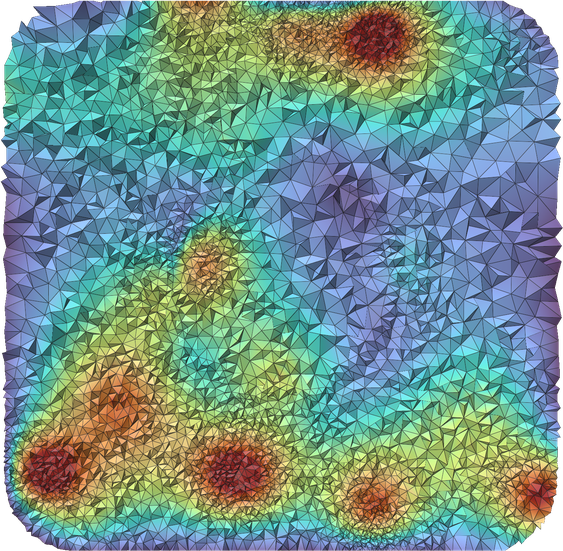} & 
            \includegraphics[width=0.15\linewidth]{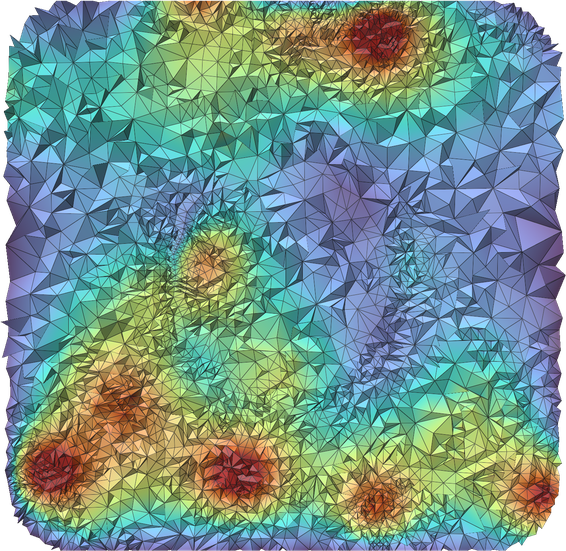} & 
            \includegraphics[width=0.15\linewidth]{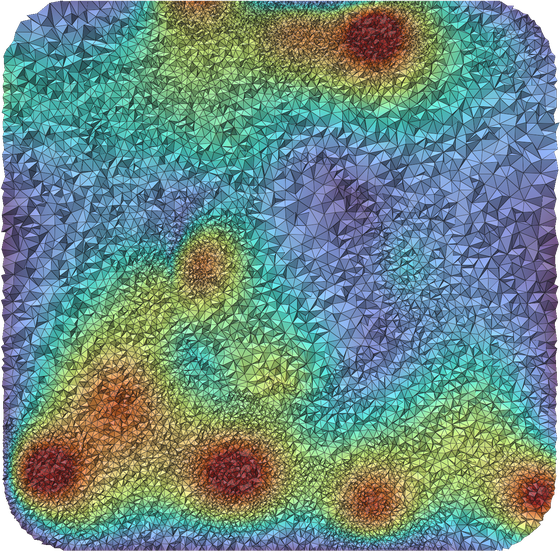} & 
            \includegraphics[width=0.15\linewidth]{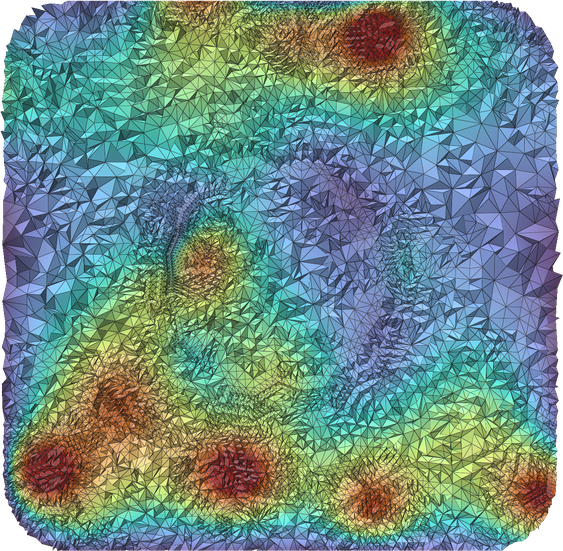} \\ 

        \end{tabular}
        
        \vspace{-2mm}
        \caption{
        We display the final meshes generated by LAMG (odd columns) and the standard AMR baseline (even columns) given the same target error thresholds $\epsilon \in \{5\times 10^{-2}, 1\times 10^{-2}, 3\times 10^{-3}\}$. The rows display three distinct test geometries. Visually, LAMG produces discretization patterns that closely mirror the AMR results, correctly identifying and refining regions of high error as the tolerance tightens. Notably, because LAMG predicts based on coarse solution, the resulting meshes exhibit more uniform element sizing compared to the sometimes noisy, localized refinement of the iterative AMR.
        }
        \label{fig:qualitative_[relative]}
    \endgroup

    \begingroup
        \renewcommand{\arraystretch}{0.9} 
        \setlength{\tabcolsep}{4pt}      
        
\begin{tabular}{l|c|c|c|c|c|c}
              & LAMG ($5\times 10^{-2}$) & AMR ($\times 10^{-2}$) & LAMG ($1\times 10^{-2}$) & AMR ($1\times 10^{-2}$) & LAMG ($3\times 10^{-3}$) & AMR ($3\times 10^{-3}$) \\ \hline

mesh1 $L_2$   & 0.0130                & 0.0135               & 0.0036               & 0.0041              & 0.0016               & 0.0013              \\
mesh2 $L_2$   & 0.0247                & 0.0238               & 0.0064                & 0.0070               & 0.0027               & 0.0025              \\
mesh3 $L_2$   & 0.0341                & 0.0333               & 0.0074               & 0.0109              & 0.0029               & 0.0032             \\
mesh1 \#Verts & 8284                 & 5530                & 32728                & 34654               & 120059               & 153288              \\
mesh2 \#Verts & 12541                & 9788                & 65406                & 71852               & 223469               & 294246              \\
mesh3 \#Verts & 17923                & 13255               & 105167               & 79315               & 372541               & 317619              \\
mesh1 Time(s) & 3.29                 & 1.95                & 7.98                 & 17.55               & 29.58                & 104.44              \\
mesh2 Time(s) & 3.87                 & 3.09                & 16.56                & 43.58               & 62.26                & 197.43              \\
mesh3 Time(s) & 5.67                 & 5.48                & 28.52                & 44.35               & 116.5                & 251.8               \\
\end{tabular}

        \captionof{table}{
        We report $L_2$ error, vertex count and runtime (s) for the examples in Figure 5, comparing both methods at the same input tolerances (defined in the bracket). At loose tolerances ($\epsilon=5e{-2}$), LAMG performs on par with AMR. However, as the tolerance tightens to $\epsilon=3 e{-3}$, LAMG demonstrates significant computational efficiency—achieving up to 3.5$\times$ speedups—while maintaining an similar error level as the AMR baseline. This confirms that our method successfully avoids the diminishing returns associated with the final, expensive iterations of standard adaptive refinement.
        }
        \label{tab:relative_poisson_perf}
    \endgroup
    \vspace{-4mm}

\end{figure*}

We present qualitative results for the relative model in Figure \ref{fig:qualitative_[relative]}, comparing the predicted meshes against the AMR solutions across varying target error tolerances ($\epsilon \in \{5\times 10^{-2}, 1\times 10^{-2}, 3\times 10^{-3}\}$). Visually, LAMG predicts discretization patterns that closely mirror the ground truth, correctly identifying and refining regions of high error while maintaining coarser elements elsewhere. This confirms that the network successfully learns the mapping from error tolerance to local mesh density. Additionally, because the predictions are derived from a coarse initial solution, the resulting LAMG meshes tend to be more uniform than the AMR, effectively filtering out localized noise while preserving the global refinement structure.

Quantitative metrics are reported in Table \ref{tab:relative_poisson_perf}. While the AMR achieves a lower absolute error at the strictest tolerance ($3\times 10^{-3}$), this marginal gain comes at a steep computational price—taking nearly $3.5\times$ longer to compute. In contrast, LAMG provides a highly accurate approximation that is sufficient for most practical engineering fidelity requirements, offering a more favorable Pareto-optimal balance between runtime and precision.

\begin{figure}[htbp!]
	\centering
	\renewcommand{\arraystretch}{0.8} 
	\begin{tabular}{@{}c@{}c@{}}
		\includegraphics[width=0.50\columnwidth]{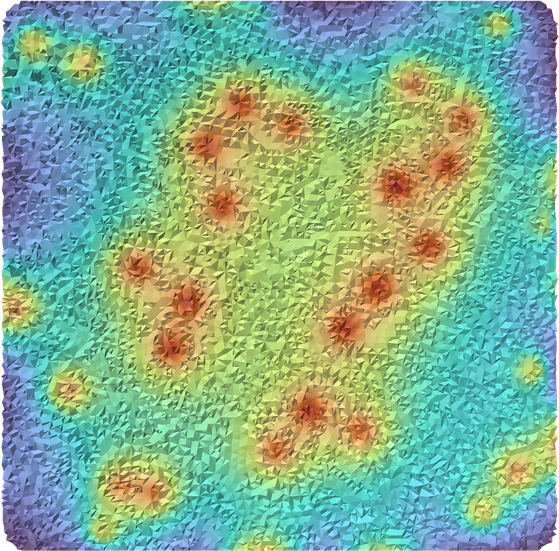} & 
		\includegraphics[width=0.50\columnwidth]{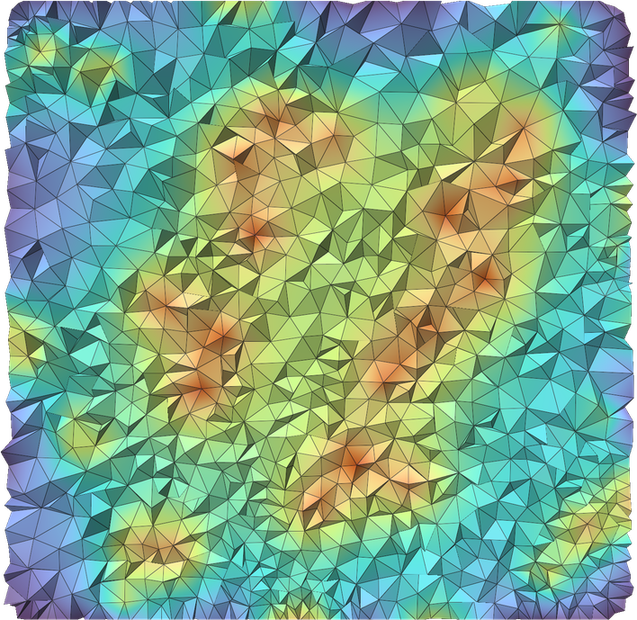}     \\
        reference & $\epsilon = 5\times 10^{-2}$\\
        \includegraphics[width=0.50\columnwidth]{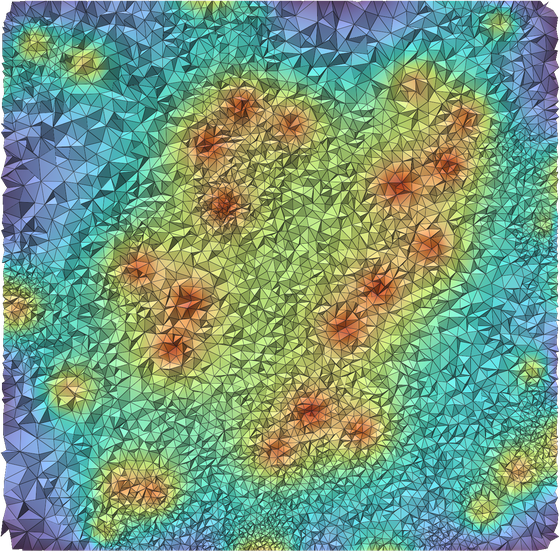} & 
		\includegraphics[width=0.50\columnwidth]{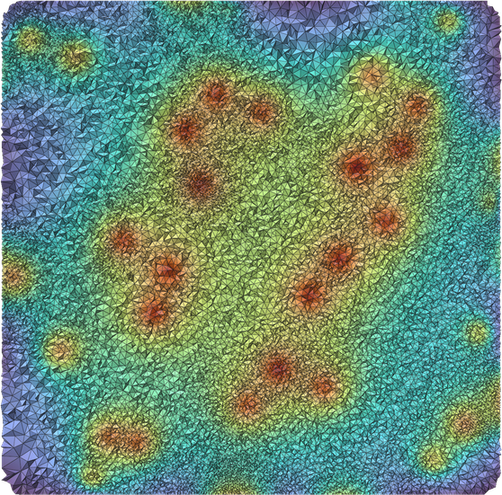}     \\
        $\epsilon = 1\times 10^{-2}$ & $\epsilon = 3\times 10^{-3}$\\

	\end{tabular}
    \vspace{-1mm}
    \caption{Visualization of the high-frequency oscillatory problem. Our method (LAMG) successfully detects the high-frequency wave patterns and adapts the mesh density locally to resolve them.}
\label{fig:relative:highfreq}
\vspace{-4mm}
\end{figure}

To further demonstrate the robustness of our method, we evaluate it on a problem exhibiting explicit high-frequency spatial oscillations, as visualized in Figure \ref{fig:relative:highfreq}. We present the reference solution alongside our method's predictions for target error tolerances $\epsilon \in \{5\times 10^{-2}, 1\times 10^{-2}, 3\times 10^{-3}\}$. The method achieves $L_2$ errors of 0.029, 0.0071, and 0.0027 with corresponding runtimes of 12.8s, 29.8s, and 108s, respectively. These results confirm that our approach effectively handles high-frequency problems, correctly identifying and refining regions of rapid solution variation to meet stringent error bounds.

\section{Extended experiments: LAMG with MC and LAMG for linear elastic deformation } \label{sec:otherexp}
In this section, we investigate the performance of LAMG when the coarse solution is obtained via Monte Carlo using a Walk on Spheres (WoS) algorithm~\cite{Sawhney:2020:MCG} and when LAMG is used to solve the elliptic PDE governing linear elastic deformation. 

\subsection{Sizing fields from coarse WoS instead of FEM} \label{sec:wos}
\begin{figure}[htbp!]
	\centering
	\setlength{\tabcolsep}{2pt} 
	\renewcommand{\arraystretch}{0.8} 
	\begin{tabular}{@{}c@{}c@{}}
		\multicolumn{2}{c}{\includegraphics[width=0.5\columnwidth]{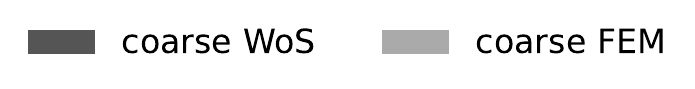}}\\
		\includegraphics[width=0.45\columnwidth]{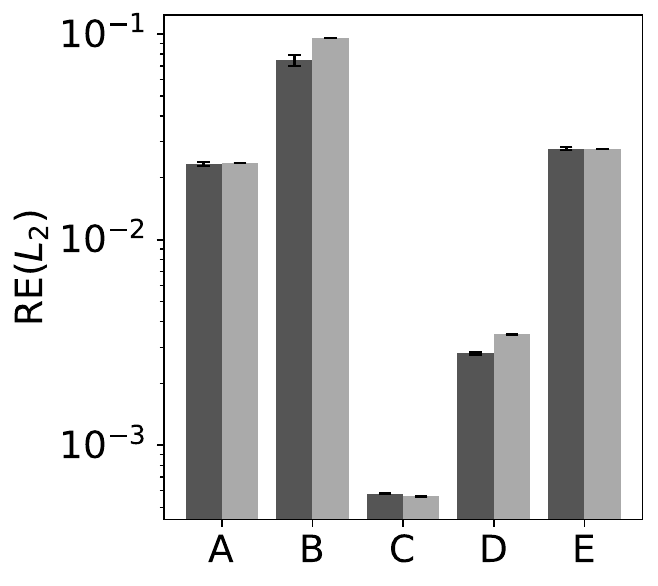} & 
		\includegraphics[width=0.45\columnwidth]{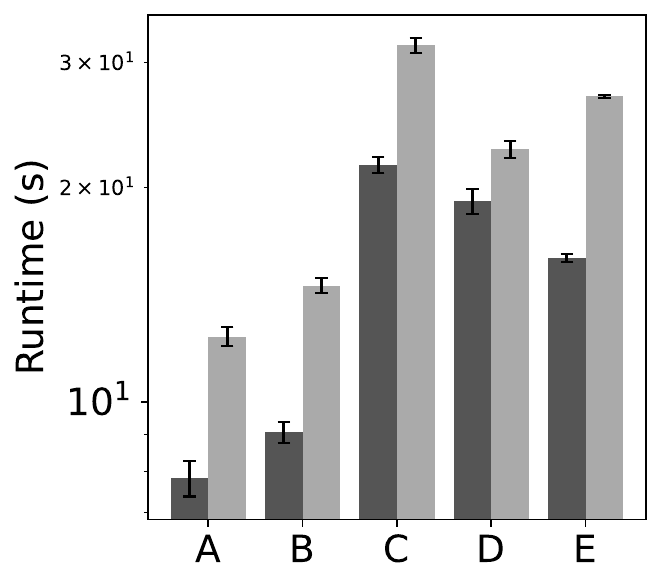}     \\
	\end{tabular}
    \vspace{-1mm}
    \caption{We compare the final $L_2$ error (left) and runtime (right) of our LAMG framework when using a sparse WoS input versus a coarse FEM input on a small set of meshes (A, B, C, D, E): cube, pillar, turbine, sausage and moldGenerator. The WoS input is faster for similar accuracy.}
\label{fig:wosvsfem}
\vspace{-4mm}
\end{figure}

We compare the performance of using WoS instead of coarse FEM in $\texttt{GenerateCoarseSolution}(\cdot)$ within Algorithm~\ref{alg:LAMG}. Figure~\ref{fig:wosvsfem} shows that both input modalities yield a similar final accuracy, but the WoS-based approach is consistently faster. The grid-free WoS solver avoids the overhead of generating an initial coarse tetrahedral mesh making it beneficial for complex geometries.

\begin{figure}[htbp]
	\centering
	\setlength{\tabcolsep}{2pt} 
	\renewcommand{\arraystretch}{0.8} 
    
	\begin{tabular}{@{}c@{}c@{}c@{}}
		\multicolumn{3}{c}{\includegraphics[width=1.0\columnwidth]{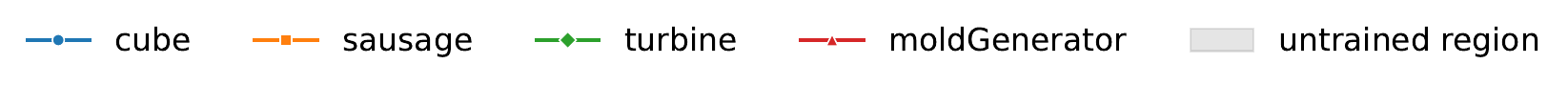}}\\
		\includegraphics[width=0.32\columnwidth]{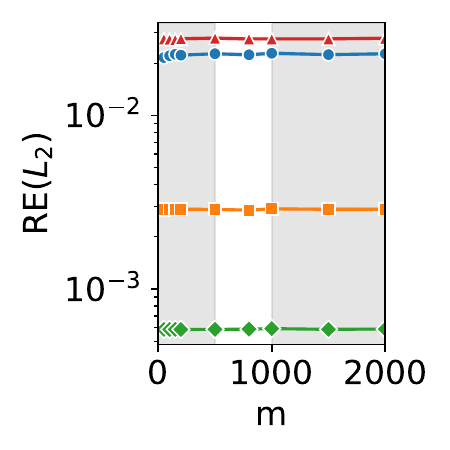} & 
		\includegraphics[width=0.32\columnwidth]{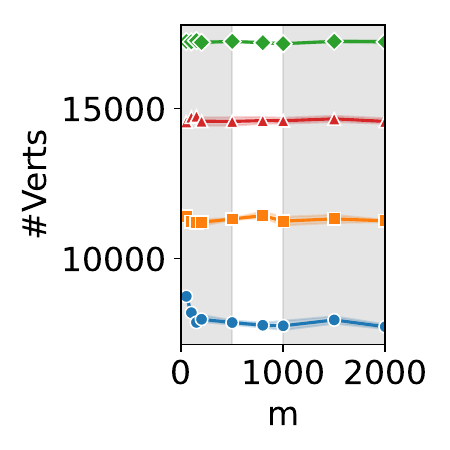} & \includegraphics[width=0.32\columnwidth]{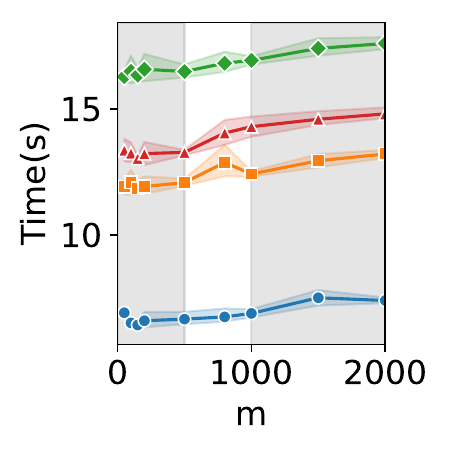}   \\
		\includegraphics[width=0.32\columnwidth]{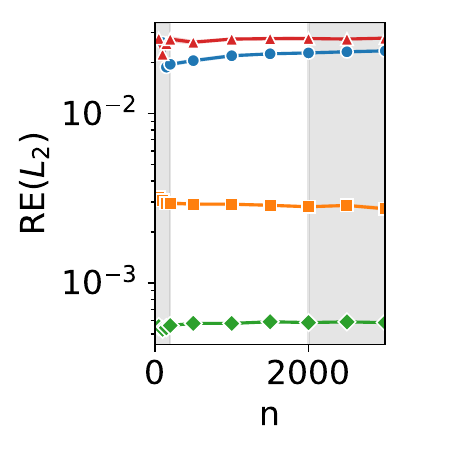} & 
		\includegraphics[width=0.32\columnwidth]{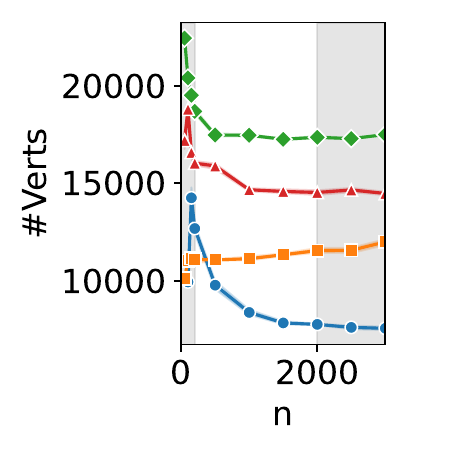}& \includegraphics[width=0.32\columnwidth]{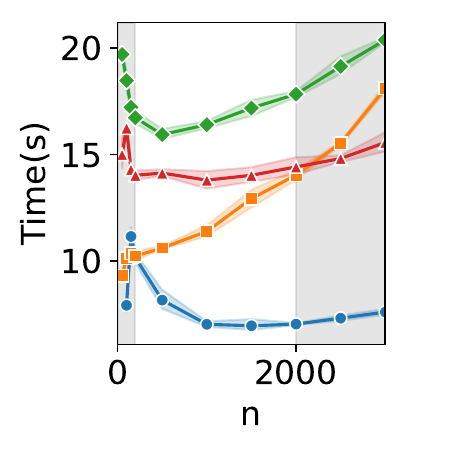}   
		
	\end{tabular} 
    \vspace{-4mm}
    \caption{
    We plot relative $L_2$ error (left column), final mesh vertex count (middle), and total runtime (right) for four different test geometries. The top row shows the effect of varying the number of walks per point ($m$), while the bottom row varies the number of initial query points ($n$). The shaded regions indicate parameter ranges that were not seen during training. The analysis shows that the final solution error is remarkably stable to changes in both parameters, demonstrating the robustness of our network to noisy or sparse inputs.
}
\label{fig:robustness}
\end{figure}

We evaluated LAMG's robustness to the quality of the sparse input, when using a coarse WoS solver for solving for steady-state heat. Figure~\ref{fig:robustness} plots the measured error, mesh vertex count and runtime while varying the number of Monte Carlo query points ($n$) and walks per point ($m$). The results show that the final solution error is stable across a wide range of both $n$ and $m$, even for values outside the training distribution. While runtime is naturally affected by these parameters (more points or walks increase the initial solve time), the stability of the final error demonstrates that the network has learned the structure of the solution even from very sparse or noisy inputs.

\subsection{Linear Elastic Deformation} \label{sec:elasticity-direct}

\begin{figure}[htbp!]
	\centering
	\setlength{\tabcolsep}{2pt} 
	\renewcommand{\arraystretch}{0.8} 
	\begin{tabular}{@{}c@{}c@{}}
		\multicolumn{2}{c}{\includegraphics[width=0.70\columnwidth]{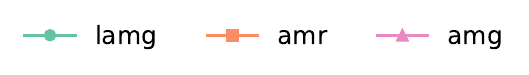}}\\
		\includegraphics[width=0.45\columnwidth]{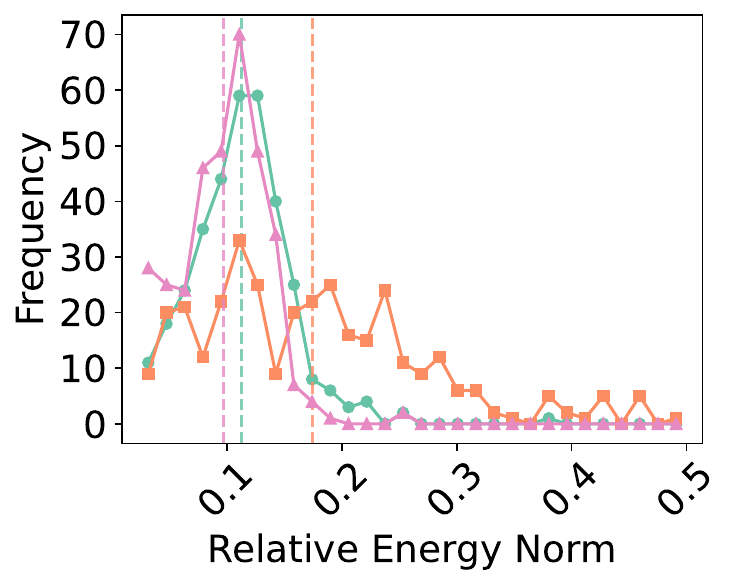} & 
		\includegraphics[width=0.45\columnwidth]{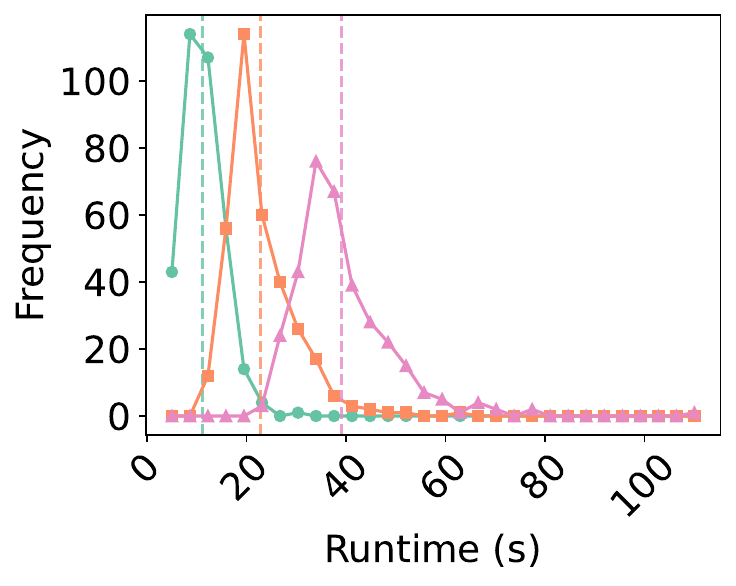}     \\
	\end{tabular}
    \vspace{-1mm}
    \caption{Generalization to homogeneous linear elasticity. Frequency distributions of (Left) Relative Energy Norm and (Right) Runtime comparing our method (LAMG) against AMR and AMG baselines. Dashed vertical lines indicate mean values. While iterative AMR often stagnates at stress singularities—resulting in excessive localized refinement and higher relative errors—LAMG and AMG leverage coarse estimates to produce more uniform, better-conditioned meshes. Consequently, LAMG matches the accuracy of AMG while significantly outperforming both methods in computational speed.
}
\label{fig:direct_elasticity}
\end{figure}

We evaluate LAMG's ability to solve linear elastic deformation with 339 PDEs with homogeneous mesh properties (uniform Young's modulus and Poisson's ratio). We use element-wise von Mises stress as input rather than the 3D direct degrees of freedom to avoid increasing the complexity of the network. Figure \ref{fig:direct_elasticity} compares the relative energy norm (left) and runtime (right) of our method (LAMG) against AMR and AMG baselines. The error distributions show that LAMG performs similarly with AMG and has a better profile than AMR. One cause for this is that iterative AMR suffers from over-refinement at stress singularities. Since LAMG derives sizing fields from coarse estimates, there is some implicit smoothing, resulting in more uniform element distributions that better resolve global features. The runtime analysis confirms that LAMG is consistently faster than AMR and AMG. This efficiency could be valuable for interactive design, as it provides rapid feedback on problematic setups, such as excessive displacements or singular constraints, allowing users to adjust boundary conditions without waiting for the costly convergence of iterative refinement. 

For these experiments, we generated ground truth data on 17 engineering-focused geometries (normalized dimensions), including beams, plates with holes, and mechanical brackets. Since iterative AMR tends to stagnate at stress singularities, we used converged uniform dense solutions as the reference ground truth. A zero Dirichlet condition ($\boldsymbol{u}=0$) is applied to the left face ($x=0$), and a random uniform displacement $\boldsymbol{u} \sim \mathcal{U}(-0.05, 0.05)$ is applied to the right face ($x=8$). We sampled Young's modulus $E \sim \mathcal{U}(5, 50)$ and Poisson's ratio $\nu \sim \mathcal{U}(0.1, 0.45)$. We train the direct method on 7 of these shapes with 1400 boundary conditions and test on the 10 held-out set. The model and loss are identical to those used for the Poisson equation, but we assess accuracy using the relative energy norm since, unlike the standard $L_2$ norm, the energy norm accounts for the material stiffness, providing a physically meaningful measure of the strain energy error. Given the global stiffness matrix $K$, the relative error is defined explicitly as:
\begin{equation}
e_{\text{elas}} = \frac{|| \U_a - \hat{\U} ||_K}{|| \hat{\U} ||_K} = \frac{\sqrt{( \U_a - \hat{\U})^\top K ( \U_a - \hat{\U})}}{\sqrt{{\hat{\U}}^\top K \hat{\U} }}
\end{equation}

\begin{figure*}[t!]
    \centering
    
    \begingroup
        \setlength{\tabcolsep}{2pt} 
        \renewcommand{\arraystretch}{0.8}
        \begin{tabular}{ccccccc}
            mesh & reference & sizing field & $\eta=1$ & $\eta=0.8$ & AMR \\ 
            \includegraphics[width=0.15\linewidth]{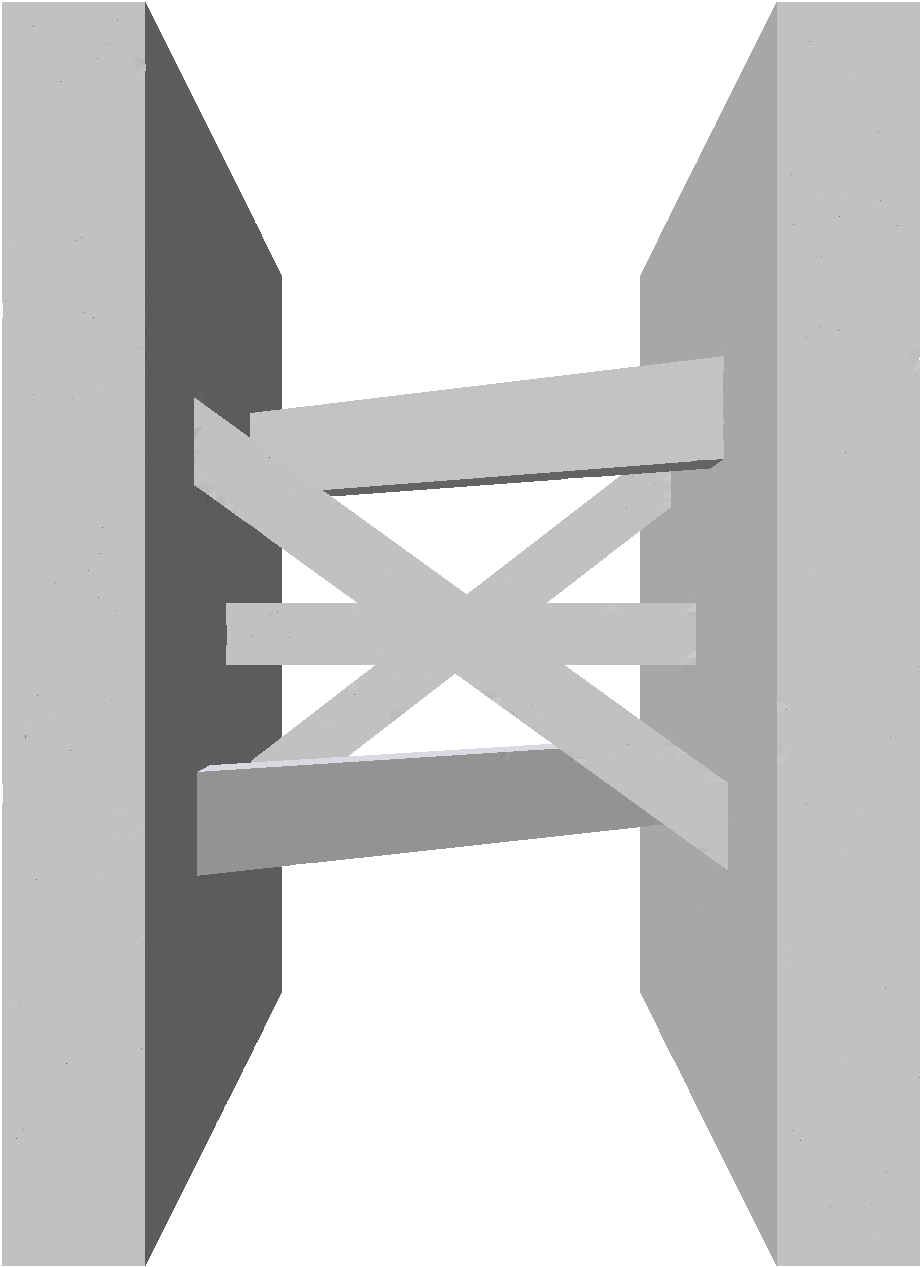} & 
            \includegraphics[width=0.15\linewidth]{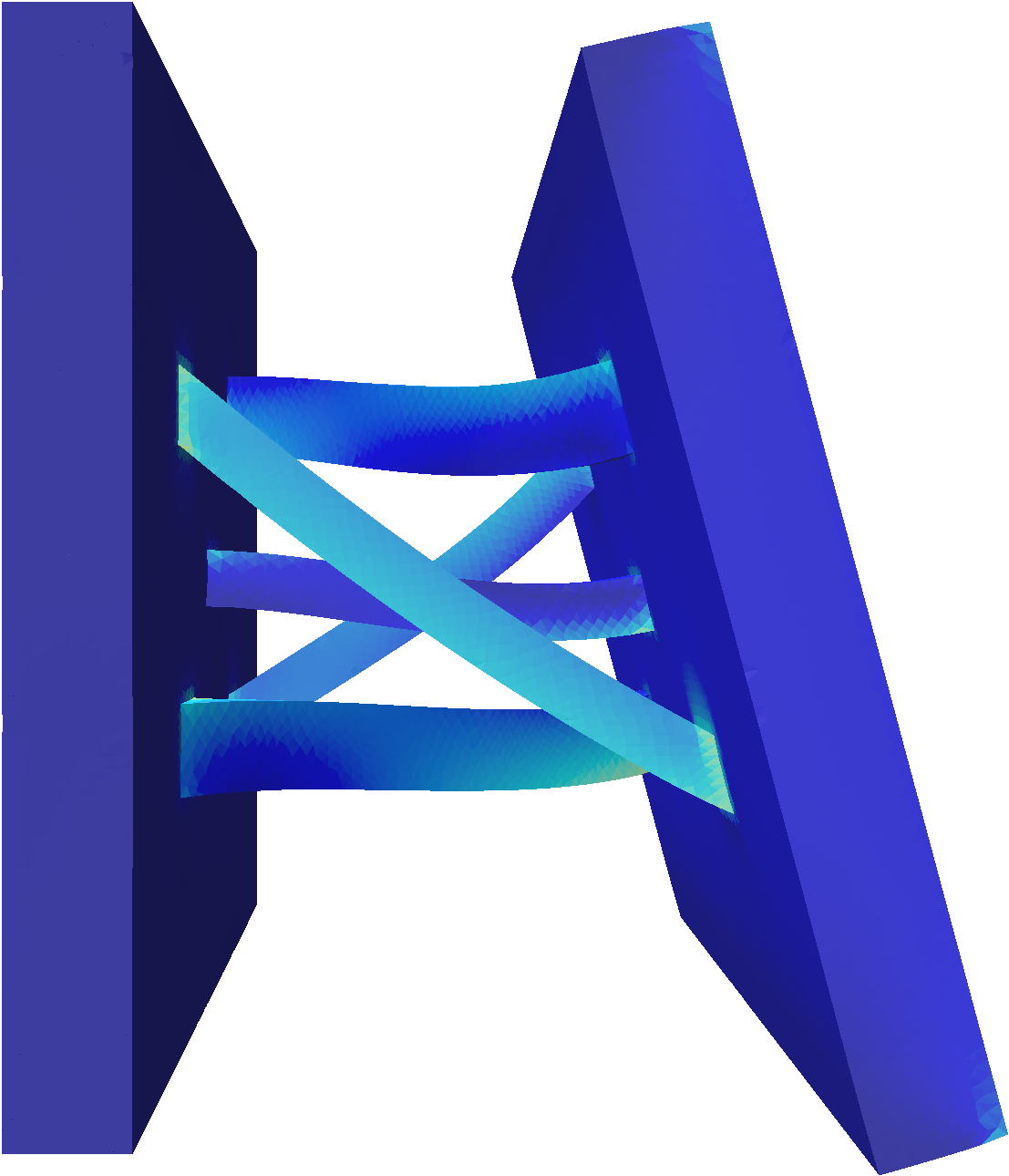} & 
            \includegraphics[width=0.15\linewidth]{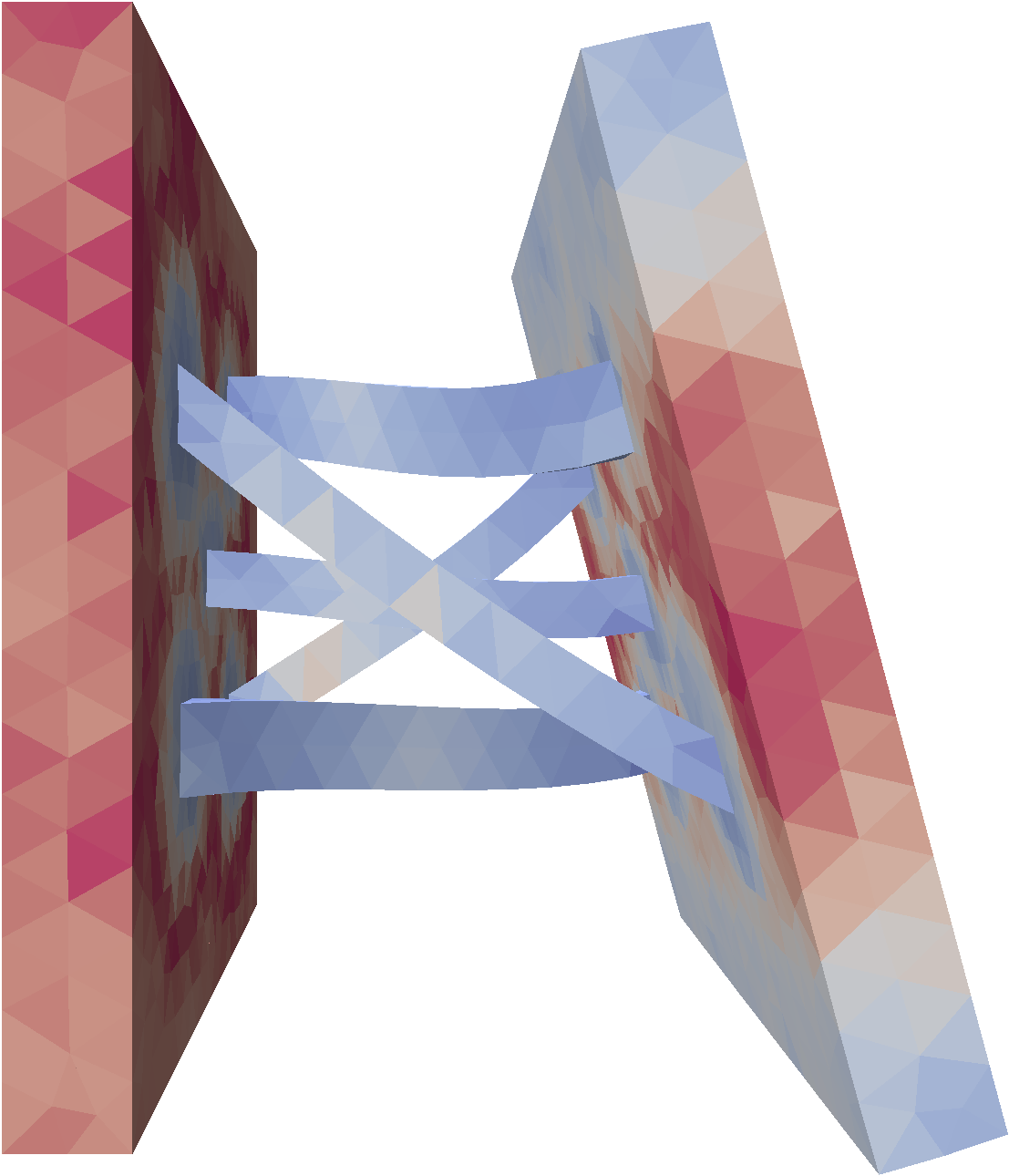} & 
            \includegraphics[width=0.15\linewidth]{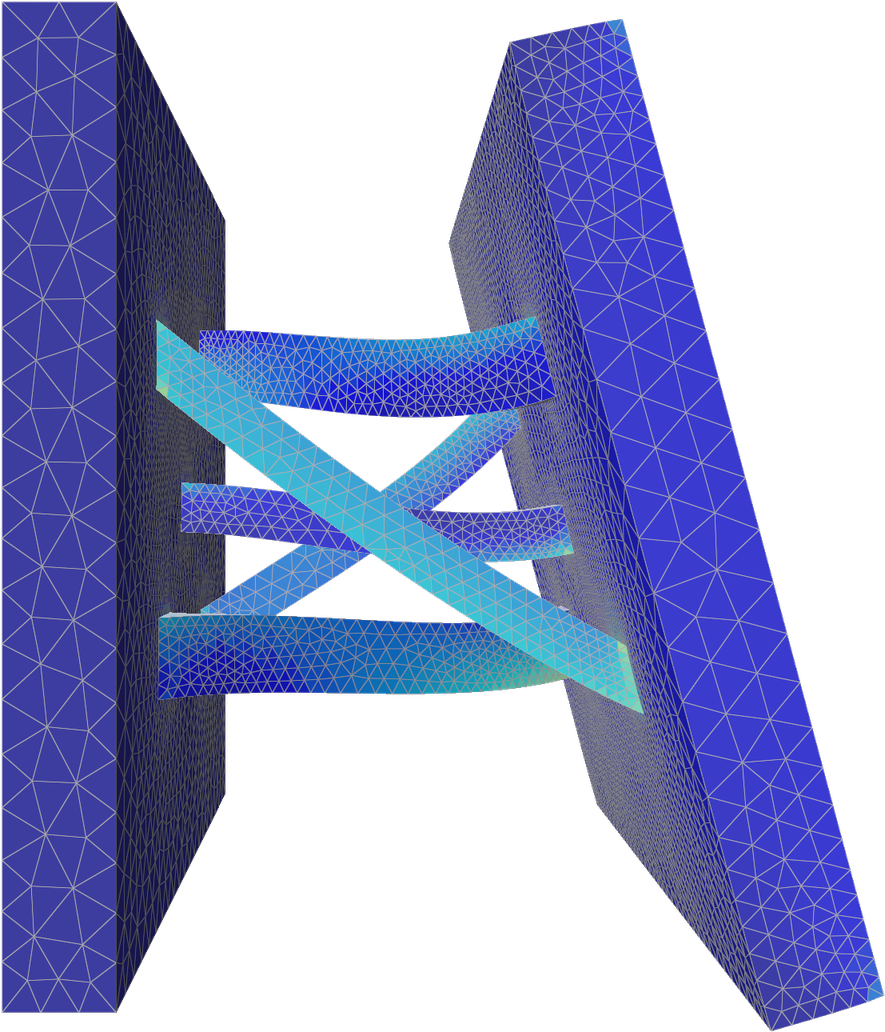} & 
            \includegraphics[width=0.15\linewidth]{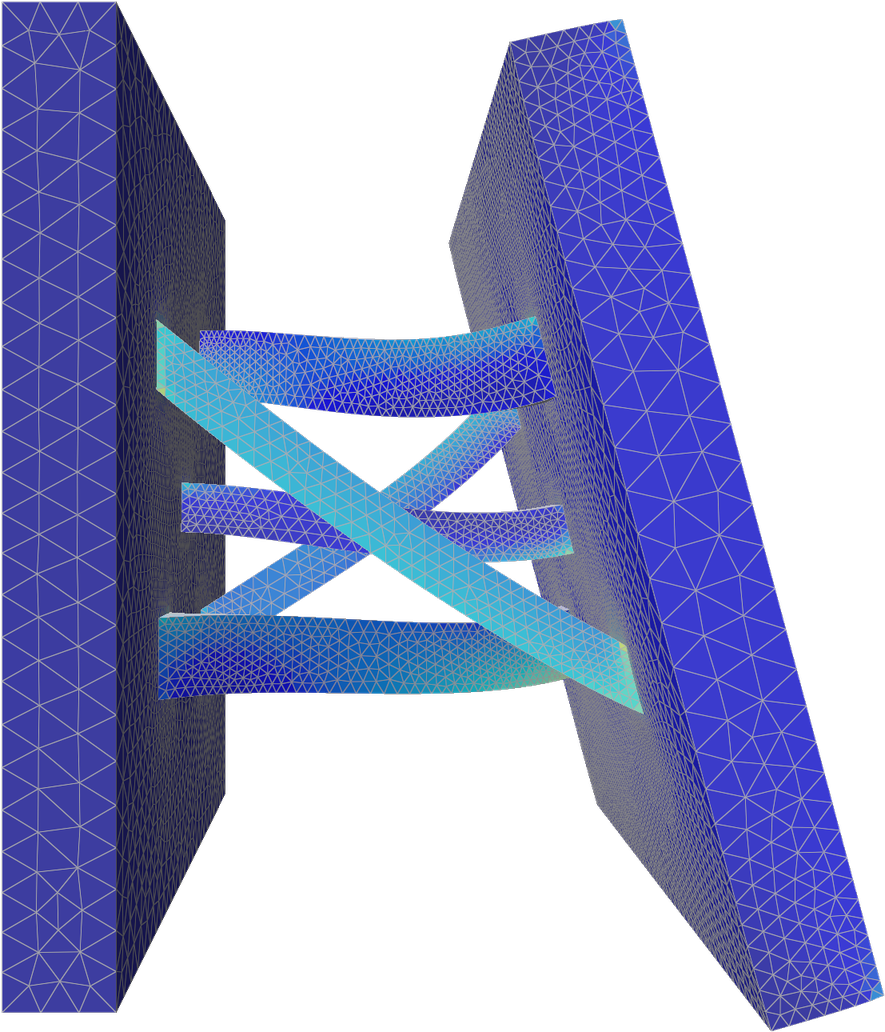} & 
            \includegraphics[width=0.15\linewidth]{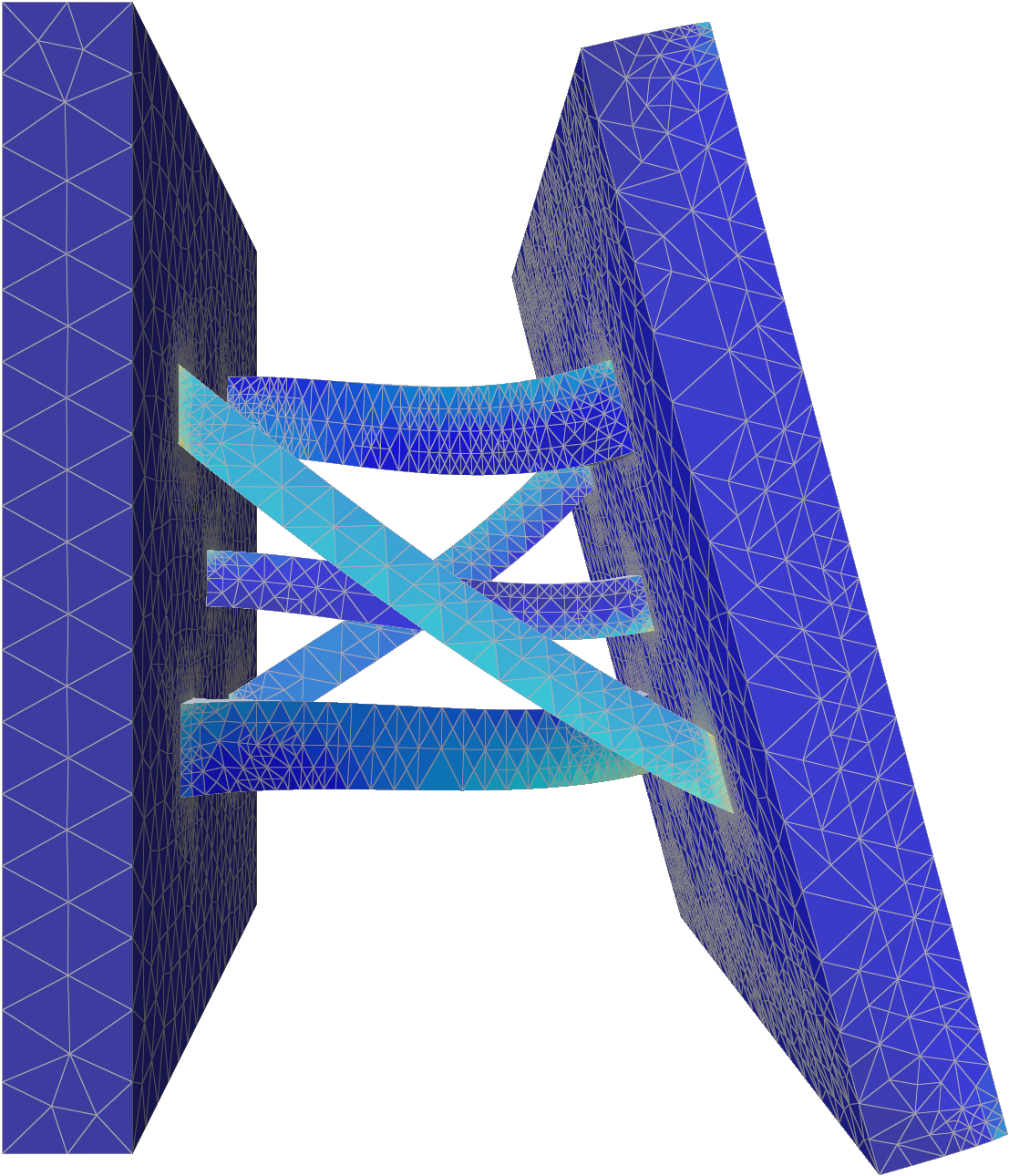} \\ 
            
            \includegraphics[width=0.15\linewidth]{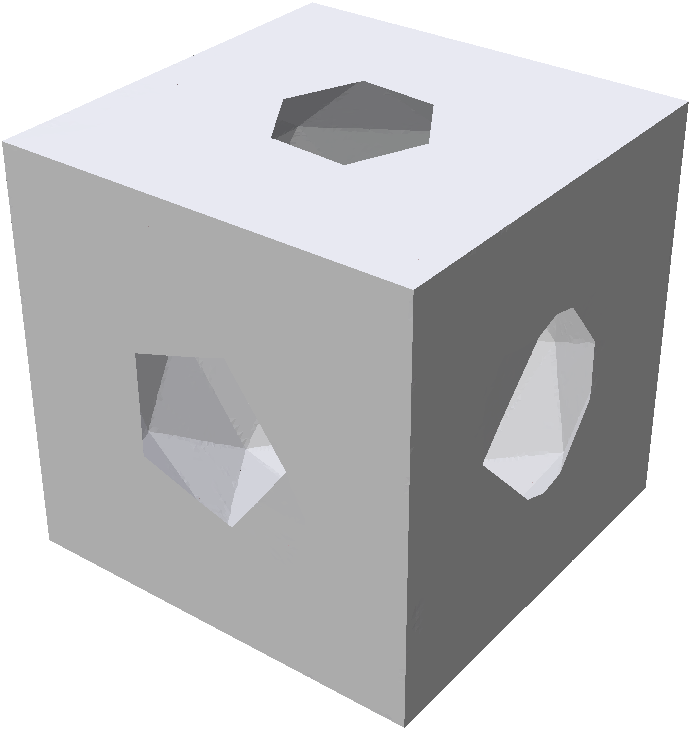} & 
            \includegraphics[width=0.15\linewidth]{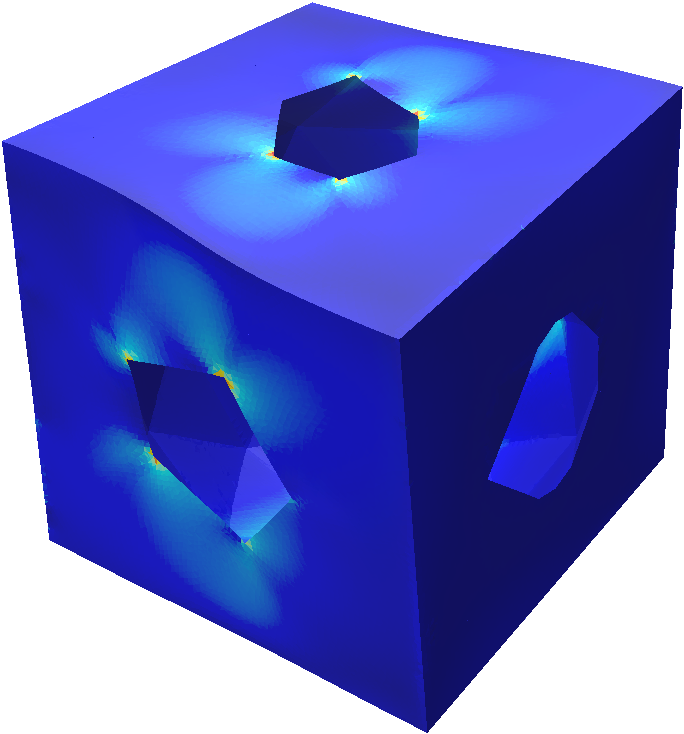} & 
            \includegraphics[width=0.15\linewidth]{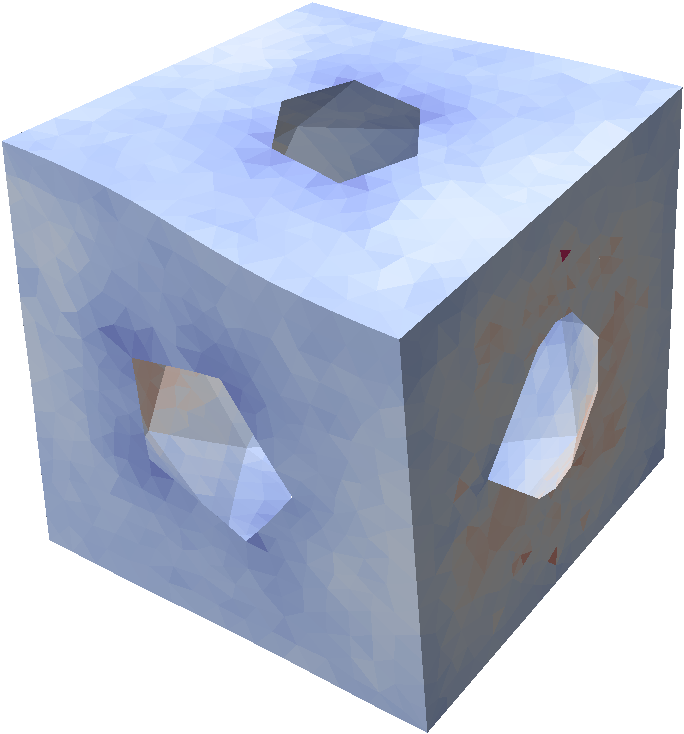} & 
            \includegraphics[width=0.15\linewidth]{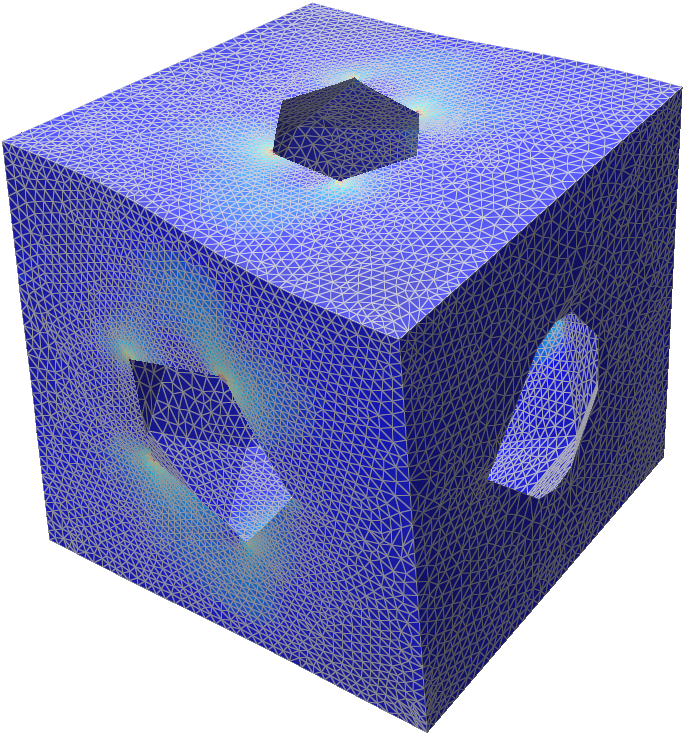} & 
            \includegraphics[width=0.15\linewidth]{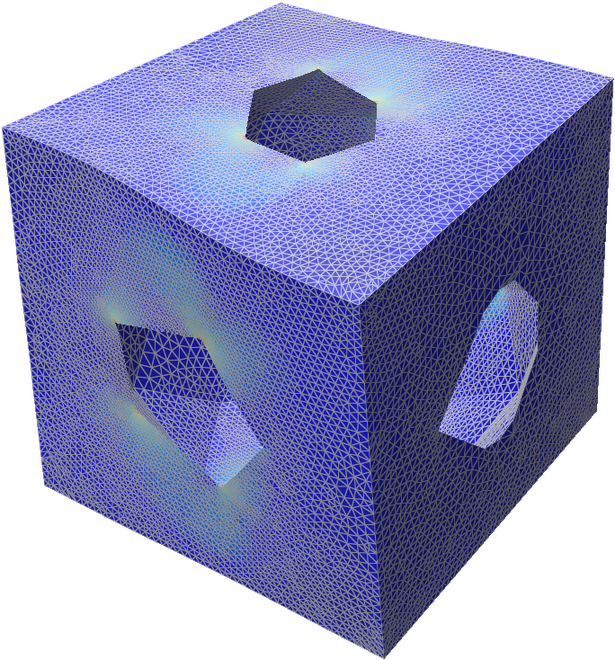} & 
            \includegraphics[width=0.15\linewidth]{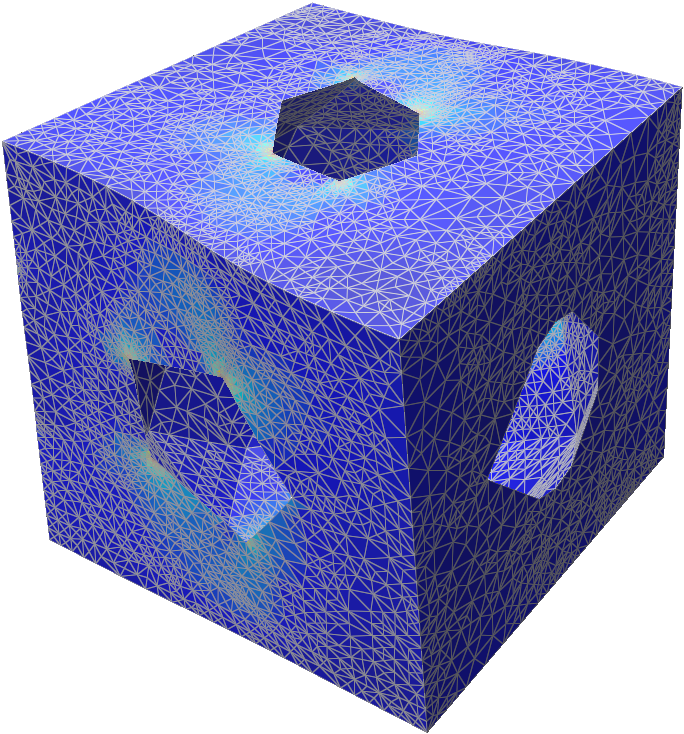} \\ 
            
            \includegraphics[width=0.15\linewidth]{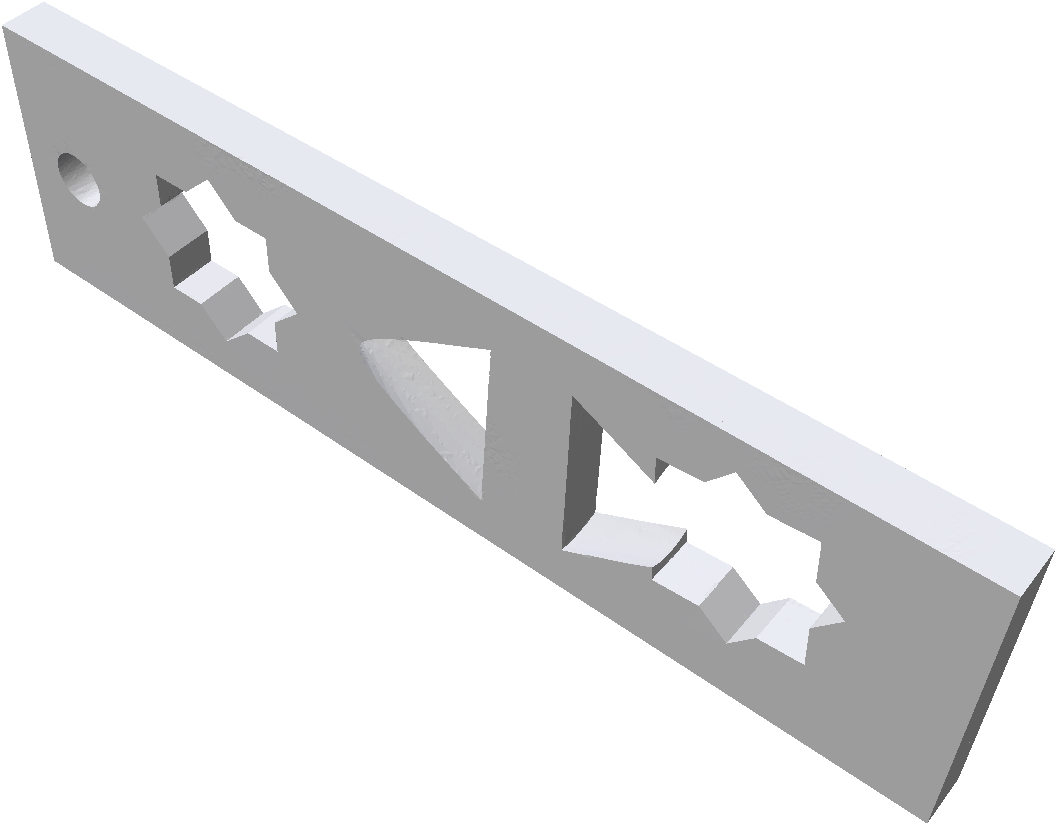} & 
            \includegraphics[width=0.15\linewidth]{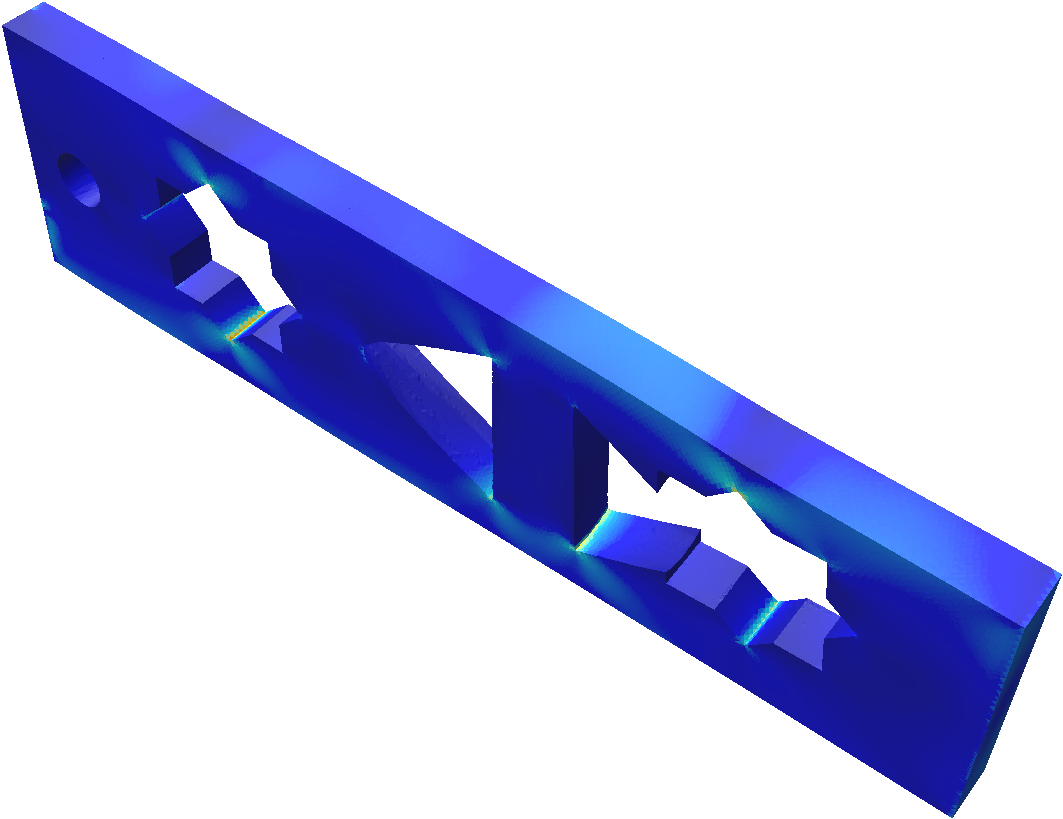} & 
            \includegraphics[width=0.15\linewidth]{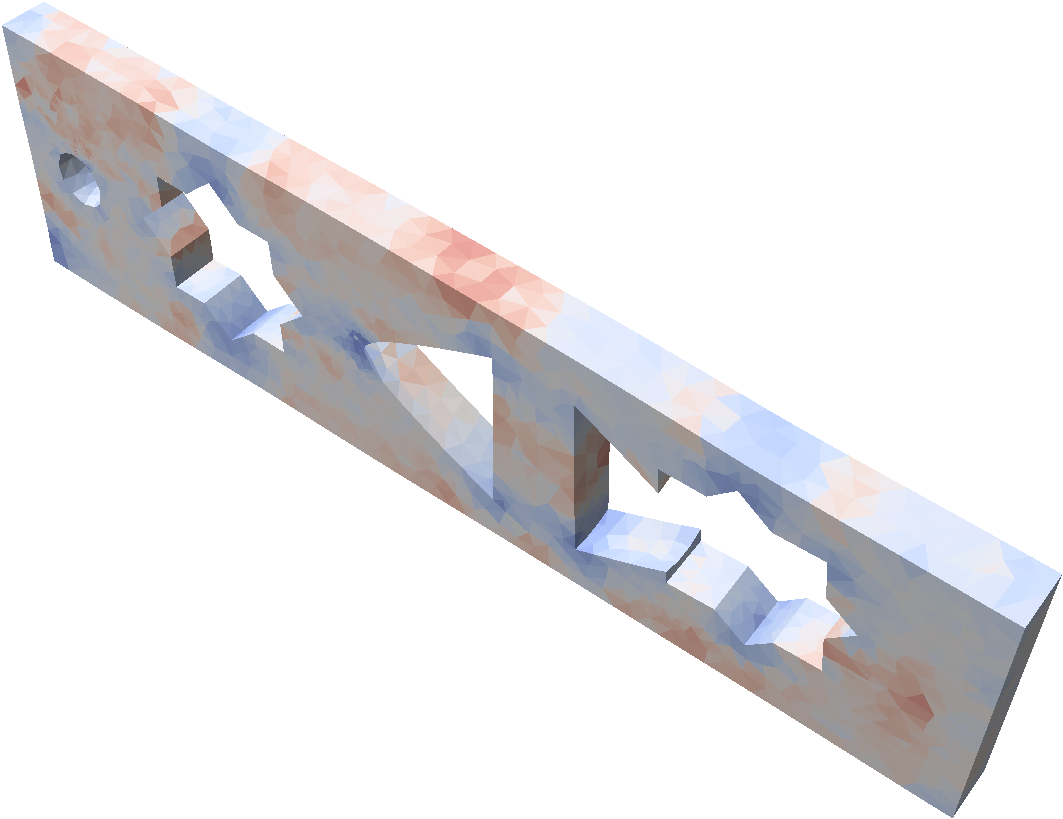} & 
            \includegraphics[width=0.15\linewidth]{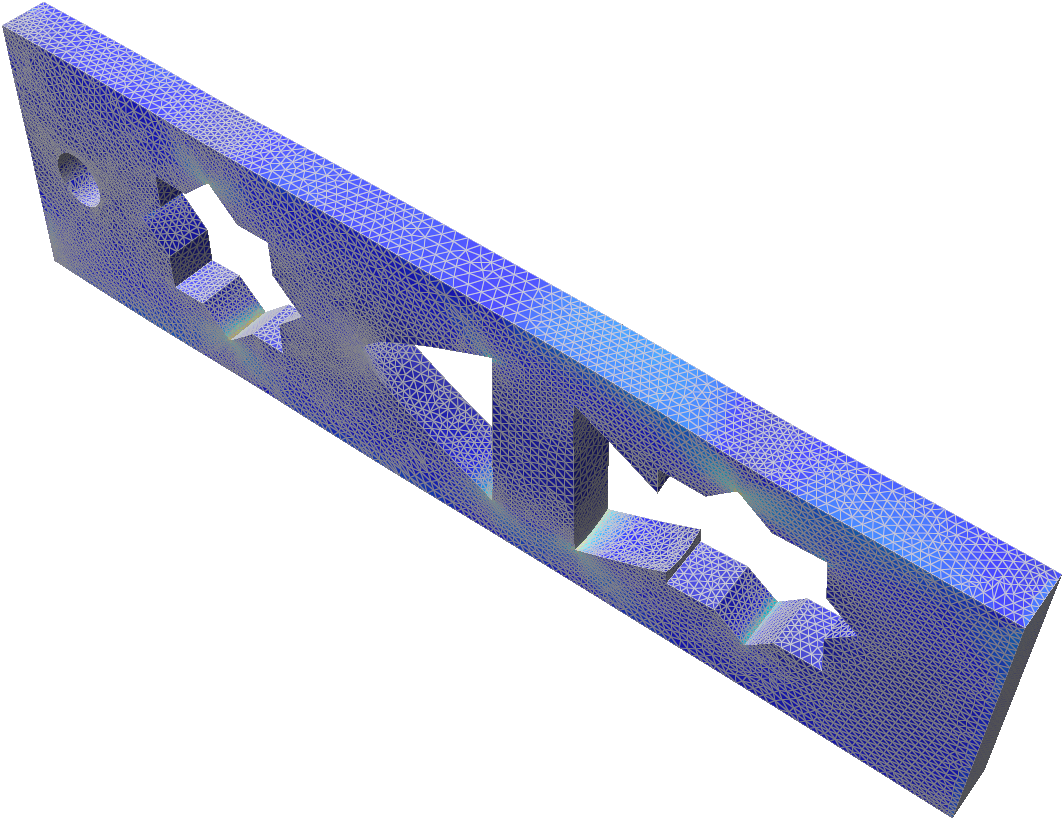} & 
            \includegraphics[width=0.15\linewidth]{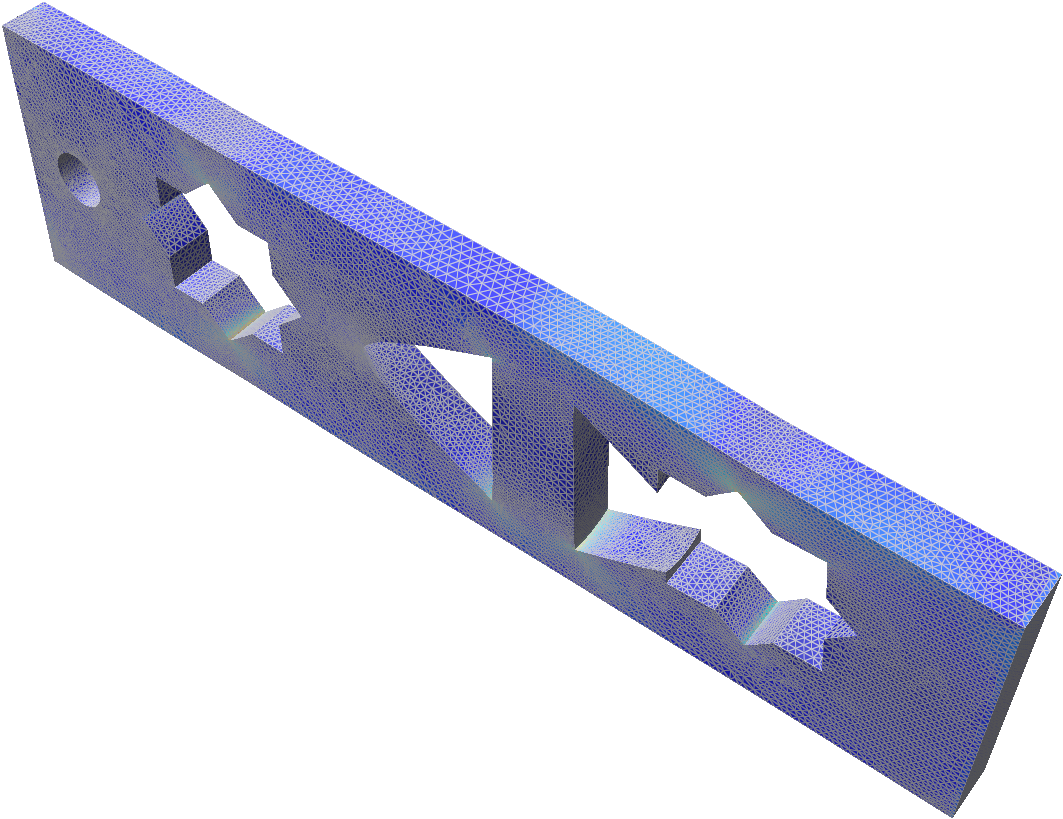} & 
            \includegraphics[width=0.15\linewidth]{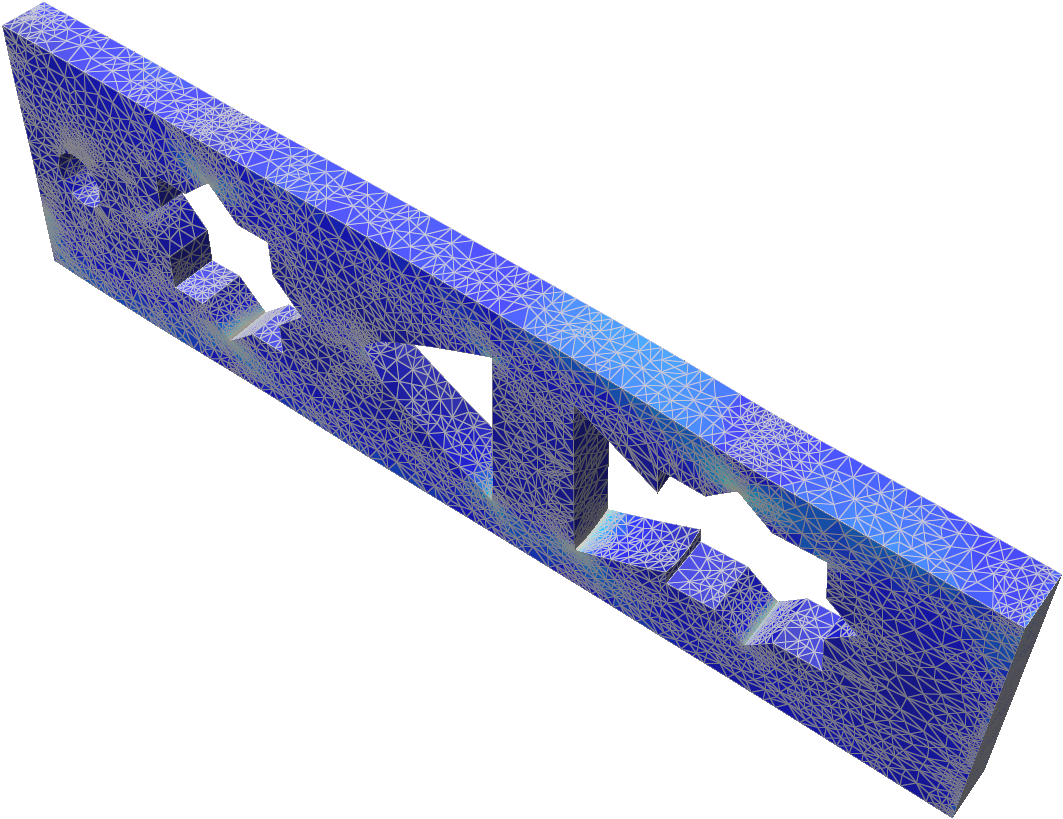} \\ 
            
            \includegraphics[width=0.15\linewidth]{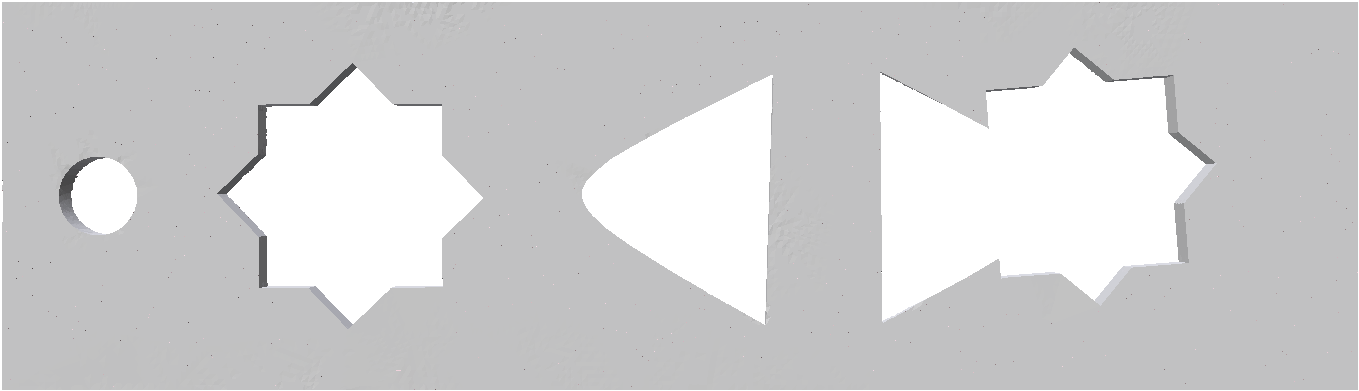} & 
            \includegraphics[width=0.15\linewidth]{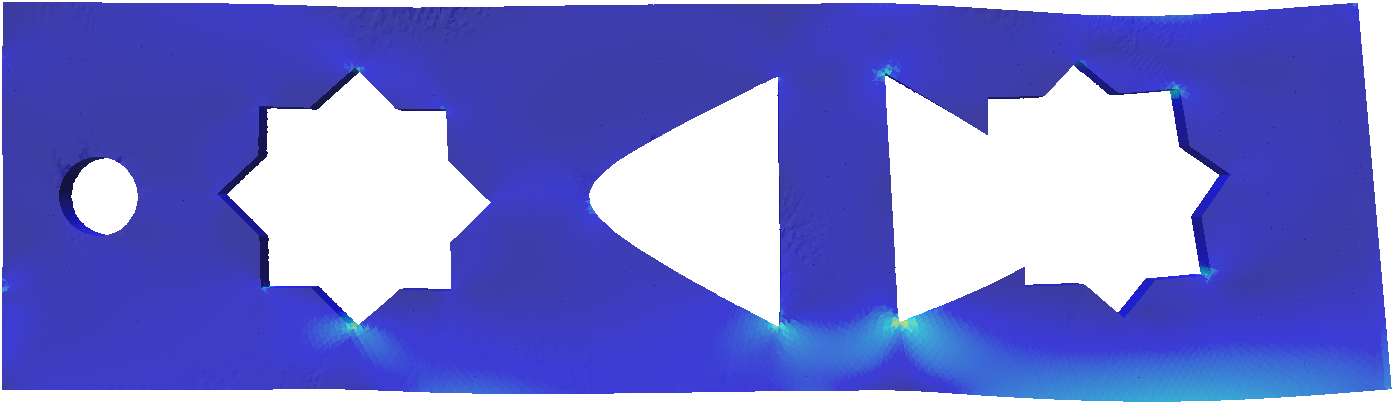} & 
            \includegraphics[width=0.15\linewidth]{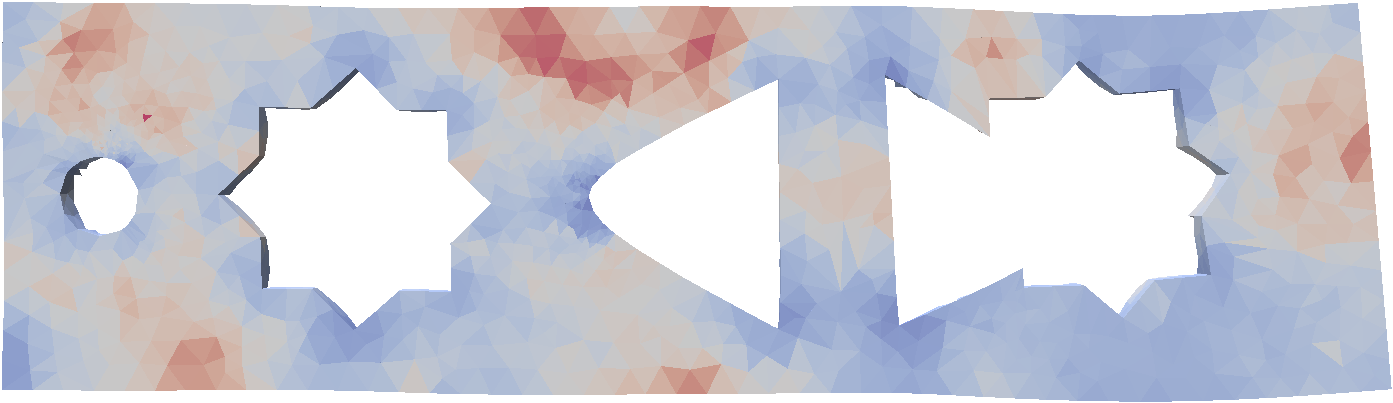} & 
            \includegraphics[width=0.15\linewidth]{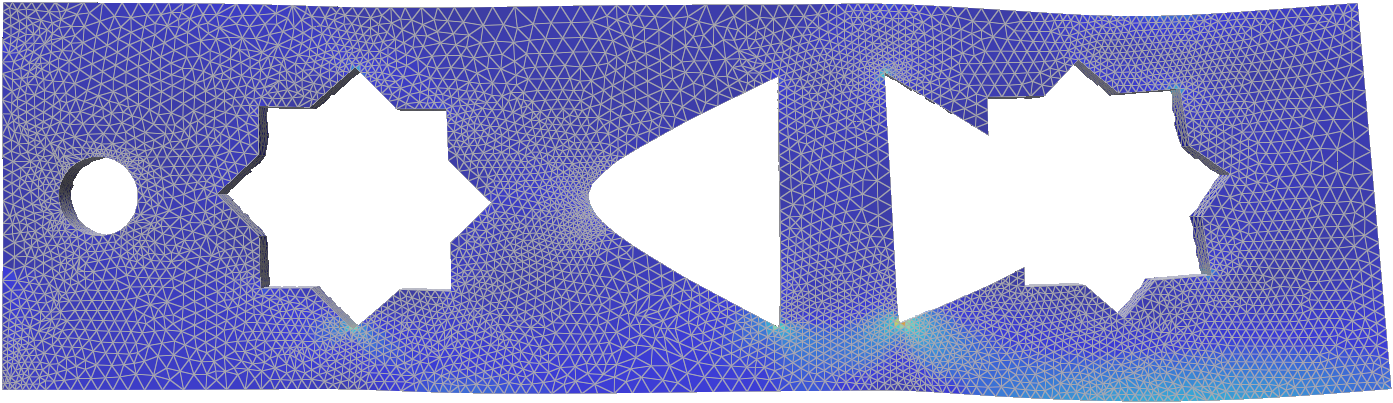} & 
            \includegraphics[width=0.15\linewidth]{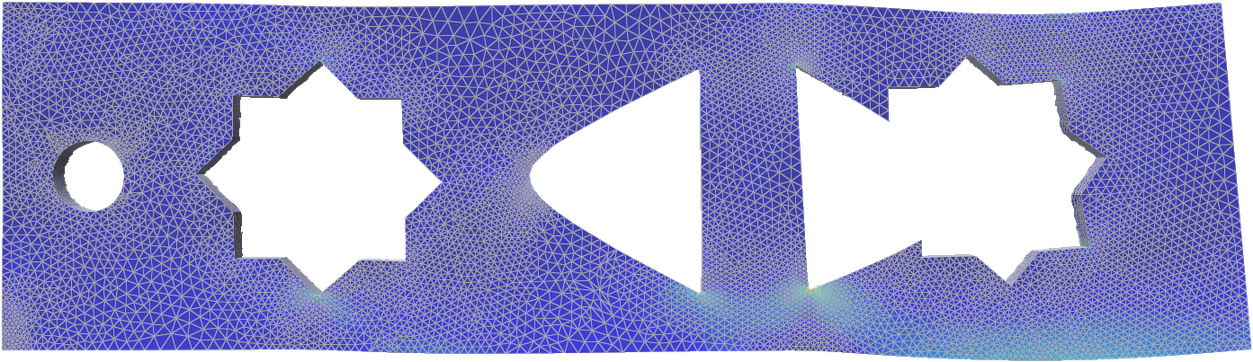} & 
            \includegraphics[width=0.15\linewidth]{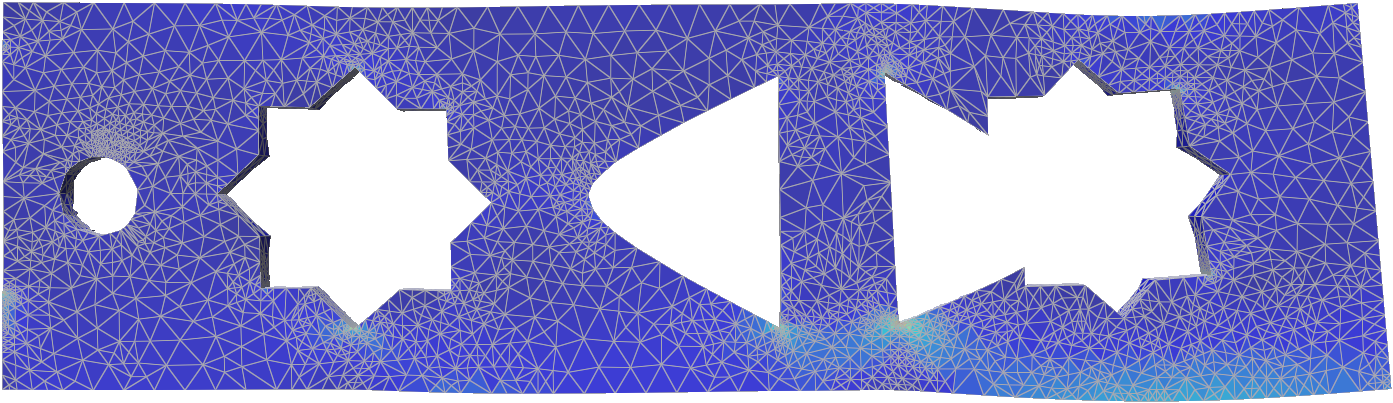} \\ 
        \end{tabular}
        
        \vspace{-2mm}
        \caption{
        Each row presents a different challenging case. From left to right: input mesh, reference solution, predicted sizing field, our final mesh ($\eta=1.0, 0.8$), and the iterative AMR baseline. These visuals demonstrate our one-shot method produces quality comparable to iterative AMR.
        }
        \label{fig:qual_elasticity}
    \endgroup

    \begingroup
        \renewcommand{\arraystretch}{0.9} 
        \setlength{\tabcolsep}{4pt}      
        
\begin{tabular}{l |c |c |c |c |c| c|c|c|c|c}
    Mesh & $\eta_{1}$ \textit{EN} & $\eta_{0.8}$ \textit{EN} & AMR \textit{EN} & $\eta_1$ \#Vert & $\eta_{0.8}$ \#Vert & AMR \#Vert & $\eta_1$ Time (s) & $\eta_{0.8}$ Time (s) & AMR Time (s)&Speedup \\
    \hline
    Mesh1   & 0.1284 & 0.0991 & 0.1050 & 14612 & 24926 & 31331 & 3.92 & 6.58 & 15.94 & 2.42\\
    Mesh2   & 0.2824 & 0.1992 & 0.4923 & 20180 & 33603 & 30134 & 5.12 & 8.42 & 16.31 & 1.93\\
    Mesh3 x & 0.4791 & 0.2247 & 0.4988 & 33294 & 57732 & 43846 & 15.53 & 34.33 & 65.33 & 1.90\\
    Mesh3 z & 0.4046 & 0.3314 & 0.5652 & 30060 & 52484 & 30744 & 13.88 & 29.76 & 42.22 & 2.14\\
\end{tabular}
        \captionof{table}{
        Quantitative performance for the elasticity examples in Figure~\ref{fig:qual_elasticity}, confirming significant speedup over AMR for similar error levels. The \textit{EN} is for Energy Norm. we see that for 4 meshes, our method can achieve more than $1.9 \times$ speedup when $\eta ={0.8}$.
        }
        \label{tab:qual_elasticity_perf}
    \endgroup
    \vspace{-4mm}

\end{figure*}

\subsubsection{Qualitative results for elasticity}

Figure~\ref{fig:qual_elasticity} visualizes the results from LAMG on four unseen engineering geometries under different boundary conditions, such as rotations, stretching, bending and compressions. For example, in 'Mesh3',  the same geometry is tested under two different displacement boundary conditions (Mesh3 x and Mesh3 z). Our network results in two distinct sizing fields, correctly identifying unique areas of high stress concentration induced by the specific load. The final adaptive mesh is appropriately refined in those areas, capturing the relationship between the coarse stress solution and the required mesh density. In this visualization, the peak von Mises stress is consistently observed at the sharp corners. The corresponding quantitative results (where \textit{EN} is the energy norm), detailed in Table~\ref{tab:qual_elasticity_perf}, confirm that these  results are also achieved with a speedup of more than $1.9\times$.

\subsection{Heterogeneous Linear Elasticity}\label{sec:heterogeneous}
Textbook problems with typical boundary conditions (fixed support, distributed load) introduce kinematic singularities at re-entrant corners and Dirichlet-Neumann transitions. In these regions, the theoretical stress is infinite, causing standard error estimators to diverge leading to a saturation effect where the global error norm plateaus despite refinement. 
To evaluate the method despite these artifacts, we introduce heterogeneous materials with spherical inclusions. Unlike boundary singularities, the stress jumps at material interfaces are finite and resolvable. This creates a robust test regime where the method must resolve physical features (inclusions) while mimicking the reference strategy's behavior at unavoidable numerical singularities.

\begin{figure}[htbp]
    \centering
    \includegraphics[width=0.5\columnwidth]{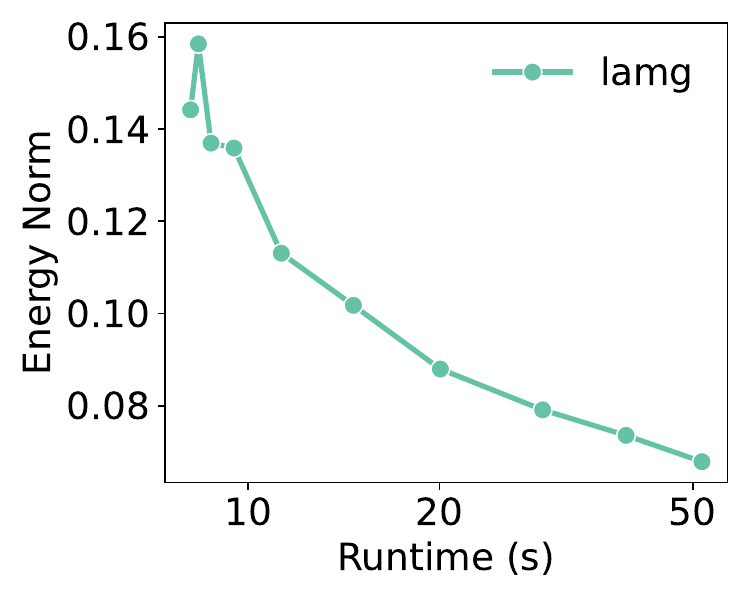} 
    \vspace{-2mm}
    \caption{We plot the Energy Norm error against total runtime (s) for varying input tolerances. The decrease of Energy Norm along increasing of runtime confirms that the Relative Method ($h_{\text{relative}}$) successfully targets increasingly strict error thresholds, effectively trading computational cost for higher fidelity even in the presence of sharp internal Young's modulus contrasts.}
    \label{fig:hetro}
    \vspace{-4mm}
\end{figure}

\begin{figure*}[t!]
    \centering

        \setlength{\tabcolsep}{2pt} 
        \renewcommand{\arraystretch}{0.8}
        \begin{tabular}{ccccccccccc}
            Young's Modulus &reference & $\epsilon=3.50e-4$  &$\epsilon=5.85-5$ & $\epsilon=2.00e-5$ \\ 
            \includegraphics[width=0.2\linewidth]{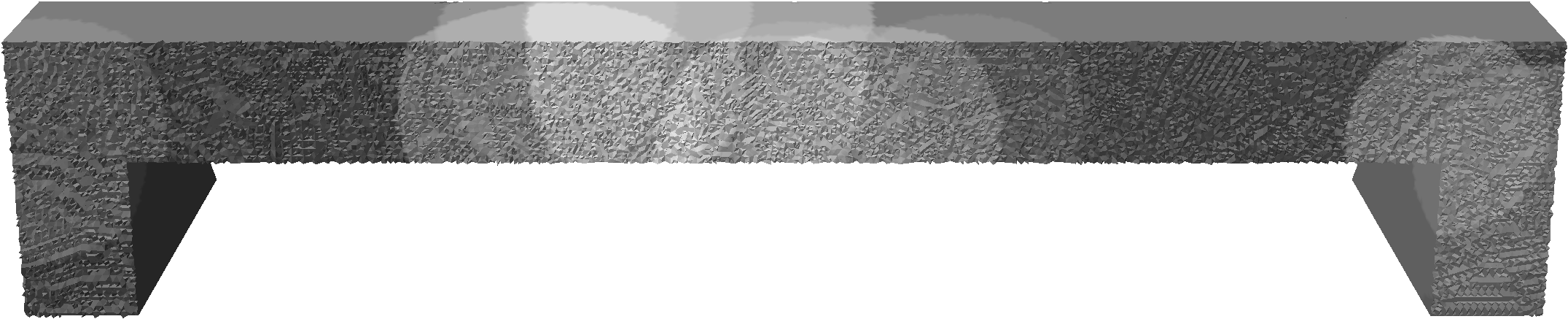} & 
            \includegraphics[width=0.20\linewidth]{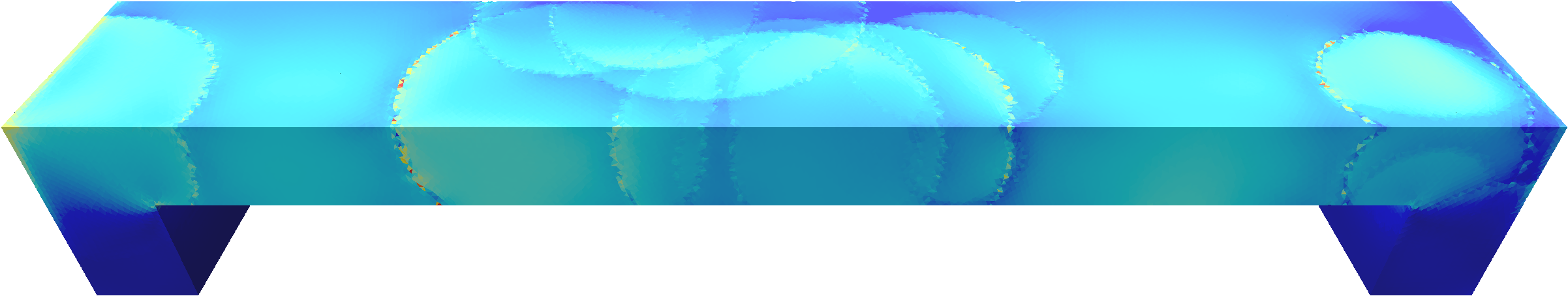} & 
            \includegraphics[width=0.2\linewidth]{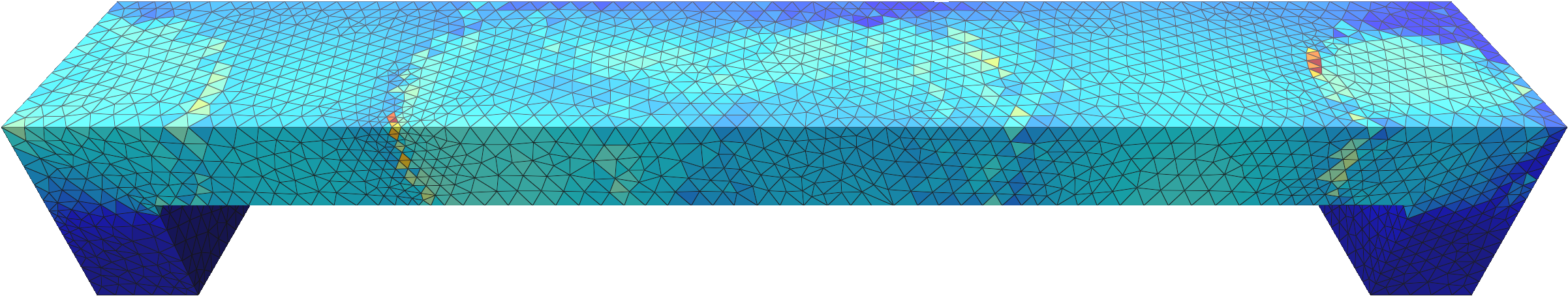} & 
            \includegraphics[width=0.20\linewidth]{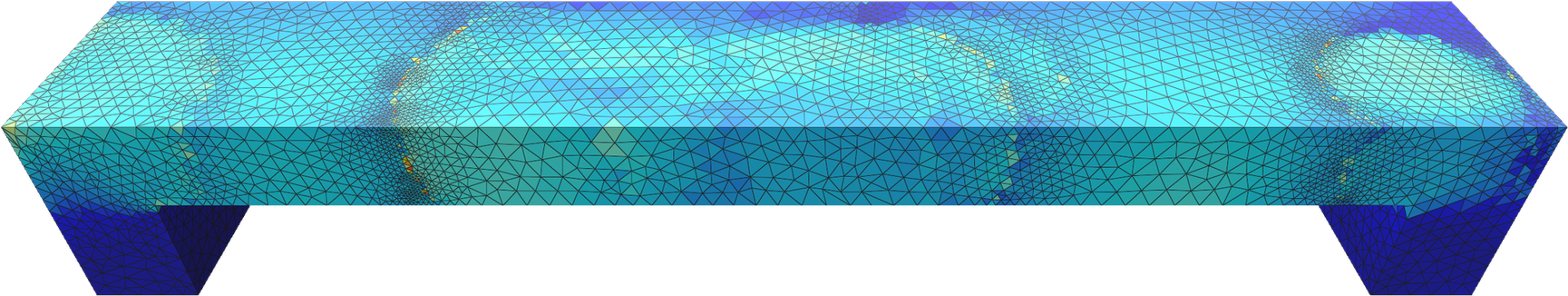} & 
            \includegraphics[width=0.20\linewidth]{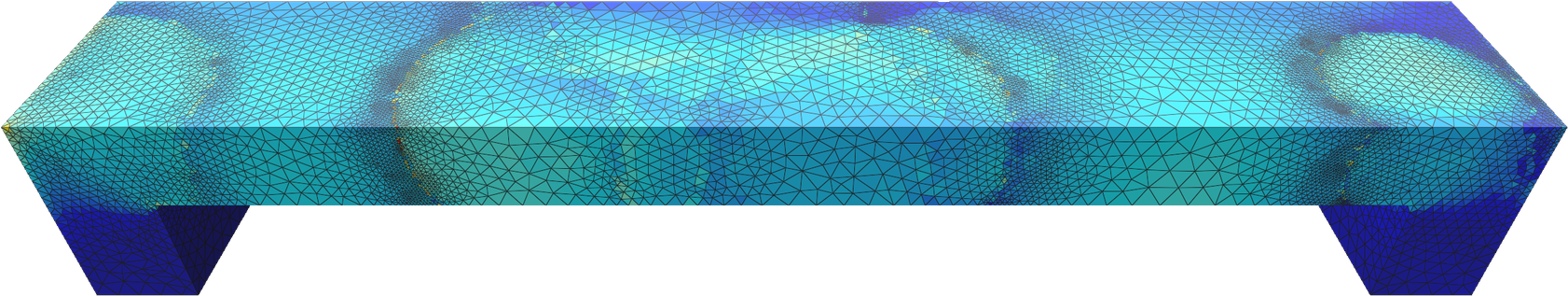} &

            
        \end{tabular}
        
        \vspace{-2mm}
        \caption{
Qualitative evolution of the mesh for heterogeneous elasticity. The first image the input Young's modulus distribution in grayscale, where brighter regions indicate higher stiffness values. The second image is the reference stress field. Subsequent show the LAMG predicted meshes for decreasing error tolerances ($\epsilon$). The method automatically identifies and refines the internal material interfaces (spherical inclusions) where stress gradients are high, demonstrating that the network has learned to respond to physical solution features rather than relying solely on domain geometry.
        }
        \label{fig:qual_elasticity_hetero}

\end{figure*}

 We show 1 example where we create Young's modulus with 12 random spherical constant field with amptitude in range $[5,30]$ with radius $[0.5,1.0]$ and uniform Poisson's ratio at $0.3$. inclusions with contrasting stiffness to create sharp internal stress gradients. We train the Relative Method ($h_{\theta R}$) here to capture the complex error landscape. We show the energy norm against runtime plot in Figure \ref{fig:hetro} and the visualization of the stress field in Figure \ref{fig:qual_elasticity_hetero}. We see the energy norm drops in magnitude when reduce the input error tolerance. And in the visualization we see that clear increasing density of element when stress changes faster and other region remain less dense. Also, in this example we are not seeing stress concentration only at sharp corners and boundaries.

We generate a test case by modulating the Young's modulus $E$ via 12 random spherical inclusions with radii in the range $[0.5, 1.0]$. These inclusions are assigned random Young's modulus amplitudes in $[5, 30]$ against the background material, with a uniform Poisson's ratio of $\nu=0.3$. 
To ensure a rigorous evaluation, we train the Relative Method ($h_{\text{relative}}$) on the same dataset used for the direct elasticity tasks, excluding this geometry. Since the material properties across the dataset are randomly generated, the network must learn to capture the complex error landscape induced by internal interfaces without simply memorizing a specific configuration.

Figure \ref{fig:hetro} presents the Pareto frontier of energy norm error versus runtime. We observe a reduction in error as the runtime increases (corresponding to tighter input tolerances), confirming the method's stability. 
Qualitative results are shown in Figure \ref{fig:qual_elasticity_hetero}. As the input tolerance $\epsilon$ decreases, the method correctly increases the mesh density in regions where the stress field changes rapidly—specifically around the spherical inclusions—while keeping the mesh coarse in the homogeneous background regions. Notably, the refinement is no-longer only appears at corners and boundaries in contrast to homogeneous case.

\section{Analysis}\label{sec:discussion}
While Sections~\ref{sec:exp} and~\ref{sec:otherexp} presented results of our method, this section presents an exploratory analysis of our design considerations. For example how relative LAMG may be adapted for use with WoS, a discussion on the benefit of adaptive sampling, analyzing the effects of the size of the network used and the performance of LAMG when limited to specific domains (but subject to unknown boundary conditions).

\subsection{Efficiency versus Uniform Refinement} 
\begin{figure}[htbp]
	\centering
	\setlength{\tabcolsep}{2pt} 
	\renewcommand{\arraystretch}{0.8} 
	\begin{tabular}{@{}c@{}c@{}}
		\includegraphics[width=0.45\columnwidth]{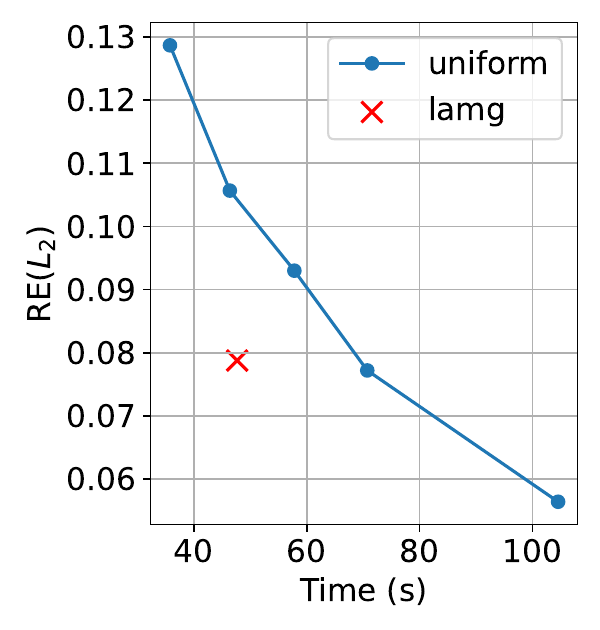} & 
		\includegraphics[width=0.45\columnwidth]{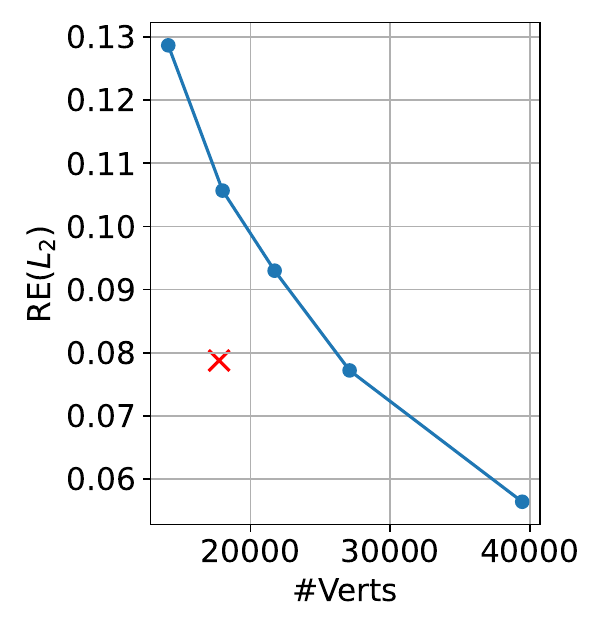}     \\
	\end{tabular}
    \vspace{-2mm}
	\caption{\label{uniform} Although large uniform meshes can achieve lower error than LAMG (with $\eta=1$), the computation time is significantly larger. The plots show RE ($L_2$) vs time (left) and number of vertices (right) for uniform meshing compared to LAMG.}
    \vspace{-3mm}
\end{figure}

Uniform sampling of the domain is a reasonable option when the boundary is known but not the boundary conditions or source terms. Given the latter, uniform meshes are usually sub-optimal with respect to computation time. For the same computation time, uniform meshes usually result in larger error than adaptive meshes that consider boundary conditions. For example, Figure~\ref{uniform} plots the RE (L2) vs time as well as the number of vertices for uniform meshing of a mesh in Thingi10k (moldgenerator). The result of applying LAMG to the same boundary mesh is shown with a cross. Although larger uniform meshes may be generated with lower error, the computational time grows rapidly. In cases where the domain has volumetrically thin features or high (surface) curvature, this gap widens further. The plot is reassuring that LAMG's behavior is consistent with classical adaptive methods such as AMR. 

\subsection{Relative LAMG with WoS}\label{sec:wosvsFEM_relative}
Recall that relative method requires element-wise error estimates, whereas grid-free methods like Walk-on-Spheres (WoS) provide sparse pointwise solution values $u(\mathbf{x})$. To bridge this gap, we trained $h_{\theta2}$ to approximate the local error feature $\phi_i$ directly from the solution value $u_i$. Training pairs are generated via coarse FEM solves, pairing WoS solution values with computed ZZ-error estimates on element centers. During inference, we use this network to predict the estimated error from sparse input and feed it to the size field prediction. 

\begin{figure}[htbp!]
	\centering
	\renewcommand{\arraystretch}{0.8} 
	\begin{tabular}{@{}c@{}c@{}}
		\multicolumn{2}{c}{\includegraphics[width=0.8\columnwidth]{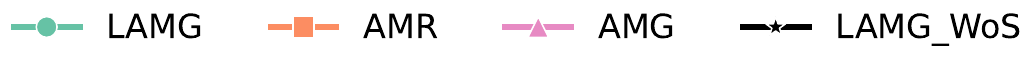}}\\
		\includegraphics[width=0.50\columnwidth]{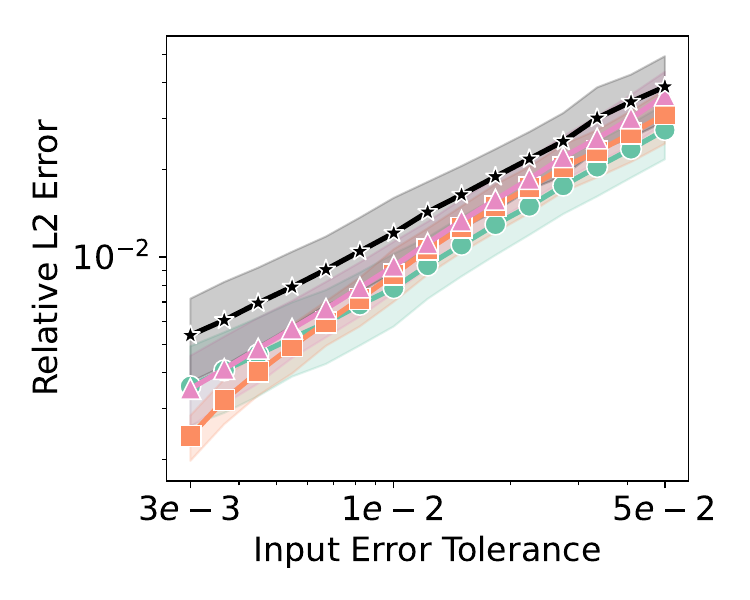} & 
		\includegraphics[width=0.50\columnwidth]{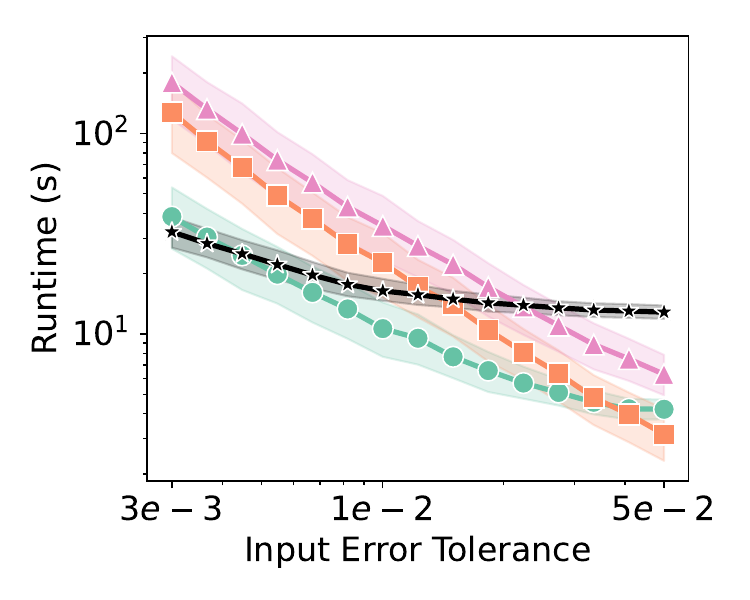}     \\
	\end{tabular}
    \vspace{-1mm}
    \caption{caption
}
\label{fig:relative:PoissonWOS}
\vspace{-4mm}
	\caption{\label{fig:wos_comparison} Figure \ref{fig:wos_comparison}: Generalization to grid-free inputs (LAMG-WoS). Comparison of (Left) Relative L2 Error and (Right) Runtime across varying input error tolerances. The WoS-driven approach (black dashed line) successfully recovers the linear convergence rate of standard FEM-based methods, albeit with a higher error offset due to the approximation of the sizing field from sparse samples. Crucially, the runtime for LAMG-WoS remains stable across all tolerances, contrasting with the steep cost increase observed in iterative AMR for high-fidelity targets.}
\end{figure}

Figure \ref{fig:wos_comparison} demonstrates the feasibility of this "LAMG-WoS" approach. While the method exhibits a higher L2 error compared to FEM-based baselines due to the approximation variance, it successfully recovers the expected linear convergence behavior with respect to the input tolerance. Furthermore, the runtime remains stable regardless of target fidelity, confirming the versatility of our approach to operate in mesh-free settings where traditional element-wise error estimates are not naturally available.

\subsection{Network Size} \label{modelspecs}
We experimented with several network models for the steady-state heat problem trained on various numbers of shapes, each time on several PDEs (varying boundary conditions). The specifications for all five models are summarized in Table~\ref{modelspecs}.

\begin{table}[h!] 
    \centering

    \renewcommand{\arraystretch}{0.9} 
    \setlength{\tabcolsep}{4pt}      

    \noindent\begin{tabular}{cccccc}
        \hline
        & \textbf{MLP}  & \textbf{GNN}  & \textbf{Learnables} & \textbf{\#Train Shapes} & \textbf{\#PDEs} \\
        \hline
        $h_{\theta 1}$ & 4 & 2 & 1393 & 1   & 1000 \\
        $h_{\theta 2}$ & 4 & 2 & 1393 & 5   & 1000 \\
        $h_{\theta 3}$ & 4 & 2 & 1393 & 100 & 1000 \\
        $h_{\theta 4}$ & 4 & 2 & 5345 & 500 & 5000 \\
        $h_{\theta 5}$ & 6 & 3 & 22491 & 2500 & 25000 \\
        \hline
    \end{tabular}
    \caption{Specifications for all trained network models.\label{modelspecs}}
    \vspace{-4mm}
\end{table}

\begin{figure}[ht]
	\centering
	\setlength{\tabcolsep}{2pt} 
	\renewcommand{\arraystretch}{0.8} 
	\begin{tabular}{@{}c@{}c@{}}
		\multicolumn{2}{c}{\includegraphics[width=0.70\columnwidth]{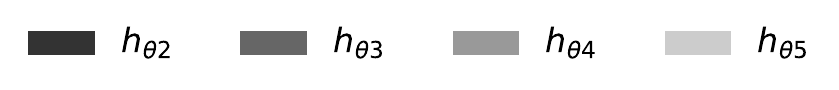}}\\
		\includegraphics[width=0.45\columnwidth]{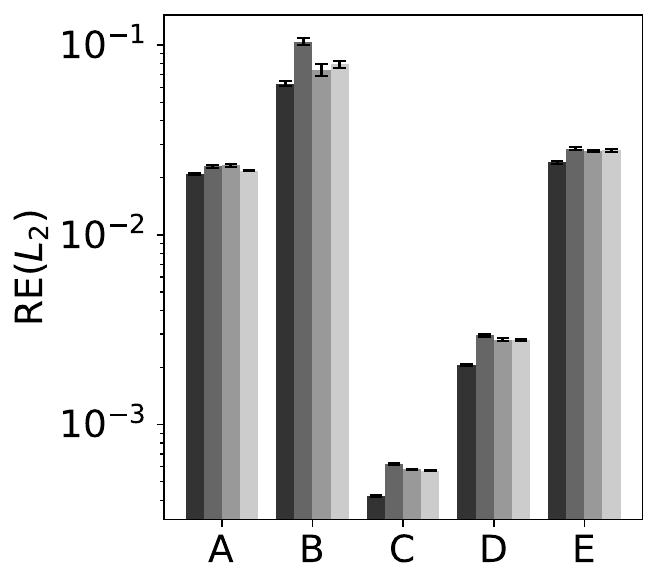} &
		\includegraphics[width=0.45\columnwidth]{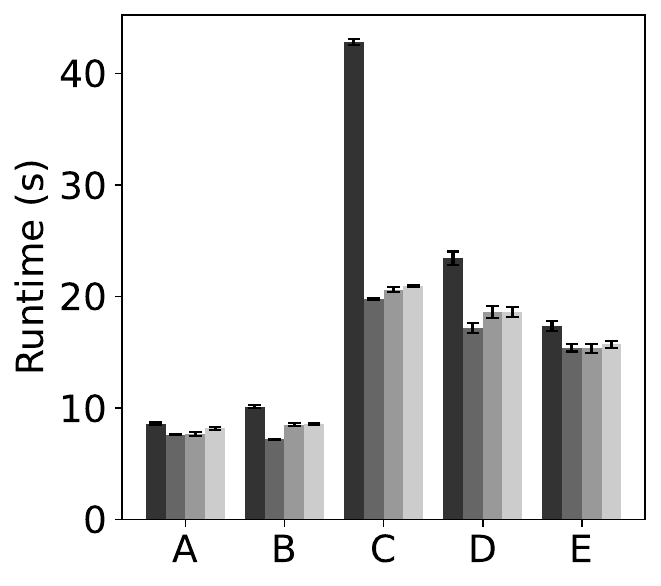} \\
	\end{tabular}
    \vspace{-1mm}
	\caption{The plots show relative error (left) and runtime (right) for four different models on a small set of meshes (A, B, C, D, E) : cube, pillar, turbine, sausage and moldGenerator. While error is similar across all models, training on a larger dataset ($h_{\theta3}$ and $h_{\theta4}$) leads to a significant runtime improvement over the model trained on only a few shapes ($h_{\theta2}$).
    }
	\label{fig:crossmodel}
\end{figure}

 Figure~\ref{fig:crossmodel} compares four of our trained models ($h_{\theta2}$ through $h_{\theta5}$) on a set of test geometries. None of the PDE problems were seen during training. 
The results show that while all models achieve a similar level of accuracy, the scale of the training data significantly impacts runtime efficiency. The model trained on only a few shapes ($h_{\theta2}$) is less efficient, as it conservatively predicts finer meshes on unseen geometries. In contrast, the models trained on larger, more diverse datasets ($h_{\theta3}$ through $h_{\theta5}$) are consistently faster. Importantly, the performance saturates with our primary model ($h_{\theta4}$), indicating that indiscriminately increasing network size ($h_{\theta5}$) yields diminishing returns. This confirms that a moderately sized network trained on a sufficiently diverse dataset provides the best trade-off.

\subsection{Overfitting to known domains}
We analyzed LAMG's application to cases where the network needs to generalize across boundary conditions and source terms for a fixed set of geometries (say in a video game where the assets are known).  We first test generalization to new boundary conditions on a single cube, where the model is trained and tested on the same geometry but evaluated on unseen, randomly generated boundary conditions. Next, we test generalization to new shapes using a small set of 7 diverse geometries; the model is trained on 5 of these and tested on the full set, including the 2 unseen shapes. The test boundary conditions and source terms for all 7 are unseen during training.

\begin{figure}[htbp!]
	\centering
	\setlength{\tabcolsep}{2pt} 
	\renewcommand{\arraystretch}{0.8} 
	\begin{tabular}{@{}c@{}c@{}}
		\multicolumn{2}{c}{\includegraphics[width=0.90\columnwidth]{figs/4_1_3legends.pdf}}\\
		\includegraphics[width=0.50\columnwidth]{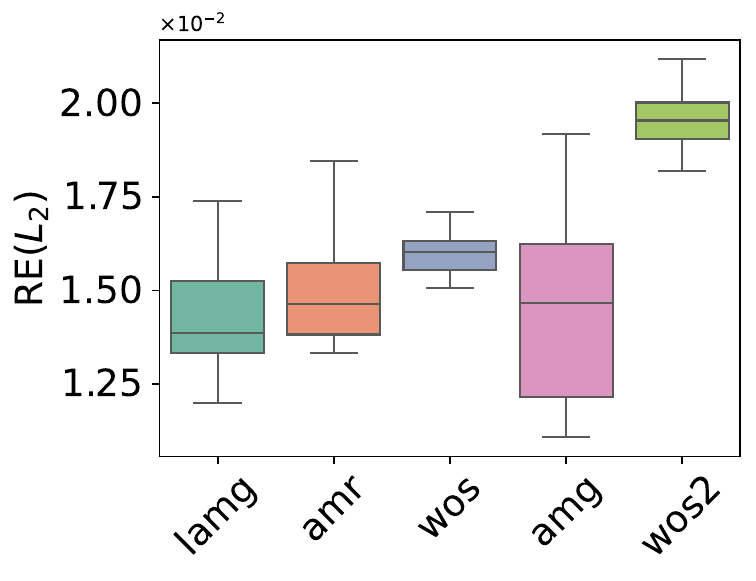} & 
		\includegraphics[width=0.50\columnwidth]{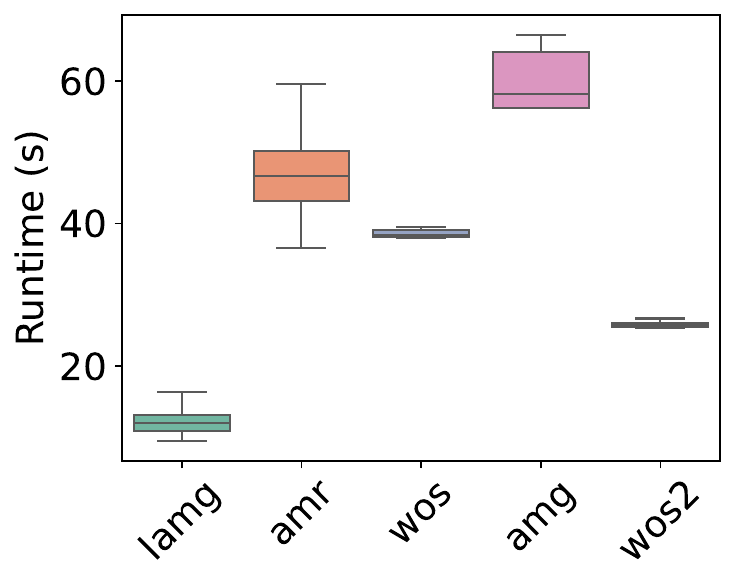}     \\
        \includegraphics[width=0.50\columnwidth]{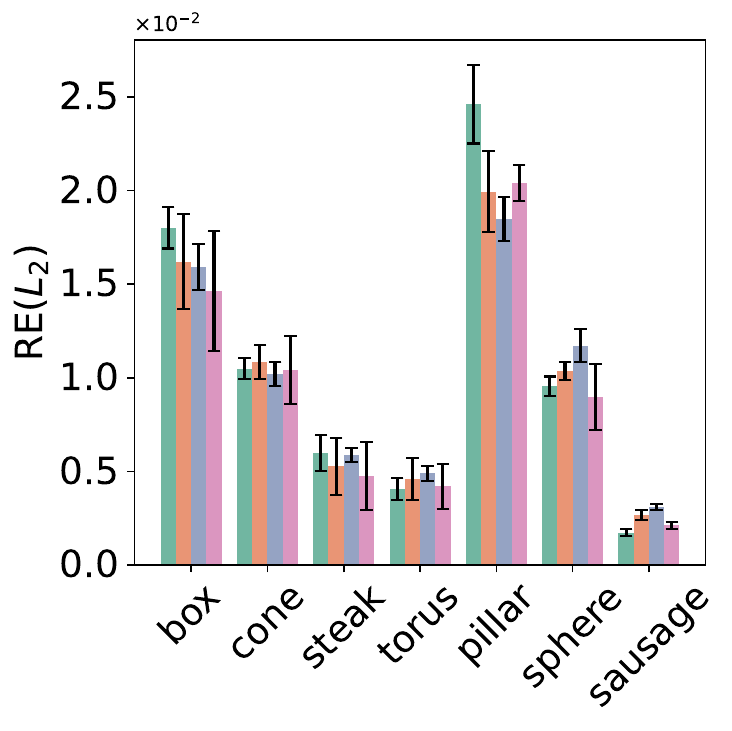} & 
		\includegraphics[width=0.50\columnwidth]{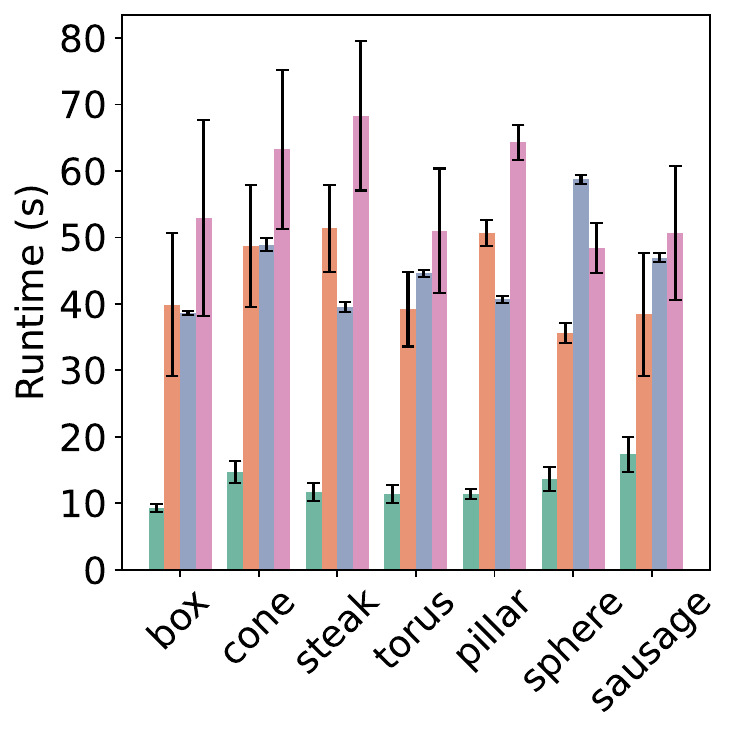}     \\
	\end{tabular}
    \vspace{-1mm}
    \caption{Performance comparison for the Direct Poisson solving, plotting relative error (top row) and execution time (bottom row) for our method (LAMG) against AMR, AMG, and WoS baselines. top row: Generalization to unseen boundary conditions on a single cube (model $h_{\theta1}$). bottom row: Performance on a small set including seen and unseen geometries (model $h_{\theta2}$). Across all scenarios, the plots show that for a similar error, LAMG is consistently faster.
}
\label{fig:direct:Poisson_cube_smallset}
\vspace{-4mm}
\end{figure}

On the single cube using model $h_{\theta1}$ (Figure~\ref{fig:direct:Poisson_cube_smallset} (top row)), we use box plots to summarize performance. The results show LAMG achieves an error comparable to AMR, AMG and WoS, while being lower than WoS2 (WoS: $n=18\text{k}, m=5000$; WoS2: $n=12\text{k}, m=5000$, where $n$ is number of query points and $m$ is number of walks). Critically, LAMG demonstrates a 2$\times$ to 4$\times$ speedup over the other methods.

This trend continues on the small set of geometries evaluated with model $h_{\theta2}$ (Figure~\ref{fig:direct:Poisson_cube_smallset} (bottom row)), which includes five shapes from the training distribution alongside two completely unseen test shapes ('box' and 'Sausage'). While all methods exhibit similar error profiles, the runtime plots confirm a consistent 2$\times$ to 4$\times$ speed advantage for our approach. Importantly, this efficiency gain holds for both the seen and unseen geometries.

\subsection{Runtime analysis}

\begin{figure}[htbp]
    \centering
    \includegraphics[width=1.0\columnwidth]{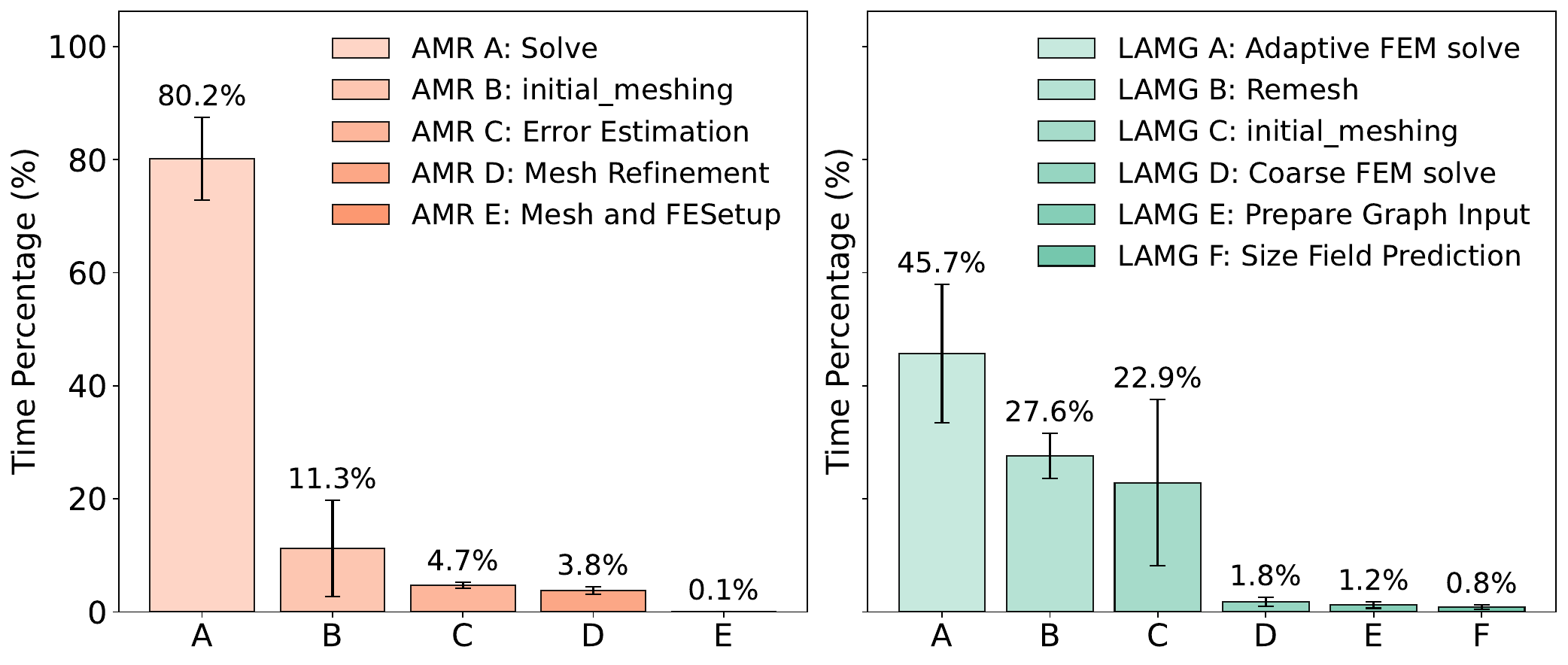} 
    \vspace{-2mm}
    \caption{Runtime breakdown comparison (mean percentage $\pm$ standard deviation).
(Left) The standard AMR workflow is dominated by iterative assembly of the FEM and its solving (80.2\% with average of 12 iterations), including the accumulating cost of integrating complex source terms at each refinement iteration. (Right) In contrast, LAMG incurs negligible overhead for Size Field Prediction (0.8\%). By predicting the target mesh in a single shot, LAMG eliminates redundant assembly steps, shifting the computational profile such that the final Adaptive FEM solve (45.7\%) is the primary cost.}
    \label{fig:breakdown}
    \vspace{-4mm}
\end{figure}

We profiled the computational cost in Figure \ref{fig:breakdown}, which reveals that the efficiency gains of LAMG are primarily the avoidance of iterative assemblies and solves of the FEM linear system. for AMR, about 80\% of total runtime is spent on this with an average of 12 iterations (with meshes of increasing size). This time includes the cost of repeatedly evaluating and integrating the mixtures-of-Gaussians source terms (as described earlier) over the domain at refinement iterations. In contrast, LAMG predicts the target density in a "one-shot" inference step (costing only $0.8\%$ of runtime), meaning it incurs this expensive fine-grid assembly cost only once for the final mesh.

\subsection{Limitations and future work}

Although LAMG offers modest gains in efficiency, it also imposes some constraints. For example, even though the architecture is largely kept the same, $h_\theta$ learns a PDE-specific mapping and therefore needs to be retrained for linear elastic deformation. The relative method is inherently suited to using coarse FEM inputs. Although we learned a mapping from degrees of freedom to error for WoS, this introduces some variance. For some problems, such linear elasticity Monte Carlo methods are still actively being researched and so we only used FEM. As future work, we plan to investigate the suitability of LAMG also for time-dependent systems governed by parabolic equations, as there is no obvious reason why our approach will not extend.

\section{Conclusion}
We have presented Learned Adaptive Mesh Generation (LAMG), a novel framework that accelerates the solution of elliptic PDEs by replacing the expensive, iterative loop of traditional refinement with a direct "one-shot" prediction of optimal sizing fields. By learning to map coarse solution proxies derived from either low-resolution FEM or sparse Walk-on-Sphere (WoS) samples to high-fidelity discretization densities, our method supports diverse application needs through two distinct variants. The direct method, controlled by a scaling factor, allows users to strictly manage the computational budget (element count), while the relative method, controlled by an input tolerance, is designed for applications where achieving a specific error bound is paramount. Our experiments on steady-state heat and linear elasticity demonstrate that LAMG generalizes across diverse geometries and boundary conditions, offering robust error control and significant computational efficiency. Notably, for small error tolerances, our method achieves a $2\times$ to $4\times$ speedup compared to standard baselines. Ultimately, this approach bridges the gap between the robustness of adaptive meshing and the efficiency required for interactive design, providing rapid, reliable feedback on physical behavior.

\bibliographystyle{ACM-Reference-Format}
\bibliography{tex_S26/lamg26}


\end{document}